\documentclass[11pt]{article}
\usepackage{etex}
\usepackage{amscd,amsmath,amssymb,amsfonts,xspace,mathrsfs,mathtools,amsthm}
\usepackage{silence}
\usepackage{xr-hyper}
\usepackage[dvipsnames]{xcolor}
\usepackage[utf8]{inputenc}
\usepackage{bbm}
\usepackage{graphicx}
\usepackage{tabu} 
\usepackage[color,matrix,arrow]{xy}
\usepackage{wasysym} 
\usepackage{stmaryrd}
\usepackage[shortlabels]{enumitem}
\usepackage{array,amsmath}
\usepackage{slashed}
\usepackage{tensor}
\usepackage{caption}
\usepackage{subcaption}
\usepackage{enumitem}
\usepackage{cellspace, hhline}
\usepackage{multirow}
\usepackage{svg}
\usepackage{jheppub}
\usepackage{booktabs, cellspace, hhline}
\usepackage{tikz}
\usepackage{tikz-cd}
\usepackage{diagbox}
\usetikzlibrary{positioning}
\usetikzlibrary{calc}
\usetikzlibrary{decorations.pathreplacing,calligraphy}

\usepackage{xstring}
\usetikzlibrary{decorations.pathmorphing} 
\usetikzlibrary{decorations.markings} 
\usetikzlibrary{arrows} 
\usetikzlibrary{shapes} 
\usetikzlibrary{matrix} 
\usetikzlibrary{positioning} 
\usepackage[english]{babel} 
\usepackage[autostyle]{csquotes}
\usetikzlibrary{positioning,fit}
\usepackage[dvipsnames]{xcolor}

\tikzset{->-/.style={decoration={
  markings,
  mark=at position #1 with {\arrow[scale=1.7]{>}}},postaction={decorate}}}

\newcommand{\IW}{\textbf{I}_{\text{W}}}
\DeclarePairedDelimiter{\bra}{\langle}{\rvert}%
\DeclarePairedDelimiter{\ket}{\lvert}{\rangle}%
\DeclarePairedDelimiterX\braket[2]{\langle}{\rangle}{#1\delimsize\vert\mathopen{}#2}%
\DeclareMathOperator{\sign}{sign}

\renewcommand{\Gamma}{\varGamma}
\renewcommand{\epsilon}{\varepsilon}
\renewcommand{\bar}{\overline}
\renewcommand{\hat}{\widehat}
\renewcommand{\leq}{\leqslant}
\renewcommand{\geq}{\geqslant}

\renewcommand{\b}{\bar }

\newcommand{\res}{\text{Res}}

\newcommand{\be}{\begin{equation}} 
\newcommand{\ee}{\end{equation}} 
\newcommand{\bea}{\begin{equation} \begin{aligned}}
\newcommand{\eea}{\end{aligned} \end{equation}}
\newcommand{\bes}{\begin{equation*}}
\newcommand{\ees}{\end{equation*}}

\newcommand{\h}{\hat}
\newcommand{\ov}{\over}

\newcommand{\nn}{\nonumber}

\newcommand{\bH}{\mathbb{H}}
\newcommand{\mat}[1]{\begin{pmatrix} #1 \end{pmatrix}}

\newcommand{\CA}{\mathcal{A}}

\newcommand{\CE}{\mathcal{E}}  
\newcommand{\CF}{\mathcal{F}} 
 
\newcommand{\CH}{\mathcal{H}}

\newcommand{\CK}{\mathcal{K}}

\newcommand{\CN}{\mathcal{N}}
\newcommand{\CO}{\mathcal{O}} 

\newcommand{\CQ}{\mathcal{Q}}

\newcommand{\CS}{\mathcal{S}}
\newcommand{\CT}{\mathcal{T}} 

\newcommand{\CV}{\mathcal{V}}

\DeclareSymbolFont{greekletters}{OML}{cmm}{m}{it}
\DeclareMathSymbol{\lunateepsilon}{\mathord}{greekletters}{"0F}
\newcommand{\eps}{\lunateepsilon}

\newcommand{\Tr}{\text{Tr}}

\renewcommand{\t}{\widetilde }
\renewcommand{\d}{\partial }
\newcommand{\Z}{\mathbb{Z}}
\newcommand{\C}{\mathbb{C}}

\newcommand{\R}{\mathbb{R}}

\newcommand{\half}{{\frac{1}{2}}}

\newcommand{\FR}{\mathfrak{R}}
\newcommand{\Fg}{\mathfrak{g}}

\newcommand{\schur}{\mathfrak{s}} 
\newcommand{\by}{\boldsymbol{y}}
 \newcommand{\zinf}{\overset{0}{\infty}}

\begin{document}

\baselineskip=18pt  
\numberwithin{equation}{section}  
\allowdisplaybreaks  


%
%


\thispagestyle{empty}
\setcounter{tocdepth}{2}

\vspace*{0.8cm} 
\begin{center}
{\huge  
Dualities and trialities in $\CN=2$ \\ \medskip 
 supersymmetric gauged quantum mechanics
}

 \vspace*{1.5cm}
Cyril Closset and  James Wynne

 \vspace*{0.7cm} 

 {   School of Mathematics, University of Birmingham,\\ 
Watson Building, Edgbaston, Birmingham B15 2TT, UK}\\
 \end{center}

\vspace*{0.5cm}

\noindent We study new Seiberg-like dualities for 1d $\mathcal{N}=2$ supersymmetric gauge theories -- that is, supersymmetric gauged quantum mechanics -- with unitary gauge group and (anti)fundamental matter in chiral and fermi multiplets, and with non-zero Fayet--Iliopoulos parameter. Similarly to its higher-dimensional analogues, this 1d Seiberg duality is an infrared duality: the supersymmetric ground states of the dual gauge theories match exactly. We provide strong evidence for the dualities, including the matching of the flavoured Witten indices, a Higgs-branch derivation in terms of dual Grassmannian manifolds, and a detailed study of the Coulomb-branch ground states in the abelian case. We study how the supersymmetric ground states, in either dual description, depend on the sign of the Fayet--Iliopoulos parameter, and we explore the corresponding wall-crossing phenomenon. For some special values of the discrete parameters defining the unitary gauge theory, the dualities, combined with trivial wall-crossing, enhances to a triality. This includes, as a special case, the dimensional reduction to 1d of the 2d $\mathcal{N}=(0,2)$ Gadde--Gukov--Putrov triality.

\medskip 
\noindent \today
\newpage

\tableofcontents



\section{Introduction}

Infrared dualities are central to our understanding of supersymmetric gauge theories. Two distinct quantum field theories (QFT) are said to be infrared dual if they are isomorphic in the low-energy limit~\cite{Seiberg:1994pq}. This is to be contrasted with exact dualities, where two theories that might look distinct semi-classically are actually exactly the same quantum system at all scales --- the prototypical example being the Montonen--Olive duality of 4d $\CN=4$ SYM~\cite{Montonen:1977sn, Sen:1994yi}. To avoid calling any effective field theory description an `infrared dual theory',  one generally reserves the `duality' monicker to those pairs of theories which are both well-defined in the UV ({\it i.e.} at least asymptotically free).  Those pairs are often known as Seiberg-like dual theories, since  Seiberg duality in 4d $\CN=1$ SQCD is the prototypical infrared duality~\cite{Seiberg:1994pq}. By now, there are many well-established examples of Seiberg-like dualities between supersymmetric field theories in four, three and two space-time dimensions (see~\cite{Intriligator:1995ne, Kutasov:1995np, Aharony:1997gp, Giveon:2008zn,  Hori:2006dk} for a very partial list), and even between non-supersymmetric 3d gauge theories~\cite{Aharony:2012nh, Seiberg:2016gmd,Karch:2016sxi}.  

In this paper, we introduce and study new infrared dualities in 1d $\CN=2$ supersymmetric gauge theories --- that is, in gauged $\CN=2$ supersymmetric quantum mechanics (SQM). In the context of quantum mechanics, an infrared duality is simply the claim that two distinct systems have isomorphic ground states. Here, the low-energy limit singles out the supersymmetric ground states, which have zero energy.  For this notion of an infrared duality in SQM to really make sense, it is crucial that the SQM be gapped --- there is a finite energy gap between the zero-energy states and any other excited state.%
\footnote{It would be very exciting to understand what is the proper notion of an infrared duality in a gapless SQM, but this would go much beyond the scope of the present work.}

We will study gauged $\CN=2$ SQM with a unitary gauge group $U(N_c)$ coupled to fundamental and antifundamental matter in chiral and fermi multiplets, which we call 1d SQCD. Note that the $\CN=2$ supersymmetry we use here is the `holomorphic' realisation of two supercharges, to be distinguished from the `real' version discussed {\it e.g.} in~\cite{Witten:1982im}.%
\footnote{If we were to uplift to a 2d supersymmetry algebra, this would correspond to considering 2d $\CN=(0,2)$ instead of $\CN=(1,1)$ supersymmetry.} 
One-dimensional gauge fields are non-dynamical and only impose some version of Gauss's law --- restricting the physical states to be gauge-invariant. The $\CN=2$ vector multiplet also include an auxiliary field $D$ which couples to the chiral multiplets and imposes standard $D$-term constraints, so that the semi-classical Higgs branch is naturally a K\"ahler manifold. 

Importantly, we will turn on a Fayet--Ilopoulos (FI) parameter~\cite{Fayet:1974jb}, denoted by $\zeta$, which is crucial to obtain a gapped system. We find two basic dualities depending on the sign of the FI parameter, which we call a right mutation  for $\zeta>0$:
\be
U(N_c)_{q_c}~, \, \zeta>0 \qquad  \qquad \longleftrightarrow \qquad U(n_1-N_c)_{-q_c}~, \; \zeta'<0
\ee
and a left mutaton for $\zeta<0$:
\be
U(N_c)_{q_c}~, \, \zeta< 0 \qquad  \qquad \longleftrightarrow \qquad U(n_2-N_c)_{-q_c}~, \; \zeta''>0
\ee
The two mutation operations are inverse of each other. 
Here $n_1$ is the number of fundamental chiral multiplets, $n_2$ is the number of antifundamental chiral multiplets, while $q_c$ is a 1d Chern--Simons level --- also known as background charge. Our 1d $\CN=2$ SQCD also contains $n_3$ fundamental fermi multiplets. Finally, we will also introduce additional gauge-neutral fields $\Gamma$ coupled to the gauged sector so that the theory has normalisable ground states --- that theory with those additional coupling we call $\Gamma$-SQCD.

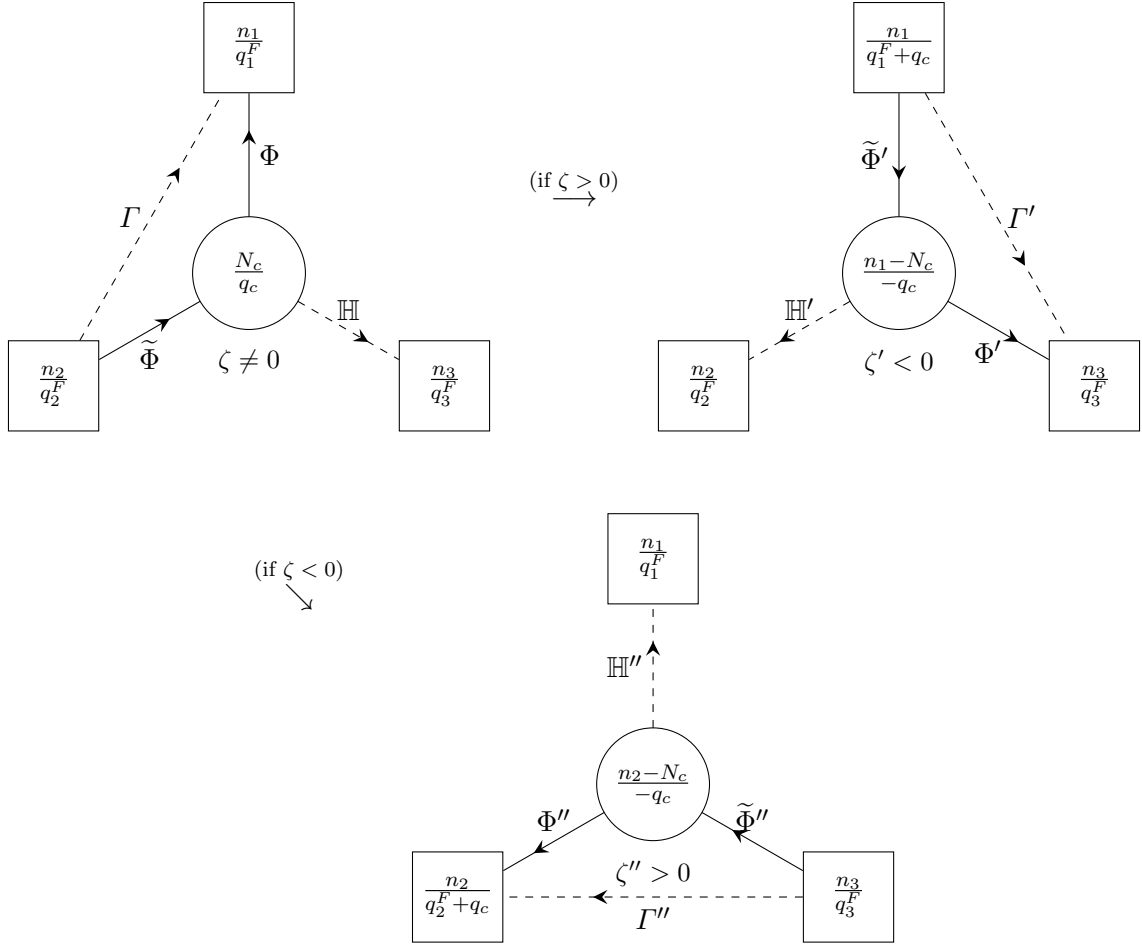
\begin{figure}
    \centering
\begin{tikzpicture}[>=stealth,scale=0.75]
 \def\SquareSize{1.2}  
    \def\CircleRadius{1}
    \def\NodeRadius{4}
    \draw (0,0) circle (\CircleRadius cm);

    \node[] (Nc) at (0,0) {${N_c\ov q_c}$};
    \node[draw, rectangle, shape aspect=1, minimum width= \SquareSize cm, minimum height=\SquareSize cm] (N1) at (90:\NodeRadius cm) {${n_1\ov q_1^F}$};
    \node[draw, rectangle, shape aspect=1, minimum width=\SquareSize cm, minimum height=\SquareSize cm] (N3) at (-30:\NodeRadius cm) {${n_3\ov q_3^F}$};
    \node[draw, rectangle, shape aspect=1, minimum width=\SquareSize cm, minimum height=\SquareSize cm] (N2) at (210:\NodeRadius cm) {${n_2\ov q_2^F}$};
    \node[below =0.48cm of Nc] {\small $\zeta \neq  0$};

    \draw[->-=0.7] (90:\CircleRadius cm) --node[midway, right] {$\Phi$} (N1);
    \draw[->-=0.7] (N2) -- node[midway, below]{${\t\Phi}$} (210:\CircleRadius cm);

    \draw[->-=0.7, dashed] (-30:\CircleRadius cm) -- node[midway, above]{$\bH$} (N3);
    \draw[->-=0.7, dashed] (N2) --node[midway,left]{$\Gamma$} (N1);

\end{tikzpicture}
\quad
\raisebox{3cm}{$\overset{\text{(if $\zeta>0$)}}{\longrightarrow}$}
\quad
\begin{tikzpicture}[>=stealth,scale=0.75]
 \def\SquareSize{1.2}  
    \def\CircleRadius{1}
    \def\NodeRadius{4}
    \draw (0,0) circle (\CircleRadius cm);

    \node[] (Nc) at (0,0) {${n_1-N_c\ov -q_c}$};
    \node[draw, rectangle, shape aspect=1, minimum width= \SquareSize cm, minimum height=\SquareSize cm] (N1) at (90:\NodeRadius cm) {${n_1\ov q_1^F+q_c}$};
    \node[draw, rectangle, shape aspect=1, minimum width=\SquareSize cm, minimum height=\SquareSize cm] (N3) at (-30:\NodeRadius cm) {${n_3\ov q_3^F}$};
    \node[draw, rectangle, shape aspect=1, minimum width=\SquareSize cm, minimum height=\SquareSize cm] (N2) at (210:\NodeRadius cm) {${n_2\ov q_2^F}$};
    \node[below =0.48cm of Nc] {\small $\zeta' < 0$};

    \draw[->-=0.7] (-30:\CircleRadius cm) --node[midway, below] {$\Phi'\;\;\,$} (N3);
    \draw[->-=0.7] (N1) -- node[midway, left]{${\t\Phi}' $} (90:\CircleRadius cm);

    \draw[->-=0.7, dashed] (210:\CircleRadius cm) -- node[midway, above]{$\bH'$} (N2);
    \draw[->-=0.7, dashed] (N1) --node[midway,right]{$\Gamma'$} (N3);

\end{tikzpicture}
\\
\phantom{empty}
\\
\qquad\qquad\qquad
\quad
 
\raisebox{4.5cm}{$\overset{\text{(if $\zeta<0$)}}{\searrow}$}
\qquad
 \begin{tikzpicture}[>=stealth,scale=0.75]
 \def\SquareSize{1.2}  
    \def\CircleRadius{1}
    \def\NodeRadius{4}
    \draw (0,0) circle (\CircleRadius cm);

    \node[] (Nc) at (0,0) {${n_2-N_c\ov -q_c}$};
    \node[draw, rectangle, shape aspect=1, minimum width= \SquareSize cm, minimum height=\SquareSize cm] (N1) at (90:\NodeRadius cm) {${n_1\ov q_1^F}$};
    \node[draw, rectangle, shape aspect=1, minimum width=\SquareSize cm, minimum height=\SquareSize cm] (N3) at (-30:\NodeRadius cm) {${n_3\ov q_3^F}$};
    \node[draw, rectangle, shape aspect=1, minimum width=\SquareSize cm, minimum height=\SquareSize cm] (N2) at (210:\NodeRadius cm) {${n_2\ov q_2^F+q_c}$};
    \node[below =0.48cm of Nc] {\small $\zeta'' > 0$};

    \draw[->-=0.7] (210:\CircleRadius cm) --node[midway, above] {$\Phi''$} (N2);
    \draw[->-=0.7] (N3) -- node[midway, above]{${\t\Phi}'' $} (-30:\CircleRadius cm);

    \draw[->-=0.7, dashed] (90:\CircleRadius cm) -- node[midway, left]{$\bH''$} (N1);
    \draw[->-=0.7, dashed] (N3) --node[midway,below]{$\Gamma''$} (N2);

\end{tikzpicture}
    \caption{ Right and left mutations for unitary $\Gamma$-SQCD, with the notation ${n\ov q}\equiv U(n)_q$. Round  and square nodes denote gauge and flavour symmetries, respectively. Ordinary arrows denote bifundamental chiral multiplets and dashed arrows denote bifundamental fermi multiplets. Note the shift of the flavour 1d CS levels $q_I^F$ under the duality. Mutation dualities also include an overall shift in the fermion number which we will explain in the main text. }
    \label{fig:right left mutations GammaSQCD}
\end{figure}

The $\Gamma$-SQCD theory and its infrared-duals are conveniently captured by quiver diagrams, as shown in figure~\ref{fig:right left mutations GammaSQCD}. Indeed, one can think of embedding our SQCD theory into a larger 1d $\CN=2$ unitary quiver theory, in which case our duality can be understood as a 1d instance of a mutation for such graded quivers~\cite{Franco:2017lpa, Closset:2018axq}.%
\footnote{Our graded quiver (with grading $m=2$ in the notation of~\protect\cite{Franco:2017lpa}) contain two types of arrows, corresponding to the bifundamental matter fields in chiral and fermis multiplets, respectively.} Our approach goes beyond the algebraic definition of a graded quiver mutation, however, and takes seriously the duality as a statement about the supersymmetric ground states.%
\footnote{Note also that the more commonly discussed (ordinary, $1$-graded) mutations of 1d $\CN=4$ quivers~\protect\cite{Denef:2002ru, Berenstein:2002fi, Manschot:2013dua} are also 1d Seiberg-like dualities, which from the $\CN=2$ SQM perspective involve additional adjoint $\CN=2$ chiral multiplets, and therefore do not fit into the class of dualities we are discussing here. (It is interesting to discuss these 1d $\CN=4$ dualities in a similar language, however; we hope to explain this elsewhere.) }

The simplest observable in any $\CN=2$ quantum mechanics is the (flavoured) Witten index~\cite{Witten:1981nf, Hori:2014tda}, which is the trace over all states counted with signs:
\be\label{IW intro}
\IW = \Tr_{\mathscr{H}}\left((-1)^{\rm F} q^{\bf H} y_F^{{\bf J}^F} \right) =  \Tr_{\mathscr{H}_0}\left((-1)^{\rm F}  y_F^{{\bf J}^F} \right)~. 
\ee
In the second equality, we used the assumption that the Hamiltonian ${\bf H}$ is gapped, so that only the supersymmetric ground states $\ket{\Psi} \in \mathscr{H}_0$ contribute --- therefore, the Witten index of two infrared-dual models must agree.  Here, ${\bf J}^F$ denotes flavour symmetries ({\it i.e.} conserved charges that commute with the supercharges), and $y_F$ denotes the associated fugacities. Note that turning on generic fugacities $y_F\neq 1$ in SQCD leads to a gapped system even in the absence of the $\Gamma$ fields, while the ground states of $\Gamma$-SQCD are well-defined even in the limit $y_F=1$, in which case $\IW$ is an integer --- the `proper' Witten index~\cite{Witten:1981nf}. For large classes of 1d $\CN=2$  gauge theories including 1d SQCD, Hori, Kim and Yi~\cite{Hori:2014tda} derived a powerful supersymmetric localisation formula for the flavoured Witten index as a Jeffrey-Kirwan (JK) residue~\cite{jeffrey1995localization,brion1999arrangement}, similarly to related supersymmetric partition functions in higher space-time dimensions~\cite{Benini:2013xpa, Benini:2015noa, Closset:2015rna, Benini:2016hjo,Closset:2016arn}. Using the JK residue formula, we show that the flavoured Witten indices of 1d SQCD match exactly across the dualities.

In quantum field theory (QFT) with $d>1$, the main checks of conjectured dualities are performed through the matching of supersymmetric partition functions, which includes the matching of various supersymmetric indices as well~\cite{Romelsberger:2005eg,Pestun:2007rz, Dolan:2008qi, Kapustin:2009kz, Jafferis:2010un,Hama:2011ea, Kapustin:2010xq, Kapustin:2010mh, Benini:2011mf, Closset:2017zgf, Closset:2017bse, Closset:2019hyt}. In supersymmetric quantum mechanics, we should be able to understand the ground states directly, which should give us a physics proof of the duality --- in this sense, SQM gives us tractable toy models of infrared dualities in supersymmetric QFT. In this paper, we carry out this ground-state analysis using two complementary approaches, which we dub {\it Higgs branch} and {\it Coulomb branch} approaches, following standard terminology in higher dimensions. Recall that we do not have a moduli space of vacua in quantum mechanics --- the ground-state wavefunction spreads over the semi-classical target space instead. What the target space looks like depends on tunable parameters, and the Higgs and Coulomb descriptions appear in two distinct scaling limits.

\medskip
\noindent
{\bf Higgs phase geometry and dualities from Grassmannians.}  The Higgs phase is obtained in the limit $|\zeta|\rightarrow \infty$, wherein the 1d gauge group is fully Higgsed and we flow to a 1d non-linear sigma model (NLSM) onto the Higgs branch. For $\Gamma$-SQCD with $y_F=1$, the target space of the NLSM is a particular vector bundle $\CE_\pm$ over the complex Grassmannian:
\be
\CE_\pm \rightarrow X_\pm~, \qquad \text{with } \, X_+\equiv {\rm Gr}(N_c, n_1)\,  \text{ and }    \, X_-\equiv {\rm Gr}(N_c, n_2)~.
\ee
Here $\pm = \sign(\zeta)$. The detailed form of $\CE_\pm$ will be derived in the main text. The supersymmetric ground states correspond to cohomology classes of the target space~\cite{Witten:1982im, Alvarez-Gaume:1983zxc}:
\be
{\mathscr{H}_0}\cong  H^\bullet(X_\pm, \CE_\pm)~.
\ee
It turns out that these sheaf cohomology groups can be computed very explicitly using the Borel--Weil--Bott theorem~\cite{bott1957homogeneous}. This gives us, in particular, a completely explicit formula for the flavoured Witten index of $\Gamma$-SQCD as a polynomial in $y_F^\pm$. (Turning on $y_F\neq 1$ gives us the equivariant cohomology.)  We will then show, following~\cite{Jia:2014ffa}, that the mutation dualities follow directly from the Grassmannian duality, which is the geometric isomorphism between $N_c$-planes in $\C^{n_f}$ and the orthogonal $(n_f-N_c)$-planes:
\be
 {\rm Gr}(N_c, n_f)\cong   {\rm Gr}(n_f- N_c, n_f)~.
\ee
This proves our Seiberg-like dualities for $\Gamma$-SQCD in the Higgs phase. Note that this approach naturally derives subtle but important shifts of the fermion number and of background 1d Chern--Simons levels for the flavour symmetry that are induced by the mutations.

\medskip
\noindent
{\bf Coulomb phase and perturbative ground states.} The Coulomb phase is obtained in the limit $\zeta \rightarrow 0$ while keeping the Coulomb-branch scale $M_C\equiv \zeta^2 g^2$ fixed, where $g$ is the 1d gauge coupling~\cite{Hori:2014tda}. In this limit, we obtain an effective description in terms of the Coulomb-branch scalar $\sigma$ and gauginos that are partnered to the 1d gauge field by supersymmetry. In this paper we will focus on the abelian case, $N_c=1$, for simplicity. Then the Coulomb branch is a line $\R\cong \{\sigma\}$, and we generally have several perturbative sectors on the $\sigma$-line separated by singular loci at which some chiral multiplet become massless. We work out the perturbative ground states in each region, incorporating the $E$-term potential associated with the $\Gamma$ fields at first order in perturbation theory. We carefully match these perturbative ground states to the flavoured Witten, pointing out that distinct perturbative sectors on the $\sigma$-line correspond to taking specific residues of the matter integrand that appears in the JK-residue formula for the Witten index. In general, there are non-trivial cancellations in the full index between states in distinct perturbative sectors. These cancellations occur precisely betweeen pairs of states $\ket{\Psi}$, $\b\lambda\ket{\Psi'}$ that would be lifted by SQM instantons if the effective field theory had a smooth potential with several zero-energy minima~\cite{Witten:1982im}. In our model, we have a singular potential with the singularities due to massless chiral multiplets, and we therefore conjecture that  the gauge-theory high-energy completion  leads to a physically equivalent effect.

It would be interesting to extend this Coulomb-phase analysis to the non-abelian theory, in which case we have a $N_c$-dimensional Coulomb branch, and the gauginos play a more crucial role. We leave this as an interesting challenge for future work. More generally, one can presumably extend our analysis to many other infrared dualities between 1d gauge theories with other gauge groups. One systematic approach would be to derive 1d dualities from 3d $\CN=2$ infrared dualities~\cite{Aharony:1997gp, Giveon:2008zn, Benini:2011mf, Nii:2020ikd, Amariti:2021snj, Closset:2023vos, Closset:2023jiq, Benvenuti:2024seb, Benvenuti:2025huk} by compactifying the 3d theories on a sphere with a partial topological twist~\cite{Closset:2013sxa,Gadde:2015wta, Closset:2017xsc, Bullimore:2018jlp, Bullimore:2019qnt, Bullimore:2021auw, Jiang:2024ifv}; see also~\cite{Aharony:2013dha,Aharony:2016jki} for related discussions.

\medskip
\noindent
{\bf Wall-crossing and trialities.} An interesting aspect of 1d $\CN=2$ gauge theories is that they can undergo wall-crossing~\cite{Hori:2014tda, Kim:2011sc, Cordova:2014oxa}. That is, the Hilbert space of ground states can change discontinuously as the FI parameters are varied, which is efficiently detected as a jump in the value of the flavoured Witten index. Throughout, we explore some aspects of wall-crossing in our 1d SQCD models. While wall-crossing is generic, some of the most interesting 1d theories are those for which the wall-crossing is trivial, so that we can smoothly interpolate between positive and negative FI parameter $\zeta$. We identify a sufficient condition for trivial wall-crossing, which we also conjecture to be necessary (except in trivial cases). Combining our dualities with a trivial wall-crossing transition (that is, flipping the sign of $\zeta$ at no cost), we can extend dualities to trialities --- that is, a set of three theories with isomorphic supersymmetric ground states. In a very special set of cases with $q_c=0$, this triality becomes precisely the naive dimensional reduction to SQM of the Gadde--Gukov--Putrov triality of 2d $\CN=(0,2)$ supersymmetric gauge theories~\cite{Gadde:2013lxa}. More generally, dualities and trialities allow us to easily identify models which are `confining', by which we mean that they can be described by free fields. This has interesting applications to the study of line defects in 3d $\CN=2$ supersymmetric gauge theories~\cite{Hwang:2017kmk, Hosomichi:2021gxe, Closset:2023bdr, CGKSZZ25}; in particular, confinement trialities elucidate some duality relations first discussed in that context~\cite{Closset:2023bdr}.

The existence of many unitary gauge theories with trivial wall-crossing also opens a tantalising possibility,  that there exists genuine 1d fixed points on the wall. Indeed, at $\zeta=0$ the Coulomb-branch scalar potential of such theories looks like
\be
U(\sigma) ={1\ov\sigma^2}~,
\ee
which is the potential for the best-known conformal model in quantum mechanics~\cite{deAlfaro:1976vlx}. It should be possible to directly establish some version of superconformal invariance in some `infrared' scaling limit of the 1d gauge theories considered here. We leave this very important question for future work.

\medskip
\medskip
\noindent
This paper is organised as follow. In section~\ref{sec:SQM}, we review 1d $\CN=2$ supersymmetric gauge theories, setting up our conventions and reviewing some elementary facts on quantising free fields before discussing our general strategy for studying the abelian Coulomb branch. In section~\ref{sec:dualities}, we define our 1d SQCD and $\Gamma$-SQCD theories, and we precisely formulate the dualities they admit. We also show that the Witten index of dual theories matches exactly. In section~\ref{sec:Higgs}, we study the Higgs phase of $\Gamma$-SQCD and prove the mutation dualities in this geometric context. In section~\ref{sec:Coulomb}, we study the Coulomb phase in much detail in the special case of an abelian gauge group. In section~\ref{sec:wc}, we study wall-crossing and how trivial wall-crossing combined with our dualities gives us interesting triality relations, including non-trivial confinement dualities between gauge theories and free theories. Finally, in section~\ref{sec:ab expls}, we explore in detail a series of instructive examples in the abelian case, in order to illustrate our general results in concrete terms.

\section{Gauged supersymmetric quantum mechanics}\label{sec:SQM}

In this section, we review 1d $\CN=2$ supersymmetric models -- that is, supersymmetric quantum mechanics (SQM) with two real supercharges $Q_1$ and $Q_2$, in order to set up our notation and conventions. We focus on 1d gauged linear sigma models (GLSM) --- that is, free matter multiplets coupled to vector multiplets, which can engineer interesting non-linear sigma models (NLSM) at low energy~\cite{Witten:1993yc}.

\subsection{Supersymmetry algebra,  superfields and Lagrangians}
In terms of the complex supercharges $\CQ= Q_1+i Q_2$ and $\bar\CQ=Q_1-i Q_2$, the supersymmetry algebra reads:
\be
\CQ^2~, \qquad \b\CQ^2=0~, \qquad \{ \CQ, \b \CQ\} = 2 (H + Z)~.
\ee
Here $\b \CQ\equiv \CQ^\dagger$, $H$  is the Hamiltonian and $Z$ denotes a real central charge. In this section we work in real time, denoted by $t$. 
 For our purpose, it will be convenient to work in terms of the component fields of supermultiplets. Let us denote a generic supersymmetry variation by:
\be
\delta = \delta_\eps + \delta_{\b\eps}= i \eps \CQ- i \b\eps \b\CQ~,
\ee
with $\eps$, $\b\eps$ some Grassmannian parameters. 
A general, unconstrained supermultiplet contains two bosonic and two fermionic components:
\be
\CS= (C, \chi, \t\chi, v)~,
\ee 
which transform as:
\bea
&\delta C&=&\;  i \eps \chi + i \b \eps \t\chi~,\\
&\delta \chi&=&\; \b\eps\left(\d_t C+ i v\right)~,\\
&\delta \t\chi&=&\; \eps\left(\d_t C- i v\right)~,\\
&\delta v&=&\; \eps \d_t \chi - \b\eps \d_t \b\chi~.\\ 
\eea
Note that use the notation $\d_t\equiv {d\ov dt}$. 
We are most interested in the matter multiplets, called chiral and fermi multiplets, and in the vector multiplets under which the matter can be charged. 

\medskip
\noindent
{\bf Vector multiplet.} The 1d $\CN=2$ vector multiplet for a real compact gauge group $G$ contains a 1d gauge field $A_t$, a real scalar $\sigma$, the gauginos $\lambda$, $\bar \lambda$, and an auxiliary field $D$,   all transforming in the adjoint representation of the Lie algebra $\Fg={\rm Lie}(G)$:
\be
\CV= (A_t, \sigma, \lambda, \b \lambda, D)~.
\ee
Their supersymmetry transformations in Wess-Zumino (WZ) gauge read:
\bea
&\delta A_t  = i \eps \b\lambda +i \b\eps \lambda~,\\
&\delta \sigma = i \eps \b\lambda +i \b\eps \lambda~,\\
& \delta \lambda =  \eps  \left( D_t \sigma -i D\right)~,\\
& \delta\b\lambda =\b \eps  \left( D_t \sigma+i D\right)~,\\
& \delta D = \eps  (D_t \b\lambda+i [\sigma, \b\lambda])- \b\eps (D_t \lambda+i [\sigma, \lambda])~.\\
\eea
Here $D_t$ denotes the gauge-covariant derivative, which acts as $D_t\varphi = \d_t \varphi - i [A_t, \varphi]$ for any field $\varphi$ in the adjoint representation.  
Importantly, $A_t-\sigma$ is a supersymmetric combination:
\be
\delta(A_t-\sigma)=0~.
\ee
These transformations realise the gauge-covariant supersymmetry algebra
\be
\{\CQ, \b\CQ\}\varphi= 2i D_t\varphi -2[\sigma, \varphi ]~,
\ee
on any adjoint field $\varphi$ in $\CV$.

\medskip
\noindent
{\bf Chiral and fermi multiplets.}  The 1d $\CN=2$ chiral multiplet consists of a complex scalar and a complex fermion,
\be
\Phi=(\phi, \psi)~, \qquad \b\Phi=(\b\phi,\b\psi)~.
\ee
The 1d $\CN=2$ fermi multiplet consists of a single complex fermion, together with an auxiliary field $G$:
\be
\bH= (\eta, G)~, \qquad \b\bH = (\b\eta, \b G)~.
\ee
 Each fermi multiplet comes equipped with two holomorphic functions of the chiral multiplets,
 \be E= E(\Phi)~, \qquad  J=J(\Phi)~,
 \ee
  which determine the so-called $E$- and $J$-term interactions to be introduced momentarily. We will sometimes call these holomorphic functions the 1d $\CN=2$ superpotentials. Using the language of graded quivers~\cite{Franco:2017lpa, Closset:2018axq}, one can encode the $E$- and $J$-terms in a `superpotential' that is the following flavour and gauge-invariant fermionic term:
  \be\label{W N02}
W= \Tr\left(\bH J +\b\bH E\right)~,
  \ee
such that $E= {\d W\ov \d \b\bH}$ and $J= {\d W\ov \d \bH}$; while not being technically a superpotential, it still correctly encodes the physics, including in particular the fact that a quadratic term such as $W= \mu \bH \Phi$ corresponds to a complex mass $\mu$ for both the fermi and the chiral multiplet involved.

Consider a chiral multiplet $\Phi$ charged under the vector multiplet, transforming in a (reducible) representation $\FR_\phi$ of $\Fg$ (and hence the antichiral multiplet $\b\Phi$ transforms in the conjugate representation $\b\FR_\phi$). Similarly, let the fermi multiplet $\bH$ transform in some representation $\FR_{\eta}$ of $\Fg$. Then gauge-invariance and supersymmetry constrain the functions $E$ and $F$ to be valued in $\FR_\eta$ and  $\b\FR_{\eta}$, respectively, while we have:
\be\label{EJeq0 rel}
\Tr(E J)=0~,
\ee
with the natural trace over $\FR_\eta$. 
The  supersymmetry transformations for $\Phi$ and $\bH$ read:
\bea
&\delta \phi  = {\sqrt 2} \eps \psi \qquad \qquad\qquad && \delta \b\phi=- {\sqrt 2}\b \eps \b\psi~,\\
&\delta \psi = {\sqrt2} i  \b\eps (D_t +i \sigma) \phi~, \qquad &&
\delta\b\psi = -{\sqrt2} i \eps (D_t -i \sigma)\b\phi~,\\
&\delta \eta  = {\sqrt 2} \eps  G + {\sqrt2} \b\eps  E \qquad \qquad && \delta \b\eta={\sqrt2} \eps  \b E+  {\sqrt 2}\b \eps  \b G~,\\
&\delta  G = {\sqrt2}i \b\eps \left((D_t+ i \sigma) \eta+i \psi_E\right)~, \qquad &&
\delta\b G = {\sqrt2}i \eps \left((D_t- i \sigma) \b\eta-i \b\psi_{\b E}\right)~,
\eea
where we defined $\psi_E\equiv {\d E\ov \d \phi} \psi$ and $\b\psi_{\b E}\equiv \b\psi {\d \b E\ov \d \b \phi}$, where the sum over representation indices is left implicit. 
While the $J$-term does not appear in the off-shell transformations, we will have $G= \b J$ on-shell, and we then have an on-shell symmetry between the fermi and antifermi fields:
\be\label{symmetry fermis}
(\eta, \b\eta, E, J, \b E, \b J)\quad \leftrightarrow\quad  (\b\eta, \eta,  J, E, \b J, \b E)~.
\ee

\medskip
\noindent
{\bf The GLSM action.} The off-shell supersymmetric action of the 1d GLSM,
\be
S= \int dt\,  L_{\rm GLSM}~,
\ee
is given in term of a sum of supersymmetric Lagrangians:
\be\label{LGLSM}
L_{\rm GLSM}= L_{\CV} + L_{\rm FI}+L_{\rm CS}+  L_{\b\Phi \Phi}+ L_{\b\bH \bH} + L_{J}+ L_{\b J}~.
\ee
The first three terms govern the vector multiplet. The 1d `super-Yang-Mills' term reads:
 \be
 L_{\CV}  ={1\ov g^2}\Tr\left( \half D_t \sigma D_t \sigma  -i\b \lambda D_t \lambda + \half D^2 +  \b\lambda [\sigma, \lambda]\right)~.
 \ee
 Note that the gauge coupling $g^2$ has mass dimension $[g^2]=3$. 
The next two terms in~\eqref{LGLSM} are the Fayet--Iliopoulos and 1d Chern--Simons terms, respectively. They are given schematically by:
\be
L_{\rm FI} = \zeta\, \Tr(D)~, 
\ee
and
\be\label{contact term Q}
 L_{\rm CS} =  Q\,  \Tr(A_t-\sigma)~.
\ee
More precisely, there will be as many FI parameters $\zeta$ and 1d CS levels $Q$ as there are distinct abelian factors inside $G$. In this paper we will focus on $G=U(N_c)$, which has a single abelian factor. Note that $[\zeta]=-1$ and $[Q]=0$. The bare CS level $Q$, also known as the bare gauge charge, needs to be integer-quantised:
\be
Q\in \Z~.
\ee
This ensures that the exponentiated action of the theory on a circle (in Euclidean time) is invariant under large gauge transformations. The existence of a supersymmetric 1d CS term is specific to $\CN=2$ supersymmetry (in particular, it is not allowed by $\CN=4$ supersymmetry). The effect of the bare CS level is to modify   Gauss's law  for the $U(1)\subset G$ factors:
\be\label{gauss law gen}
{\delta S\ov \delta A_t} =  {\bf J}_{\rm matter}  +Q  =0~.
\ee
Here ${\bf J}_{\rm matter}$ is the ordinary $U(1)$ gauge charge built out of the matter fields. Note that $A_t$ enters the 1d theory as an auxiliary field. We will discuss important vacuum contributions to this expectation value in subsection~\ref{subsec: quantize free fields}.

The matter sector consists of the kinetic terms for the chiral and fermi multiplets:
\bea\label{L chiral and fermi}
L_{\bar{\Phi}\Phi}&=D_t\bar{\phi}D_t\phi-\bar{\phi}(\sigma^2+D)\phi-i\bar{\psi}(D_t-i\sigma)\psi+i\sqrt{2}\bar{\phi}\lambda\psi-i\sqrt{2}\bar{\psi}\bar{\lambda}\phi~,\\
   L_{\b\bH\bH}&=-i\bar{\eta}(D_t+i\sigma)\eta+\bar{G}G-\bar{E}E+\bar{\psi}_E\eta+\bar{\eta}\psi_E~, \\ 
  \eea
where the vector-multiplet fields act on the chiral and fermi multiplets in their respective gauge representation, and with the overall traces left implicit. Finally, we have the $J$-terms:
\be
 L_{J}+ L_{\b J} = - G J + \eta \psi_J - \b G \b J - \b\eta\b\psi_{\b J}~,
\ee
with $\psi_J$ defined similarly to $\psi_E$ above. 
Integrating out the auxiliary fields $G$, $\b G$, the full action for the fermi multiplets becomes
\be
L_{\rm fermi}= -i \b\eta (D_t+i \sigma) \eta - \b E E- \b J J + \b\psi_E \eta + \b \eta \psi_E+ \eta \psi_J +\b\psi_{\b J} \b\eta~,
\ee
which exhibits the symmetry~\eqref{symmetry fermis}. We also have the standard $D$-term equations
\be
D= \b \phi \phi - \zeta~,
\ee
schematically. The scalar potential thus reads:
\be
V= |E|^2 + |J|^2+ {\half }D^2~,
\ee
and therefore supersymmetric ground states must satisfy $E=J=D=0$.

\subsection{Quantising free superparticles and one-loop determinants}\label{subsec: quantize free fields}
Consider a GLSM with chiral multiplets $\Phi^i$ and fermi multiplets $\bH^I$ of charges $q_i$ and $q_I$, respectively, under some  $G=U(1)$ vector multiplet (without much loss of generality). Here we first consider the free chiral and fermi multiplets in a supersymmetric background for the vector multiplet. The dynamics of the vector multiplet itself will be discussed later.

\subsubsection{Canonical quantisation of chiral and fermi multiplets} 
We consider the supersymmetric background for the abelian vector multiplet:
\be
A_t\pm \sigma= \text{constant}~, \qquad \lambda=\b\lambda=0~,\qquad D=0~.
\ee
The canonical momenta for the complex scalar and fermions is defined as $\Pi_\varphi = {\delta S\ov \delta D_t\varphi}$, hence we have:
\be\label{canonical momenta}
\Pi_{\phi^i} = D_t \b \phi_i~, \qquad \b \Pi_{\b \phi_i} = D_t  \phi^i~, \qquad \Pi_{\psi^i} = - i \b\psi_i~, \qquad  \Pi_{\eta^I} = - i \b\eta_I~.
\ee
where we used the indices $i, j, ...$ and $I, J, ...$ for the chiral and fermi multiplets, respectively. 
In our conventions, the canonical commutation relations are:
\be
[\Pi_{\phi^i}, \phi^j] = i \delta_i^j~, \qquad\quad
[\b\Pi_{\b\phi_i},\b \phi_j] = i \delta_j^i~,\qquad\quad
\{\b \psi_i, \psi^j\} =  \delta_i^j~, \qquad\quad
\{\b \eta_I, \eta^J\} =  \delta_I^J~.
\ee
Note that this corresponds to the Heisenberg equations:
\be
D_t \CO =-i  [{\bf H}, \CO]~,
\ee
and to weighting the path integral by the factor $e^{-i S}$. Here the Hamiltonian reads
\bea
&{\bf H} &=&\; \sum_i \left( \Pi_{\b \phi_i}  \b \Pi_{\phi^i} + \b\phi_i (q_i \sigma)^2 \phi^i - \psi_i (q_i \sigma)\b\psi^i \right)\\
&&&+ \sum_I  \left(\eta^I (q_I \sigma) \b\eta_I+|E_I|^2+ |J^I|^2  +  \psi_{J_I} \eta^I  +  \psi_{E^I} \b \eta_I  +\b\eta_I \b\psi_{\b J^I}+\eta^I \b\psi_{\b E_I}   \right)~.
\eea
The electric charge (that is, the background $U(1)$ gauge charge) is given by:
\be
{\bf J}=\sum_i  q_i \left( i \b \Pi_{\b \phi_i} \b \phi_i- i \Pi_{\phi^i}\phi^i -\b \psi_i  \psi^i  \right)+\sum_I q_I  \, \eta^I  \b\eta_I ~.
\ee
Note that we have $[{\bf J}, \CO]= q[\CO] \CO$. Finally, we have the supercharges: 
\bea
\CQ=- \sqrt2 \sum_i \left(\Pi_{\phi^i}\psi^i - i \b\phi^i (q_i \sigma) \psi^i \right)- \sqrt2 \sum_I \left(i \b \eta_I \b J^I + i \eta^I  \b E_I \right)~,\\
\b \CQ=- \sqrt2 \sum_i \left(\b\psi_i \b\Pi_{\b \phi_i} + i \b\psi_i(q_i \sigma) \phi^i \right)- \sqrt2 \sum_I \left(-i  J_I \eta^I - i \b \eta_I E^I\right)~,
\eea
which commute with both ${\bf H}$ and ${\bf J}$. 
These operators realise the on-shell supersymmetry algebra:
\be
\CQ^2 =0~, \qquad \b\CQ^2=0~, \qquad \{Q, \b Q\}= 2\left({\bf H} -\sigma {\bf J}\right)~.
\ee
Note that the condition \eqref{EJeq0 rel} is necessary for the supercharges to be nilpotent. The supersymmetric ground states are annihilated by both supercharges.  They contribute to the Witten index~\eqref{IW intro}, which now reads:
\be\label{IW free fields}
{\bf I}_{\rm W}= \Tr((-1)^{\rm F} x^{\bf J})~, 
\ee
with $x$ acting as a fugacity for the electric charge. In the above,  the normal-ordering constants in ${\bf H}$ and ${\bf J}$ were fixed by supersymmetry together with our conventions for defining the ground states, which we now explain.

\medskip
\noindent
{\bf Quantising the chiral multiplet.} Consider a single chiral multiplet  $\Phi$ of $U(1)$ charge $q$, with the supercharges, gauge charge and Hamiltonian given as above. Assuming $E=J=0$ for now, this is a free theory --- we simply have bosonic and fermionic harmonic oscillators related by supersymmetry.  
The two harmonic oscillators from the complex scalar $\phi$ are best expressed in terms of ladder operators:
\bea
\phi = {1\ov \sqrt{2\omega}}(a^\dagger + b)~, \qquad && \b\phi  = {1\ov \sqrt{2\omega}}(b^\dagger+a )~,\\
\pi_\phi = i  \sqrt{\omega\ov 2}(a - b^\dagger)~, \qquad && \b\pi_{\b\phi} = i \sqrt{\omega\ov 2}(b-a^\dagger)~,\\
\eea
with the non-trivial commutators being
\be
[a, a^\dagger]=1~, \qquad\qquad [b, b^\dagger]=1~.
\ee
Here we defined the frequency:
\be
\omega = |q \sigma|~.
\ee
Note that we have the electric charges ${\bf J}=-1, 1, 1, -1$ for $a$, $a^\dagger$, $b$, and $b^\dagger$, respectively. In these variables, the supercharges take the form:
\be
\CQ= \begin{cases}
2i {\sqrt{\omega}} \,b^\dagger \psi\qquad &\text{if}\;\; q\sigma >0~, \\
-2i {\sqrt{\omega}}\, a \psi\qquad &\text{if}\;\; q\sigma <0~, \\
\end{cases}\qquad\qquad
\b \CQ= \begin{cases}
-2i {\sqrt{\omega}} \,b \b\psi\qquad &\text{if}\; \;q\sigma >0~, \\
2i {\sqrt{\omega}}\, a^\dagger\b\psi\qquad &\text{if}\; \; q \sigma <0~, \\
\end{cases}
\ee
which depends on the sign of the real mass $q\sigma$. The Hamiltonian and electric charge become:
\be
{\bf H}= |q\sigma|(a^\dagger a + b^\dagger b+ 1) - q\sigma \psi \b\psi~, \qquad
 {\bf J}= q(a^\dagger a- b^\dagger b - \b\psi \psi)~.
\ee
Choosing the supersymmetric vacuum to be annihilated by $a$ and $b$, we see that we must choose a different quantisation of the fermion depending on the sign of $q\sigma$. Namely, we define a supersymmetric vacuum  $\ket{0}_\pm$ for  $\sign{(q \sigma)}=\pm$ that satisfy:
\be
a \ket{0}_\pm =0~, \qquad 
b \ket{0}_\pm =0~, \qquad 
\begin{cases} \psi\ket{0}_+ =0\qquad &\text{if}\; \;q\sigma >0~,\\
 \b\psi\ket{0}_- =0\qquad &\text{if}\; \;q\sigma <0~,
\end{cases} 
\ee
so that $\ket{0}_\pm$ is annihilated by both supercharges. Note that we chose the normal-ordering of the Hamiltonian such that:
\be
{\bf H} \ket{0}_+=0~, \qquad
{\bf J} \ket{0}_+=0~, \qquad
(-1)^{\rm F}  \ket{0}_+= \ket{0}_+~,
\ee
for $q\sigma>0$. The ground state  $\ket{0}_-$ for $q\sigma <0$ will then necessarily have non-zero energy and charge, as well as non-trivial fermion parity:
\be\label{vac charges ket0-}
{\bf H} \ket{0}_-=|q\sigma| \ket{0}_-~, \qquad
{\bf J} \ket{0}_-=-q \ket{0}_-~, \qquad
(-1)^{\rm F}  \ket{0}_-= -\ket{0}_-~.
\ee
We still have 
\be
({\bf H} - \sigma {\bf J})\ket{0}_\pm =0~,
\ee
 consistently with supersymmetry.
The non-trivial vacuum charges~\eqref{vac charges ket0-} are entirely due to the fermion; this phenomenon is generally known as a `parity anomaly' --- in quantum mechanics, it is rather a `charge conjugation anomaly'~\cite{Elitzur:1985xj}. In the path integral language, this is related to the statement that,  when integrating out a free 1d fermion with the real mass $q\sigma$, we obtain an empty theory for $q\sigma>0$ but generate a non-trivial contact terms~\eqref{contact term Q} with $Q=-1$ for $q\sigma <0$.%
\footnote{See~\protect\cite{Closset:2012vp} for a closely related discussion in 3d.}

The supersymmetric ground states of the chiral multiplet take the form:
\be
(a^\dagger)^s  \ket{0}_+~,  \qquad \qquad (b^\dagger)^t  \ket{0}_-~,\qquad \text{with }  s,t \in \Z_{\geq 0}~,
\ee
 for $q\sigma>0$ and $q\sigma<0$, respectively. Indeed, consider the states $(a^\dagger)^s (b^\dagger)^t \b \psi^f \ket{0}_+$ for $q\sigma>0$, with $f=0,1$. Then we have ${\bf H}=(s+t+f)q\sigma$ and ${\bf J}=q(s-t-f)$, and supersymmetry requires ${\bf H}-\sigma{\bf J}=2q\sigma(t+f)=0$ which implies $t=f=0$. The Witten index~\eqref{IW free fields} then takes the form:
 \be\label{IW chiral pos}
 {\bf I}_{\rm W} = 1 + x^q + x^{2 q} + x^{3 q}+ \cdots  = \sum_{s\geq 0} x^{qs}= {1\ov 1-x^q}~.
 \ee
 Similarly, if $q\sigma <0$ we have the states $(a^\dagger)^s (b^\dagger)^t  \psi^f \ket{0}_+$ with  ${\bf H}=(s+t+f+1)|q\sigma|$ and ${\bf J}=q(s-t+f-1)$, which gives us  ${\bf H}-\sigma{\bf J}=2|q\sigma|(s+f)=0$. 
 The Witten index now reads:
  \be\label{IW chiral neg}
 {\bf I}_{\rm W} =  -x^{-q}\left(1 + x^{-q} + x^{-2q} + x^{-3q}+ \cdots\right)    =-x^{-q} \sum_{t\geq 0} x^{-t q}= {1\ov 1-x^q}~,
 \ee
where the overall factor in the first equality is the contribution from the vacuum $\ket{0}_-$ as given in~\eqref{vac charges ket0-}. Of course, \eqref{IW chiral pos} and \eqref{IW chiral neg} agree as analytic functions since the Witten index is meromorphic in $x$.

\medskip
\noindent
{\bf Quantising the fermi multiplet.} Consider now a single fermi multiplet  $\bH$ of $U(1)$ charge $q$, with $E=J=0$. This is an exceedingly simple free theory  $L_{\rm fermi}= -i \b\eta (D_t+i \sigma) \eta$, with vanishing supercharges and with:
\be
{\bf H}= \eta (q\sigma)\b\eta~, \qquad\qquad {\bf J} = q\eta \b\eta~,
\ee
so that ${\bf H}-\sigma {\bf J}=0$ identically--hence, there are only two states by the Fermi exclusion principle, and they are all supersymmetric ground states. We choose to quantise the theory according to the sign of the real mass:
\be
\begin{cases} \b\eta\ket{0}_+ =0\qquad &\text{if}\; \;q\sigma >0~,\\
 \eta \ket{0}_- =0\qquad &\text{if}\; \;q\sigma <0~,
\end{cases}
\ee
which gives us ${\bf H}\ket{0}_+=0$ and ${\bf J}\ket{0}_+=0$, but also:
\be\label{vac charges ket0- fermi}
{\bf H} \ket{0}_-=q\sigma \ket{0}_-~, \qquad
{\bf J} \ket{0}_-=q \ket{0}_-~, \qquad
(-1)^{\rm F}  \ket{0}_-= -\ket{0}_-~.
\ee
 Thus, for $q\sigma>0$, we have the ground states $\eta^f \ket{0}_+$ for $f=0,1$, while for $q\sigma<0$ we have the ground states $\b\eta^f \ket{0}_-$ for $f=0,1$. This gives us the Witten index:
\be
{\bf I}_{\rm W}= 1- x^q= (-x^q)(1-x^{-q})~.
\ee

\subsubsection{Witten index, path integral and parity anomaly}
For the general set of free chiral multiplets $\Phi^i$ and fermi multiplets  $\bH^I$ charged under some background $U(1)$, we then have the Witten index:
\be\label{IW U1 theory free matter}
{\bf I}_{\rm W}= x^Q \prod_i {1\ov 1- x^{q_i}}\, \prod_I(1-x^{q_I})~,
\ee
where we allow for a bare CS level $Q\in \Z$. 
This can also be computed in the path integral formalism, as a partition function of the 1d model on the circle. This can be done using the Euclidean time $\tau = - i \tau$ (with $\tau\sim \tau+2\pi$), which gives us the Euclidean path integral weighted $e^{-S_E}$ with:
\be\nn
S_E= \int d\tau \left(\sum_i \left(D_\tau\bar{\phi_i}D_\tau\phi^i+\bar\phi_i(q_i\sigma)^2\phi^i + \bar\psi_i(D_\tau+q_i\sigma)\psi^i\right)+\sum_I  \bar\eta^I(D_\tau+q_I\sigma)\psi_I \right)~.
\ee
The Witten index is then computed as the Gaussian path integral on the  circle:
\be\label{IW path integral}
{\bf I}_{\rm W}=Z_{S^1} = \prod_i {\det(D_\tau +q_i \sigma)\ov \det(-D_\tau^2+(q_i\sigma)^2)}\, \prod_I \det(D_\tau-q_I \sigma)= {\prod_I \det(D_\tau-q_I \sigma)\ov \prod_i \det(-D_\tau+q_i \sigma)}~,
\ee
with periodic boundary conditions for the fermions. Let us define the complexified mass:
\be
u \equiv - A_\tau+  i \sigma = -i(A_t-\sigma)~,
\ee
and the fugacity:
\be
x\equiv e^{2\pi i u} = \exp{\left(- i \int d\tau (A_\tau -i \sigma) \right)}~.
\ee
Large gauge transformation along the circle corresponds to $u\rightarrow u+1$. 
We immediately notice that the 1d CS contact term~\eqref{contact term Q} contributes  $x^Q$ to the result --- this is a contact term for the background gauge field that we are free to add, with the constraint that $Q\in \Z$ to preserve background gauge invariance. We still have to compute the infinite products that appear in~\eqref{IW path integral}, which each take the simple form:
\be
Z_\eta(x) \equiv \prod_{n\in \Z} (n+ u)= \exp\left(  \sum_{n\in \Z}\log(n+u) \right) = \exp\left(c_0 + c_1  u +\log(\sin(\pi u)) \right)~. 
\ee
A classic $\zeta$-function regularisation%
\footnote{For another quick and dirty regularisation, consider $f(u)=  \sum_{n\in \Z}\log(n+u) $. One can use  the fact that $f''(u)$ formally converges to $f''(u)= -\pi^2 /\sin(\pi u)^2$ to obtain the regularised $f(u)$ by integration, which gives us the stated result including the integration constants $c_0$ and $c_1$.}
 gives us $Z_\eta(x) =2 i  \sin(\pi u)$~\cite{Hori:2014tda}, but we are free to add contact terms as indicated. We choose a scheme that preserves the invariance under large gauge transformations ($u\sim u+1$) at the expense of parity (which acts as $u\rightarrow -u$), hence we have:
\be
Z_\eta(x) = x^Q (1-x)~,
\ee
for some $Q\in \Z$. Choosing $Q=0$ gives us precisely the expected result~\eqref{IW U1 theory free matter} for the one-loop determinants of the matter multiplets.%

\medskip
\noindent
{\bf Further comments on 1d CS levels and parity anomaly.} Our conventions for quantising 1d fermions can be summarised as follows. In the UV, which corresponds to having massless chiral and fermi multiplets, we have an effective background charge:
\be\label{quv to Q}
q_{\rm uv}= \langle  {\bf J} \rangle_{\rm uv}=  \langle Q+ {\bf J}_{\rm matter}\rangle_{\rm uv} =Q -\half \sum q_i + \half \sum_I q_I~.
\ee
The classical contribution $Q\in \Z$ from~\eqref{contact term Q} is supplemented by one-loop effects from the massless fermions $\psi_i$ and $\eta_I$ (running in a tadpole diagram).  Note that the combination $q_{\rm uv}$ is the one that is generally called `the 1d CS level' of the theory--we call it UV effective CS level to distinguish it from the bare (classical) level $Q$. The two parameters $q_{\rm uv}$ and $Q$ are simply related by~\eqref{quv to Q}, and we shall characterise our theory using one or the other depending on convenience.  In the actual GLSM, the 1d CS parameter $q_{\rm uv}$  or $Q$ is part of the definition of the theory. In particular, it directly affects Gauss's law~\eqref{gauss law gen}.

Once we turn on $\sigma\neq 0$, we have the two choices of vacua $\ket{0}_\pm$ for $\sign(\sigma)=\pm$, with:
\bea
\sigma>0\; : \qquad  && \begin{cases} \psi^i\ket{0}_+ =0\qquad &\text{if}\; \;q_i >0~,\\
 \b\psi_i \ket{0}_+ =0\qquad &\text{if}\; \;q_i <0~,
 \end{cases}
 \quad \text{and}\quad 
 \begin{cases} \b\eta_I\ket{0}_+ =0\qquad &\text{if}\; \;q_I>0~,\\
 \eta^I \ket{0}_+ =0\qquad &\text{if}\; \;q_I <0~,
\end{cases}\\
\sigma<0\; : \qquad  && \begin{cases} \b\psi_i  \ket{0}_- =0\qquad &\text{if}\; \;q_i >0~,\\
 \psi^i\ket{0}_- =0\qquad &\text{if}\; \;q_i <0~,
 \end{cases}
 \quad \text{and}\quad 
 \begin{cases}  \eta^I\ket{0}_- =0\qquad &\text{if}\; \;q_I>0~,\\
 \b\eta_I\ket{0}_- =0\qquad &\text{if}\; \;q_I <0~.
\end{cases}
\eea
In these vacua, we then have:
\be
\langle {\bf J} \rangle_\pm = -\sum_{\substack{i\,\text{such that} \\   \sign(q_i)=\mp}} q_i+\sum_{\substack{I\,\text{such that} \\   \sign(q_I)=\mp}} q_I~.
\ee
The fact that this is an integral charge is important, since this computes the 1d CS level
\be
Q_{\rm ir} = \langle  Q+ {\bf J}\rangle_\pm \in \Z
\ee
in the empty effective theory in the infrared.
The statement of the parity anomaly is that $q_{\rm uv}\neq 0$, in general. However, if we are willing to choose $Q$ half-integer, one could pick
\be\label{HKY quantisation}
q_{\rm uv}=0 \qquad \Leftrightarrow \qquad Q= Q^{\rm HKY}\equiv  \half\sum_i q_i -\half \sum_I q_I~,
\ee
in which case we have
\be
\langle  Q^{\rm HKY}+ {\bf J} \rangle_{\pm}= \half \sum_i \sign(q_i \sigma) q_i -\half \sum_i \sign(q_I \sigma) q_I~,
\ee
which matches the quantisation condition generally chosen in the literature~\cite{Hori:2014tda}, although this expectation value is not an integer in general.  Indeed \eqref{HKY quantisation} is the only parity-preserving quantisation, but it is not the one we shall use. We prefer a parity-violating quantisation which preserves gauge invariance.

\subsection{Effective theories for the abelian vector multiplet}\label{subsec:U1 EFT}
We are not interested in free theories per se, but in 1d gauge theories. Once we introduce the dynamical vector multiplet, the full 1d GLSM Lagrangian~\eqref{LGLSM} becomes dominated by the gauge interactions, although we will also have to account for the $E$- and $J$-term interactions. Here we focus on the case of a single $U(1)$ vector multiplet, for simplicity, which we couple to the chiral multiplets $\Phi_i$ and fermi multiplets $\bH_I$ with electric charges $q_i$ and $q_I$, respectively. Henceforth $x$ will denote the gauge fugacity, and $y_i$ or $y_I$ denote the flavour fugacities.%
\footnote{We turn on an independent flavour fugacity for each matter multiplet, with the understanding that these are defined modulo the gauge shift $(y_i, y_I)\sim  (x^{q_i} y_i, x^{q_I} y_I)$.} Recall that $\sigma \propto -\log |x|$ while $m \propto -\log|y|$ give us real masses.

We wish to understand the supersymmetric ground states of this theory in the Coulomb-branch scaling~\cite{Hori:2014tda}:
\be
|\zeta| \rightarrow 0~, \qquad g^2\rightarrow \infty~, \qquad M_C\equiv g^2\zeta^2 \;\, \text{fixed.}
\ee
This is to be contrasted with the Higgs branch scaling $|\zeta| \rightarrow \infty$, in which chiral multiplets aquire large vevs and the vector multiplet can be integrated out at tree-level%
\footnote{Assuming the electric charges allow a complete Higgsing, at least.}, leading to a NLSM into the semi-classical Higgs branch.  
 In the Coulomb-branch scaling, on the other hand, we first assume that the real scalar $\sigma$ takes a vev that  gives a mass to all charged matter (we may also turn on the real masses $m_i$, $m_I$). The resulting Fock-space states of free particles were discussed above; we denote the corresponding supersymmetric ground states by $\CF^0$. Integrating out the charged matter, we then obtain an effective theory for the abelian vector multiplet which is valid below the Coulomb-branch scale $M_C$. This procedure is self-consistent if the vacuum obtained from the effective theory indeed freezes the scalar $\sigma$ at the needed non-zero value. We wish to proceed in three steps:
\begin{itemize}
    \item[1.] Starting with the free theory of massive chiral and fermi multiplets, impose the Gauss law~\eqref{gauss law gen} for any value of $\sigma$ (and possibly of $m_i$, $m_I$). The effect of this is to project us onto appropriate subspaces of the free-field Hilbert spaces of ground states:
    \be
     \CF^0\quad  \longrightarrow\quad  \CF^0_{\rm phys} = \left\{ \ket{\Psi}\in \CF^0 \; |\;  {\bf J} \ket{\Psi} =0  \right\}~.
    \ee
    Note that there might be different sectors at different values of $\sigma$. 
 \item[2.] For each would-be ground state obtained in the first step, write down the effective theory for the scalar $\sigma$ and the gaugino $\lambda, \b\lambda$, and determine whether the exact ground state of that effective theory actually exists. 
 \item[3.] Add in the effect of the $E$- and $J$-term interactions. This can be challenging to do, in general. In our examples, we will treat these interactions in perturbation theory.
\end{itemize}
This three-step procedure allows us to obtain Coulomb-branch states which are candidates for the true ground states of the GLSM. In general, however, the true ground states are fewer than the number of perturbative ground states obtained from this Coulomb-branch analysis. In simpler settings without gauge interactions, one also finds that perturbative ground states of a SQM with a scalar potential can be lifted by non-perturbative effects mediated by instantons~\cite{Witten:1982im}. A completely analogous phenomenon occurs here, as we shall see in later sections.

For now, let us focus on the perturbative ground states of the effective field theory for the Coulomb-branch scalar. This theory takes the form~\cite{Witten:1981nf, Hori:2014tda}:
\be
L_{\rm eff} = \half \dot\sigma^2 - i \b\lambda\dot\lambda - \half (h'(\sigma))^2 + h''(\sigma)\b\lambda\lambda~,
\ee
with $h'(\sigma)\equiv {dh \ov d\sigma}$. Here the real superpotential $h(\sigma)$ is obtained at one loop, as we will review momentarily, giving us the real potential $U= \half (h')^2$. We also rescaled the fields $\sigma$, $\lambda$ and $\b\lambda$ by a factor of $g$. The supersymmetry transformations read:
\bea
&\delta \sigma = i \eps \b\lambda +i \b\eps \lambda~,\qquad
& \delta \lambda =  \eps  \left( \dot\sigma -i {h'(\sigma)}\right)~,\qquad 
& \delta\b\lambda =\b \eps  \left( \dot\sigma+i {h'(\sigma)}\right)~.
\eea
Canonical quantisation gives us the momentum $\Pi_\sigma = \dot \sigma$ and the commutation relations:
\be
[\Pi_\sigma, \sigma]= i~, \qquad \quad \{\lambda, \b\lambda\}= 1~.
\ee
The supercharges  and the Hamiltonian are given by
\bea
&\CQ = -i (\Pi_\sigma- i {h'(\sigma)}) \b\lambda~, \qquad\qquad
\b\CQ = i (\Pi_\sigma+ i  {h'(\sigma)}) \lambda~, \\
&{\bf H}= \half \Pi_\sigma^2 + \half ( {h'(\sigma)})^2 - \half  {h''(\sigma)}(\b\lambda\lambda- \lambda\b\lambda)~.
\eea
Let us define:
\be
\Sigma(\sigma)= e^{h(\sigma)}~, \qquad \quad  \t\Sigma(\sigma)= e^{-h(\sigma)}~,
\ee
which are the wavefunctions satisfying the first-order equations:
\be
\left({d\ov d\sigma}- h'\right) \Sigma=0~, \qquad\quad
\left({d\ov d\sigma}+ h'\right) \t \Sigma=0~,
\ee
respectively. Let us choose $\lambda$ to be the fermionic annihilation operator, so it annihilates any ground state. Then, we find the two candidate supersymmetric ground states $\ket{\Sigma}$ and $\b\lambda \ket{\t\Sigma}$ --- indeed, they are annihilated by both supercharges. Last but not least, we must check whether these states are square-normalisable. While this might depend on the details of the superpotential, in the case at hand we should only trust $U(\sigma)$ near critical points (that is, near $\sigma=\sigma_\ast$ so that $h'(\sigma_\ast)=0$) or, more generally, for energies $U < M_C$. 
Performing a local analysis near a single critical point, one finds that:
\bea
&&{\ket{\Sigma}} \; \text{ is normalisable}\qquad &\Leftrightarrow &\quad & {h''(\sigma_\ast)} < 0~,\\
&&{\b\lambda\ket{\t\Sigma}} \; \text{ is normalisable}\qquad &\Leftrightarrow &\quad & {h''(\sigma_\ast)} >0~.
\eea
This gives us a good handle (or, least, as good a handle as one can hope for) on the perturbative ground states in the $U(1)$ gauge theory. 

Finally, let us discuss the explicit form of the superpotential obtained by integrating out the chiral multiplets. In a given state $\ket{\Psi}\in \CF_{\rm phys}^0$, we define the positive integers $N_{\phi_i}$ counting the number of $a_i^\dagger$ or $b_i^\dagger$ oscillators:
\be
 {\bf N}_{\phi_i} \ket{\Psi}= N_{\phi_i} \ket{\Psi}~, \qquad \qquad {\bf N}_{\phi_i} \equiv a_i^\dagger a_i +b_i^\dagger b_i + 1~. 
\ee
In other words, we have 
\be
N_{\phi_i} = s_i + t_i + 1 
\ee
if $\ket{\Psi}$ is obtained using the creation operators $(a_i^\dagger)^{s_i}(b_i^\dagger)^{t_i}$.
A one-loop computation then gives us the superpotential:
\be
h(\sigma) = -\t \zeta \sigma +\half \sum_i N_{\phi_i} q_i \sign(q_i \sigma+m_i) \log|q_i \sigma +m_i|~,
\ee
and hence the scalar potential:
\be\label{Usigma general}
U(\sigma)= \half \left(-\t\zeta +\sum_i {q_i N_{\phi_i}\ov 2|q_i \sigma+ m_i|}\right)^2~.
\ee
Here we introduced the rescaled FI parameter $\t\zeta \equiv g \zeta = \sqrt{M_C}$. Note that this shape of the potential is only reliable for $U(\sigma)< M_C$. The divergence in the potential energy near the loci $q_i \sigma+m_i=0$ has a simple and standard physical explanation --- these are the loci at which chiral multiplets become massless and therefore where the effective description is not reliable.

\section{Infrared dualities between unitary gauged quantum mechanics}\label{sec:dualities}

In this section, we propose a Seiberg-like duality for 1d $\CN=2$ unitary SQCD. We first spell out the model, using the conventions detailed in the previous section. We then give the precise statement of the duality, which depends on the sign of the FI parameter.

\subsection{One-dimensional unitary SQCD}\label{subec:defSQCD}

We are interested in 1d $\CN=2$ SQCD with a unitary gauge group. This is the supersymmetric gauged quantum mechanics with a $U(N_c)$ vector multiplet, $n_1$ fundamental chiral multiplets, $n_2$ antifundamental chiral multiplets, and $n_3$ fundamental fermi multiplets. We also need to specify the bare gauge charge (1d CS level)  $Q_c\in \Z$ under the overall $U(1)\subset U(N_c)$:
\be
S_{Q} = Q_c \int dt \, \Tr(A_t- \sigma)
\ee
 As explained in the previous section, the effective UV 1d CS level~\eqref{quv to Q} includes a one-loop effects from the matter fields:
\be\label{qc to Qc}
q_c = Q_c - {n_1-n_2-n_3\ov 2}~,
\ee
and as such it can be half-integer. We often denote the 1d gauge group together with its CS level by $U(N_c)_{q_c}$. Last but not least, the theory also admits an FI parameter $\zeta$. We will focus on the theory with $\zeta \neq 0$, where we expect a 1d NLSM interpretation of the low-energy states to be available.

\begin{table}
\renewcommand{\arraystretch}{1.2}
    \centering
    \begin{tabular}{|c||c||c|c| c||c|}
    \hline
    & $U(N_c)$& $U(n_1)$ & $U(n_2)$ & $U(n_3)$& $U(n_4)$ \\ \hline   \hline
    $\Phi^\alpha_i$ & $\square$ & $\overline{\square}$ & ${\bf 1}$ & ${\bf 1}$& ${\bf 1}$ \\ 
    $\t\Phi^j_\alpha$ &  $\overline{\square}$ & ${\bf 1}$ & $\square$ &${\bf 1}$& ${\bf 1}$ \\
    $\bH^\alpha_k$ & $\square$ & ${\bf 1}$ &${\bf 1}$  & $\overline{\square}$& ${\bf 1}$ \\ 
    \hline
    $\Gamma^j_i$ & ${\bf 1}$ &   $\overline{\square}$ & $\square$ & ${\bf 1}$& ${\bf 1}$ \\ \hline\hline
    $\t\bH^k_\alpha$ & $\overline{\square}$ & ${\bf 1}$ &${\bf 1}$ & ${\bf 1}$ & $\square$ \\ 
     \hline
\end{tabular}
    \caption{ Gauge and flavour charges of 1d SCQD (and of $\Gamma$-SQCD). Here $\alpha=1, \cdots, N_c$ is a gauge index, and $i=1, \cdots, n_1$, $j=1,\cdots, n_2$ and $k=1, \cdots, n_3$ (as well as $l=1,\cdots, n_4$) are flavour indices. (The antifundamental fermis are discussed in the main text.)}
    \label{table:matter 1dSQCD}
\end{table}

This 1d SQCD  is determined by  five integers,  namely the rank $N_c$, the non-negative integers $(\boldsymbol{n}^F)\equiv (n_I)= (n_1, n_2, n_3)$, and the bare CS level $Q_c$, together with the sign of the FI parameter. 
 In addition, in order to fully specify the UV Lagrangian we need to fix some contact terms for the flavour symmetry. Indeed, we have a flavour symmetry group%
\footnote{Or flavour symmetry {\it algebra} -- we are not being careful about the global form of the flavour group here.}
\be\label{GF def}
G_F = {U(n_1) \times U(n_2) \times U(n_3)\ov U(1)}~,
\ee
thus we should specify the corresponding bare CS levels $\boldsymbol{Q}^F= (Q^F_I)$, whose effect is simply to shift the background flavour charge of the vacuum through the linear contact term:
\be
 S_{\boldsymbol{Q}^F} =\sum_{I=1}^3 Q_I^F  \int dt \, \Tr(A_{t I}^F - \sigma^F_{I})~.
\ee
This parameterisation is convenient although it is slightly redundant, since the overall $U(1)$ background gauge field  is actually gauged -- we have that  $\sum_I \Tr(A_{t I}^F)= \Tr(A_t)$. 
  In summary, we wish to consider the SQCD theory:
\be\label{SQCD general}
\text{SQCD}\left[N_c~;\, Q_c~;\, \boldsymbol{n}^F~;\, \boldsymbol{Q}^F\right]_\pm \; \equiv \; U(N_c)_{q_c} \, \text{ with } \, n_1\,\Phi \,\oplus\, n_2\,\t\Phi\,\oplus\, n_3\,\bH\, ,\; \sign(\zeta)=\pm~,
\ee
with the matter multiplets specified in table~\ref{table:matter 1dSQCD}. 

\medskip
\noindent {\bf Defining $\Gamma$-SQCD.}
In the following, we will also study a slight variant of this theory dubbed $\Gamma$-SQCD. It consists of the SQCD theory just defined together with a gauge-neutral fermi multiplet $\Gamma$ in the fundamental of $U(n_2)$ and in the antifundamental of $U(n_1)$, as indicated in table~\ref{table:matter 1dSQCD}, and with the $E$-term given by the `meson' chiral multiplet:
\be\label{EGamma SQCD}
E_\Gamma =\t \Phi \Phi~, \qquad J_{\Gamma}= 0~, 
\ee 
where the gauge indices are contracted in the obvious way. This $E$-term lifts the flat directions in the SQM target space by setting the vacuum expectation value of the gauge-invariant meson operators $M= \t \Phi \Phi$ to zero. Then, for $\zeta \neq 0$, there will be a finite number of supersymmetric ground state.  
For future reference, we note that the flavour  $U(n_I)$ UV effective CS levels read:
\bea\label{qIF full}
&q_1^F= Q_1^F+ {N_c\ov 2}-{n_2\ov 2} + \delta q_1^F~,\\
&q_2^F= Q_2^F- {N_c\ov 2}+{n_1\ov 2} + \delta q_2^F~,\\
&q_3^F= Q_3^F- {N_c\ov 2}  + \delta q_3^F~,\\
\eea
where $\delta q_I^F$ are possible shifts from quantising any additional chiral and fermi multiplets which we might couple to the flavour group, for instance if we embed our SQCD model into a larger unitary quiver theory (we thus have $\delta q_I^F=0$ for the strict $\Gamma$-SQCD theory).

\medskip
\noindent {\bf Antifundamental fermi multiplets.} Note that, in our definition of SQCD, we considered fermi multiplets in the fundamental representation but not in the antifundamental. This is because, as reviewed around~\eqref{symmetry fermis}, the antifundamental fermi multiplet is on-shell equivalent to a fundamental fermi multiplet with the $E$- and $J$-terms swapped. However, for some purpose, it can be useful to introduce $n_4$ antifundemental hypermultiplets, as shown in table~\ref{table:matter 1dSQCD}. We easily see that the two following configurations are equivalent:
\be
 n_3 >0 \, , \, n_4 >0  \quad \longleftrightarrow \quad
 n_3'=n_3+n_4~, n_4'=0~,\text{ and } Q_c'=Q_c-n_4~,
\ee
up to a shift of the fermion number  $(-1)^{\rm F} \rightarrow (-1)^{\rm F}(-1)^{n_4}$. Thus, we can always shift between fundamental and antifundamental fermis at the price of shifting the bare CS levels.%
\footnote{Note that, according to our rules for quantising fermions, the UV effective CS level does not change:
\be\nn
q_c=Q_c- {n_1-n_2-n_3+n_4\ov 2} = Q_c'- {n_1-n_2-(n_3+n_4)\ov 2} = q_c'~.
\ee
}

\medskip
\noindent {\bf Generic $E$ and $J$ terms.} Our SQCD theory only had fundamental fermis by conventions, but one could consider more general interation terms:
\be\label{EJLambda gen}
(E_\bH)^\alpha_k = \Phi^\alpha_i \h F(\varphi)^i_k~, \qquad (J_\bH)^k_\alpha= \h G(\varphi) ^k_j\t\Phi^j_\alpha~.
\ee
This can happen when the SQCD theory is embedded into a larger theory with additional chiral multiplets, denoted collectively by $\varphi$, which are $U(N_c)$-neutral but carry the appropriate flavour charges.  Note that supersymmetry requires that
\be\label{TrEJ SQCD gen}
\Tr(E J) =\Tr(\h F \h G \t\Phi \Phi)=0~. 
\ee
More precisely, since we could have additional $E$- and $J$-terms from other fermi fields $\eta_l$ in the larger theory, we only need that $\Tr(\h F \h G \t\Phi \Phi)= -\sum_l E_{\eta_l} J_{\eta_l}$.

\subsection{The two duality moves: left and right mutations}
We are now ready to state the two Seiberg-like dualities for SQCD with $\zeta\neq 0$. There are two distinct dualities depending on the sign of the FI term for the $U(N_c)$ theory~\eqref{SQCD general}.

\subsubsection{Right mutation}
For $\zeta>0$, we have the  {\it right mutation}:
\bea\label{right mutation SQCD}
&\text{SQCD}\left[N_c~;\, Q_c~;\, (n_1, n_2, n_3)~;\, (Q_1^F, Q_2^F, Q_3^F)\right]^{\zeta>0} \quad  \longleftrightarrow  \qquad\qquad \\
&\,\text{SQCD}\left[n_1- N_c~;\, -Q_c-n_2~;\,  (n_3, n_1, n_2)~;\, (Q_3^F, Q_1^F+Q_c, Q_2^F+n_1-N_c)\right]^{\zeta<0}{ \oplus M \oplus \Gamma'}~.
\eea
In other words, this duality flips the sign of the both the gauge CS level and of the FI parameter, giving us the dual gauge groups:
\be
U(N_c)_{q_c}^{(\zeta>0)}\qquad \longleftrightarrow \qquad 
U(n_1-N_c)_{-q_c}^{(\zeta'<0)}~,
\ee
while the fundamental chirals become antifundamental chirals, the antifundamental chirals befome fundamental fermis, and the fundamental fermis become fundamental chirals. Note that the bare CS level for the gauge group transform as:
\be
Q_c \qquad \longleftrightarrow \qquad Q_c' = -Q_c-n_2~.
\ee
The bare flavour CS levels are shifted as indicated in~\eqref{right mutation SQCD}, which amounts to the following shift of the UV effective CS levels $q_I^F$ for the $U(n_I)$ flavour groups:
\be
(q_1^F~,\, q_2^F~,\, q_3^F)\qquad \longleftrightarrow \qquad  (q_1^F+q_c~,\, q_2^F~,\, q_3^F)
\ee
The definition of the fermion number also shifts under the duality, as:
\be\label{shift fermion nbr right mut}
(-1)^{\rm F} \qquad \longleftrightarrow \qquad (-1)^{\rm F'}= (-1)^{n_2(n_1-N_c)} (-1)^{\rm F}~.
\ee
\begin{table}
\renewcommand{\arraystretch}{1.2}
    \centering
    \begin{tabular}{|c||c||c|c| c|}
    \hline
    & $U(n_1-N_c)$& $U(n_1)$ & $U(n_2)$ & $U(n_3)$ \\ \hline   \hline
    ${\Phi'}^\beta_k$ & $\square$ &  ${\bf 1}$  & ${\bf 1}$&  $\overline{\square}$ \\ 
    ${\t\Phi}^{'i}_\beta$ &  $\overline{\square}$ & $\square$ & ${\bf 1}$  &${\bf 1}$\\
    ${\bH'}^\beta_j$ & $\square$ & ${\bf 1}$ & $\overline{\square}$ &${\bf 1}$  \\ 
    ${\Gamma'}^i_k$ & ${\bf 1}$ &   $\square$&  ${\bf 1}$ & $\overline{\square}$   \\
    \hline
         ${M}^j_i$ & ${\bf 1}$ &   $\overline{\square}$ & $\square$ & ${\bf 1}$  \\
     \hline
\end{tabular}
    \caption{ Gauge and flavour charges of the dual of 1d SCQD obtained by a right mutation. The meson $M$ is integrated out in the $\Gamma$-SQCD variant.}
    \label{table:matter 1dSQCD right mutated}
\end{table}
\begin{figure}
    \centering
\begin{tikzpicture}[>=stealth,scale=0.75]
 \def\SquareSize{1.2}  
    \def\CircleRadius{1}
    \def\NodeRadius{4}
    \draw (0,0) circle (\CircleRadius cm);

    \node[] (Nc) at (0,0) {${N_c\ov q_c}$};
    \node[draw, rectangle, shape aspect=1, minimum width= \SquareSize cm, minimum height=\SquareSize cm] (N1) at (90:\NodeRadius cm) {${n_1\ov q_1^F}$};
    \node[draw, rectangle, shape aspect=1, minimum width=\SquareSize cm, minimum height=\SquareSize cm] (N3) at (-30:\NodeRadius cm) {${n_3\ov q_3^F}$};
    \node[draw, rectangle, shape aspect=1, minimum width=\SquareSize cm, minimum height=\SquareSize cm] (N2) at (210:\NodeRadius cm) {${n_2\ov q_2^F}$};
    \node[below =0.48cm of Nc] {\small $\zeta > 0$};

    \draw[->-=0.7] (90:\CircleRadius cm) --node[midway, right] {$\Phi$} (N1);
    \draw[->-=0.7] (N2) -- node[midway, below]{${\t\Phi}$} (210:\CircleRadius cm);

    \draw[->-=0.7, dashed] (-30:\CircleRadius cm) -- node[midway, above]{$\bH$} (N3);

\end{tikzpicture}
\quad
\raisebox{3cm}{$\rightarrow$}
\quad
\begin{tikzpicture}[>=stealth,scale=0.75]
 \def\SquareSize{1.2}  
    \def\CircleRadius{1}
    \def\NodeRadius{4}
    \draw (0,0) circle (\CircleRadius cm);

    \node[] (Nc) at (0,0) {${n_1-N_c\ov -q_c}$};
    \node[draw, rectangle, shape aspect=1, minimum width= \SquareSize cm, minimum height=\SquareSize cm] (N1) at (90:\NodeRadius cm) {${n_1\ov q_1^F+q_c}$};
    \node[draw, rectangle, shape aspect=1, minimum width=\SquareSize cm, minimum height=\SquareSize cm] (N3) at (-30:\NodeRadius cm) {${n_3\ov q_3^F}$};
    \node[draw, rectangle, shape aspect=1, minimum width=\SquareSize cm, minimum height=\SquareSize cm] (N2) at (210:\NodeRadius cm) {${n_2\ov q_2^F}$};
    \node[below =0.48cm of Nc] {\small $\zeta' < 0$};

    \draw[->-=0.7] (-30:\CircleRadius cm) --node[midway, below] {$\Phi'\;\;\,$} (N3);
    \draw[->-=0.7] (N1) -- node[midway, left]{${\t\Phi}' $} (90:\CircleRadius cm);

    \draw[->-=0.7, dashed] (210:\CircleRadius cm) -- node[midway, above]{$\bH'$} (N2);
    \draw[->-=0.7, dashed] (N1) --node[midway,right]{$\Gamma'$} (N3);
    \draw[->-=0.7] (N2) --node[midway,left]{$M$} (N1);

\end{tikzpicture}
    \caption{ Right mutation for unitary SQCD. We use the notation ${n\ov q}\equiv U(n)_q$ to minimise clutter. Note the presence of the chiral and fermi mesons $M$ and $\Gamma'$, which couple to the other fields through the $E$- and $J$-terms as  explained in the text.}
    \label{fig:right mutation}
\end{figure}
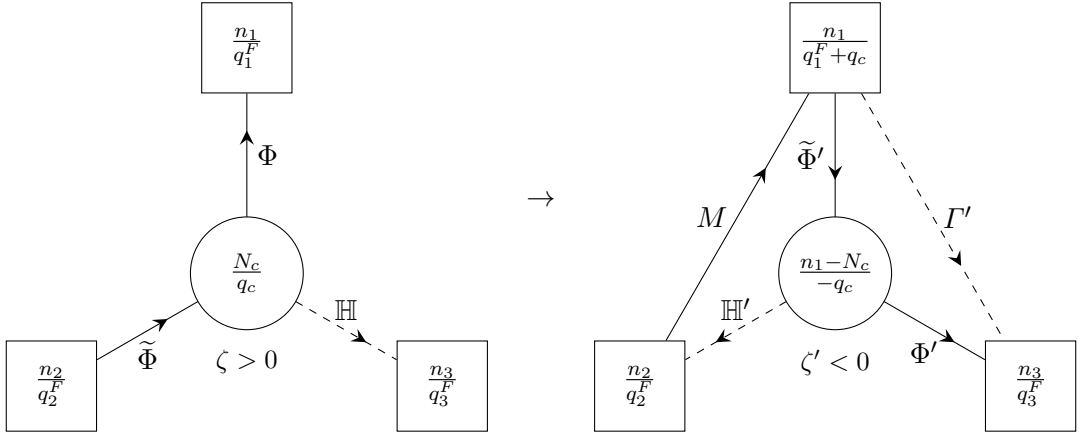

\noindent
Finally, we have the `mesons', namely $n_1 n_2$ gauge-invariant chirals  $M$ and $n_1 n_3$ fermis $\Gamma'$, which correspond to gauge-invariant operators in the original theory according to:
\be
 \t \Phi^j_\alpha \Phi^\alpha_i \;\longleftrightarrow\; M^j_i ~, \qquad \qquad
 \bar\Phi^i_\alpha \bH^\alpha_k  \;\longleftrightarrow\;{\Gamma'}^i_k~,
\ee
respectively. These mesons couple non-trivially to the gauge sector through the  $E$-term and $J$-terms:
\bea\label{EJ after right mutation simple}
&(E_{\Gamma'})^i_k= {{\widetilde{\Phi}}{}'}^{i}_\beta {\Phi'}^\beta_k~,\qquad\qquad
&&(J_{\Gamma'})^k_i= 0~,\\
&(E_{\bH'})^\beta_j= 0~,\qquad\qquad
&&(J_{\bH'})^j_\beta= M^j_i  {{\widetilde{\Phi}}{}'}^{i}_\beta~,\\
\eea
where $\beta=1, \cdots, n_1-N_c$ is a gauge index for the dual gauge group. The matter content of the dual gauge theory is given in table~\ref{table:matter 1dSQCD right mutated}. The duality is most easily pictured as a local operation on a quiver, as shown in figure~\ref{fig:right mutation}. As we explained in the introduction, the name {\it mutation} is appropriate for this Seiberg-like duality since it precisely follows the rules of mutations of graded quiver~\cite{buan2009coloured, Franco:2017lpa}.

\medskip
\noindent {\bf $\Gamma$-SQCD and general superpotential terms.} If we start with $\Gamma$-SQCD instead, we easily see that the superpotential term~\eqref{EGamma SQCD} becomes a linear term $E_{\Gamma}= M$ after the duality, which means that both $M$ and $\Gamma$ are massive and can be integrated out. Hence we obtain a more symmetric version of the duality, as depicted in figure~\ref{fig:right left mutations GammaSQCD}.

More generally, one could consider SQCD coupled to any set of fields through the more general superpotential terms~\eqref{EJLambda gen}. Then, under a right mutation, we have~\cite{Franco:2017lpa, Closset:2017xsc}: 
\bea\label{EJ after right mutation}
&(E_{\Gamma'})^i_k= {{\widetilde{\Phi}}{}'}^{i}_\beta {\Phi'}^\beta_k- {\d (E_{\bH})^\alpha_k\ov \d \Phi^\alpha_i}~,\qquad\qquad
&&(J_{\Gamma'})^k_i= - {\d (J_{\bH})^k_\alpha\ov \d \t\Phi^j_\alpha} M^j_i~,\\
&(E_{\bH'})^\beta_j= {\Phi'}^\beta_k {\d (J_{\bH})^k_\alpha\ov \d \t\Phi^j_\alpha}~,\qquad\qquad
&&(J_{\bH'})^j_\beta= M^j_i  {{\widetilde{\Phi}}{}'}^{i}_\beta~,\\
\eea
where all repeated indices are summed over. This obviously generalises~\eqref{EJ after right mutation simple} and includes $\Gamma$-SQCD as a special case. Note that we have
\be\label{susy EJ dual}
\Tr(EJ)=\Tr\left({\d E_{\bH}\ov \d \Phi} {\d J_{\bH} \ov \d \t\Phi} M\right) = \Tr( \h F \h G M)=0~,
\ee
where the last equality follows from~\eqref{TrEJ SQCD gen}.

\subsubsection{Left mutation}
For $\zeta<0$, we have the {\it left mutation}:
\bea\label{right mutation SQCD}
&\text{SQCD}\left[N_c~;\, Q_c~;\, (n_1, n_2, n_3)~;\, (Q_1^F, Q_2^F, Q_3^F)\right]^{\zeta<0}\quad  \longleftrightarrow  \qquad\qquad \\
&\,\text{SQCD}\left[n_2- N_c~;\, -Q_c-n_3~;\,  (n_2, n_3, n_1)~;\, (Q_2^F+Q_c+n_3, Q_3^F-N_c, Q_1^F)\right]^{\zeta'>0} \;{ \oplus M' \oplus \Gamma''}~.
\eea
This is the inverse of the right mutation, with the dual gauge groups:
\be
U(N_c)_{q_c}^{(\zeta<0)}\qquad \longleftrightarrow \qquad 
U(n_2-N_c)_{-q_c}^{(\zeta>0)}~.
\ee
Here the  antifundamental chirals become fundamental chirals, the fundamental chirals befome fundamental fermis, and the fundamental fermis become antifundamental chirals. The bare CS level of the dual theory is $Q_c' = -Q_c-n_3$, and we have:
\be
(q_1^F~,\, q_2^F~,\, q_3^F)\qquad \longleftrightarrow \qquad  (q_1^F~,\, q_2^F+q_c~,\, q_3^F)~,
\ee
while fermion number also shifts as:
\be
(-1)^{\rm F} \qquad \longleftrightarrow \qquad (-1)^{\rm F'}= (-1)^{n_3 N_c} (-1)^{\rm F}~.
\ee
\begin{table}
\renewcommand{\arraystretch}{1.2}
    \centering
    \begin{tabular}{|c||c||c|c| c|}
    \hline
    & $U(n_1-N_c)$& $U(n_1)$ & $U(n_2)$ & $U(n_3)$ \\ \hline   \hline
    ${\Phi''}^\beta_j$ & $\square$ &  ${\bf 1}$ &  $\overline{\square}$ & ${\bf 1}$ \\ 
    ${{\t\Phi}^{''k}_\beta}$  & $\overline{\square}$ & ${\bf 1}$  &${\bf 1}$&  $\square$\\
    ${\bH''}^\beta_i$ & $\square$ & $\overline{\square}$ & ${\bf 1}$  &${\bf 1}$  \\ 
    ${\Gamma''}^k_j$ & ${\bf 1}$ &  ${\bf 1}$ & $\overline{\square}$ &   $\square$   \\
    \hline
         ${M}^j_i$ & ${\bf 1}$ &   $\overline{\square}$ & $\square$ & ${\bf 1}$  \\
     \hline
\end{tabular}
    \caption{ Gauge and flavour charges of the dual of 1d SCQD obtained by a left mutation. Here $\beta=1, \cdots, n_2-N_c$ is a gauge index.   $M$ is integrated out in the $\Gamma$-SQCD variant.}
    \label{table:matter 1dSQCD left mutated}
\end{table}
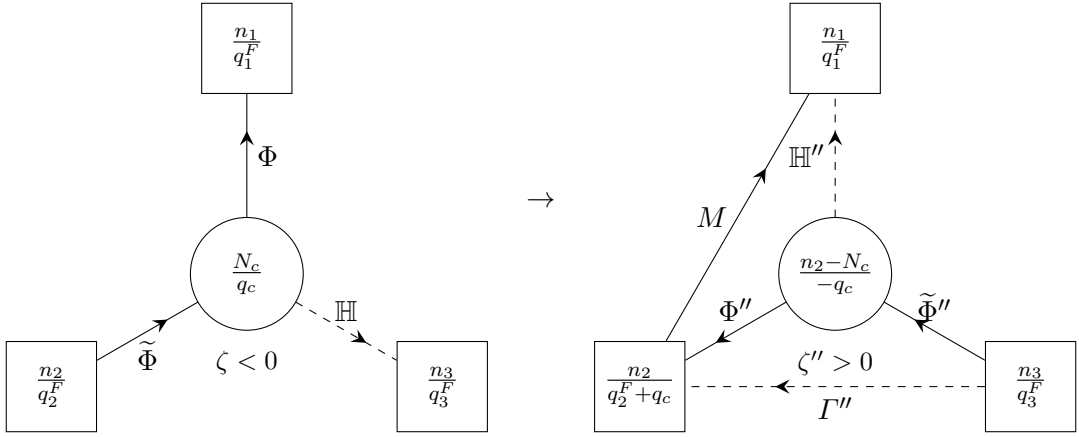
\begin{figure}
    \centering
\begin{tikzpicture}[>=stealth,scale=0.75]
 \def\SquareSize{1.2}  
    \def\CircleRadius{1}
    \def\NodeRadius{4}
    \draw (0,0) circle (\CircleRadius cm);

    \node[] (Nc) at (0,0) {${N_c\ov q_c}$};
    \node[draw, rectangle, shape aspect=1, minimum width= \SquareSize cm, minimum height=\SquareSize cm] (N1) at (90:\NodeRadius cm) {${n_1\ov q_1^F}$};
    \node[draw, rectangle, shape aspect=1, minimum width=\SquareSize cm, minimum height=\SquareSize cm] (N3) at (-30:\NodeRadius cm) {${n_3\ov q_3^F}$};
    \node[draw, rectangle, shape aspect=1, minimum width=\SquareSize cm, minimum height=\SquareSize cm] (N2) at (210:\NodeRadius cm) {${n_2\ov q_2^F}$};
    \node[below =0.48cm of Nc] {\small $\zeta < 0$};

    \draw[->-=0.7] (90:\CircleRadius cm) --node[midway, right] {$\Phi$} (N1);
    \draw[->-=0.7] (N2) -- node[midway, below]{${\t\Phi}$} (210:\CircleRadius cm);

    \draw[->-=0.7, dashed] (-30:\CircleRadius cm) -- node[midway, above]{$\bH$} (N3);

\end{tikzpicture}
\quad
\raisebox{3cm}{$\rightarrow$}
\quad
\begin{tikzpicture}[>=stealth,scale=0.75]
 \def\SquareSize{1.2}  
    \def\CircleRadius{1}
    \def\NodeRadius{4}
    \draw (0,0) circle (\CircleRadius cm);

    \node[] (Nc) at (0,0) {${n_2-N_c\ov -q_c}$};
    \node[draw, rectangle, shape aspect=1, minimum width= \SquareSize cm, minimum height=\SquareSize cm] (N1) at (90:\NodeRadius cm) {${n_1\ov q_1^F}$};
    \node[draw, rectangle, shape aspect=1, minimum width=\SquareSize cm, minimum height=\SquareSize cm] (N3) at (-30:\NodeRadius cm) {${n_3\ov q_3^F}$};
    \node[draw, rectangle, shape aspect=1, minimum width=\SquareSize cm, minimum height=\SquareSize cm] (N2) at (210:\NodeRadius cm) {${n_2\ov q_2^F+q_c}$};
    \node[below =0.48cm of Nc] {\small $\zeta'' > 0$};

    \draw[->-=0.7] (210:\CircleRadius cm) --node[midway, above] {$\Phi''$} (N2);
    \draw[->-=0.7] (N3) -- node[midway, above]{${\t\Phi}'' $} (-30:\CircleRadius cm);

    \draw[->-=0.7, dashed] (90:\CircleRadius cm) -- node[midway, left]{$\bH''$} (N1);
    \draw[->-=0.7, dashed] (N3) --node[midway,below]{$\Gamma''$} (N2);
  \draw[->-=0.7] (N2) --node[midway,left]{$M$} (N1);
\end{tikzpicture}
    \caption{ Left mutation for unitary SQCD, with the notation ${n\ov q}\equiv U(n)_q$. The couplings of the chiral and fermi mesons $M$ and $\Gamma''$ are explained in the text.}
    \label{fig:left mutation}
\end{figure}

\noindent
The mesons are the $n_1 n_2$ gauge-invariant chirals  $M$ and the $n_2 n_3$ fermis $\Gamma'$:
\be
 \t \Phi^j_a \Phi^\alpha_i \;\longleftrightarrow\; M^j_i ~, \qquad \qquad
 \t\Phi^j_\alpha \bH^\alpha_k  \;\longleftrightarrow\;{{\b\Gamma''}}^j_k~,
\ee
which couple to the gauge sector through the  $E$-term and $J$-terms:
\bea\label{EJ after right mutation simple}
&(E_{\Gamma''})^k_j= {{\widetilde{\Phi}}{}''}^{k}_\beta {\Phi''}^\beta_j~,\qquad\qquad
&&(J_{\Gamma''})^j_k= 0~,\\
&(E_{\bH''})^\beta_i=   {{\widetilde{\Phi}}{}''}^{b}_j M^j_i ~,\qquad\qquad
&&(J_{\bH''})^i_\beta=0~,\\
\eea
where $\beta=1, \cdots, n_2-N_c$ is a gauge index for the dual gauge group. The matter content of the dual gauge theory is given in table~\ref{table:matter 1dSQCD left mutated} and the duality is summarised in figure~\ref{fig:left mutation}.

\medskip
\noindent {\bf $\Gamma$-SQCD and general superpotential terms.}  Starting from the $\Gamma$-SQCD theory, we can again integrate out the meson $M$ together with the $\Gamma$ fermis in the dual theory, leading us to the simpler presentation of the duality as shown in figure~\ref{fig:right left mutations GammaSQCD}. More generally, 
starting with a larger theory with the superpotentials~\eqref{EJLambda gen}, we obtain the superpotentials:
\bea
&(E_{\Gamma''})^k_j= {{\widetilde{\Phi}}{}''}^{k}_\beta {\Phi''}^\beta_j -  {\d (J_{\bH})^k_\alpha\ov \d \t\Phi^j_\alpha}~,\qquad\qquad
&&(J_{\Gamma''})^j_k= - M^j_i {\d (E_{\bH})^\alpha_k\ov \d \Phi^\alpha_i}~,\\
&(E_{\bH''})^\beta_i=   {{\widetilde{\Phi}}{}''}^{b}_j M^j_i ~,\qquad\qquad
&&(J_{\bH''})^i_\beta=  {\d (E_{\bH})^\alpha_k\ov \d \Phi^\alpha_i} {\t\Phi}^{''k}_\beta~.\\
\eea
Supersymmetry is ensured by the same condition~\eqref{susy EJ dual} as for the right mutation. 

Finally, note that the right and left mutations are simply related to each other by a CP transformation of the full theory -- that is, the right mutation of the original SQCD theory gives us the same theory as the one obtained by a left mutation of the CP-conjugated SQCD theory.%
\footnote{A CP transformation is equivalent to a time-reversal transformation of the 1d $\CN=2$ gauge theory. It simply amounts to flipping the sign of all the gauge and flavour vector superfields simultaneously.} Moreover, the two mutation dualities are inverse of each other: a right mutation followed by a left mutation lands us back on the original theory.

\subsection{Matching the Witten index across the dualities}
The Witten index of any 1d $\CN=2$ gauge theory can be computed by supersymmetric localisation on the Coulomb branch~\cite{Hori:2014tda} -- we review some of the key results in appendix~\ref{app:Witten index}. 

In our conventions, the Witten index of the unitary SQCD theory~\eqref{SQCD general} is given by a contour integral:
\be\label{IW SQCD JK}
\IW[N_c, Q_c, \boldsymbol{n}^F]^{\pm} \equiv {(-1)^{N_c}\ov N_c!} \oint_{{\rm JK}_\pm} \prod_{\alpha=1}^{N_c}\left[ {d x_\alpha\ov 2\pi i x_\alpha} x_\alpha^{Q_c}\right] \prod_{\substack{\alpha,\beta=1 \\ \alpha\neq \beta}}^{N_c} \left(1- {x_\alpha\ov x_\beta}\right) \IW^{\rm matter}[N_c, \boldsymbol{n}^F]~,
\ee
where $\pm = \sign(\zeta)$,  the integer $Q_c$ related to $q_c$ as in~\eqref{qc to Qc}, and with the Witten index of the matter sector which reads:
\be\label{Imatter full}
\IW^{\rm matter}[N_c, \boldsymbol{n}^F] = \prod_{\alpha=1}^{N_c}\left[ {\prod_{k=1}^{n_3} \left(1- {x_\alpha \ov y_{3,k}}\right) \ov \prod_{i=1}^{n_1} \left(1- {x_\alpha \ov y_{1,i}}\right) \prod_{j=1}^{n_2} \left(1- {y_{2,j} \ov x_\alpha}\right)}\right]~,
\ee
in keeping with our conventions of subsection~\ref{subsec: quantize free fields}. We set $Q_I^F=0$ without loss of generality, since a generic contact term for the flavour symmetry simply enters as an overall factor:
\be
\IW[N_c, Q_c, \boldsymbol{n}^F,\boldsymbol{Q}^F]^\pm= \prod_{I=1}^3 \Bigg(\prod_{l_I=1}^{n_I} y_{I, l_I}\Bigg)^{Q_I^F}\,\times \, \IW[N_c, Q_c, \boldsymbol{n}^F]^\pm
\ee
The contour integral in~\eqref{IW SQCD JK} is a Jeffrey-Kirwan residue determined with the sign of the FI term. For $\zeta>0$, we pick all the residues at poles located at $x_\alpha= y_{1, i}$ for the chiral multiplets, while for $\zeta <0$ we pick all the residues at the poles located at $x_\alpha=y_{2, j}$ for the antichiral multiplets and introduce an additional overall sign $(-1)^{N_c}$. For future convenience, let us introduce the indexing set:
\be
    \CS_{n_I,  N_c}:=\{S\subset \{1, 2, \cdots, n_I\}\;:\; |S|=N_c\}~, \qquad\quad
    \left|   S_{n_I,  N_c}\right| = \binom{n_I}{N_c}~.
\ee
For $I=1$ or $2$, these sets index the distinct choices of contributing poles up to $U(N_c)$ Weyl symmetry, because the poles with $x_\alpha=x_\beta$ for $\alpha\neq \beta$ have vanishing residue. 
It is then straightforward to evaluate the JK residue explicitly, and thus to verify the dualities at the level of the index. 

Note that the Witten index of the $\Gamma$-SQCD theory is simply obtained by multiplying the SQCD index by the contribution from $n_1n_2$ free fermi multiplets, as:
\be\label{WIGammaSQCD}
\IW[N_c, Q_c, \boldsymbol{n}^F]^\pm_{\Gamma} \equiv \prod_{i=1}^{n_1}\prod_{j=1}^{n_2}\left(1-{y_{2,j}\ov y_{1, i}}\right)\times\IW[N_c, Q_c, \boldsymbol{n}^F]^\pm~.
\ee
Since the duality looks most symmetric in its $\Gamma$-SQCD formulation, this is the one we will check in what follows. Note that the parameters $q_I^F$ are then given by~\eqref{qIF full} with $\delta q_I^F=0$. 

\medskip
\noindent
{\bf Witten index for $\zeta>0$ and right mutation.} Consider first the case $\zeta>0$. We may assume that $n_1\geq N_c$, otherwise the Witten index vanishes. Let us denote the set of contributing poles, from fundamental chirals, by:
\be
S= \{i_\alpha\}_{\alpha=1}^{N_c} \in \CS_{n_1, N_c}~.
\ee
We then find the $\Gamma$-SQCD index:
\bea\label{IW explicit from JK plus}
&\IW[N_c, Q_c, \boldsymbol{n}^F]^\Gamma_+ &=&\; \sum_{S\in \CS_{n_1, N_c}} \prod_{i=1}^{n_1}\prod_{j=1}^{n_2}\left(1-{y_{2,j}\ov y_{1, i}}\right) \prod_{\alpha\neq \beta}^{N_c} \left(1-{y_{1, i_\alpha}\ov y_{1, i_\beta}}\right)\\
&&&\qquad\quad \times \prod_{\alpha=1}^{N_c} \left[{y_{1,i_\alpha}^{Q_c-1} \prod_{k=1}^{n_3} \left(1- {y_{1,i_\alpha} \ov y_{3,k}}\right) \ov \prod_{i\neq i_\alpha}^{n_1} \left(1- {y_{1,i_\alpha} \ov y_{1,i}}\right) \prod_{j=1}^{n_2} \left(1- {y_{2,j} \ov y_{1,i_\alpha}}\right)}\right]~.
\eea
and by straightforward manipulations one can check that:
 \bea\label{duality indentity right mutation}
&\IW[N_c, Q_c, (n_1, n_2, n_3),(0,0,0)]^\Gamma_+ \\
&\qquad=(-1)^{n_2(n_1-N_c)}\, \IW[n_1-N_c, -Q_c-n_2,(n_3, n_1, n_2), (0, Q_c, n_1-N_c)]^-_{\Gamma'}~.
 \eea
Namely, we precisely match the flavoured Witten index accross the right mutation defined above. The key to the proof is the realisation that we have a one-to-one match between non-zero residues by using the isomorphism between the sets $S$ and their complements $S^c$ inside $\{1, \cdots, n_1\}$:
\be
S \in \CS_{n_1, N_c} \qquad \longleftrightarrow \qquad S^c\in \CS_{n_1, n_1-N_c}~.
\ee
We refer to appendix~\ref{subsec:proof WI} for a detailed proof of the identity~\eqref{duality indentity right mutation}. 

\medskip
\noindent
{\bf Witten index for $\zeta<0$ and left mutation.} Similarly, consider the case $\zeta<0$ and the set of poles from antifundamental chirals indexed by
\be
S= \{j_\alpha\}_{\alpha=1}^{N_c} \in \CS_{n_2, N_c}~.
\ee
We then find the index:
\bea
   & \IW[N_c,Q_c,n^F]^-_\Gamma=\sum_{\mathcal{S}\in\mathcal{S}_{n_2,N_c}}\prod_{i=1}^{n_1}\prod_{j=1}^{n_2}(1-\frac{y_{2,j}}{y_{1,i}})\prod_{a\neq b}^{N_c}(1-\frac{y_{2,j_\alpha}}{y_{2,j_\beta}})\\
&\qquad\qquad    \times \prod_{\alpha=1}^{N_c}\frac{y_{2,j_\alpha}^{Q_c-1}\prod_{k=1}^{n_3}(1-\frac{y_{2,j_\alpha}}{y_{3,k}})}{\prod_{i=1}^{n_1}(1-\frac{y_{2,j_\alpha}}{y_{1,i}})\prod_{j\neq j_\alpha}^{n_2}(1-\frac{y_{2,j_\alpha}}{y_{2,j}})}~.
\eea
By similar manipulations as in the previous case, one finds the identity
 \bea\label{duality indentity left mutation}
&  \IW[N_c,Q_c,(n_1,n_2,n_3),(0,0,0)]^-_\Gamma\\
&\qquad=(-1)^{n_3 N_c}\, \IW[n_2-N_c,-Q_c-n_3,(n_2,n_3,n_1),(Q_c+n_3,-N_c,0)]^+_{\Gamma''}~,
\eea
which establishes the left mutation at the level of the index. 

\section{Ground states and dualities from the Higgs branch}\label{sec:Higgs}
For large FI parameter, the SQCD theory is a 1d GLSM that flows to a supersymmetric NLSM onto the Grassmannian manifold. In this section, we describe the supersymmetric ground states from this geometric perspective. This leads to an explicit expansion of the Witten index in terms of characters of the flavour group. It also nicely explains the mutation duality presented above as originating simply from the Grassmannian duality:
\be
{\rm Gr}(N_c, n_1)\cong {\rm Gr}(n_1-N_c, n_1)~.
\ee
We note that such a geometric perspective was first proposed in~\cite{Jia:2014ffa} to explain the 2d $\CN=(0,2)$ triality. Here we extend this approach to the more general target geometry allowed by our $\CN=2$ SQM.

\subsection{Ground states from sheaf cohomology for positive FI parameter}\label{subsec:HBpos}
Consider the case of large positive FI parameter, $\zeta>0$, with $n_1\geq N_c$. Then we have a target space
\be\label{EoverX gen}
\CE_+ \longrightarrow X_+~, \qquad \qquad X_+= {\rm Gr}(N_c, n_1)~,
\ee
where $X_+$ is the complex Grassmannian of $N_c$-planes inside $\C^{n_1}$, and $\CE_+$ is a particular vector bundle over $X_+$ to be discussed momentarily. 
By abuse of notation, we may call either $X_+$ or the total space of the bundle $\CE_+$ the `Higgs branch' of the GLSM, by contrast to the Coulomb-branch description that we will consider in later sections. By standard arguments, the supersymmetric ground states $\ket{\Psi}$ are in one-to-one correspondence with the sheaf cohomology classes:
\be\label{group states Hbullet}
\ket{\Psi} \qquad \longleftrightarrow \qquad [\Psi]\in H^\bullet(X_+, \CE_+)~.
\ee
Using the Borel--Weil--Bott theorem, we can explicitly compute this cohomology for any $GL(n_1, \C)$-equivariant bundle~\cite{fulton2013representation, weyman2003cohomology}.

\medskip
\noindent
{\bf The vector bundle $\CE_+$ for $\Gamma$-SQCD.}  The geometry~\eqref{EoverX gen} is more easily discussed in the case of $\Gamma$-SQCD, in which case the superpotential interactions render the bosonic target space effectively compact. Recall that the rank-$N_c$ tautological bundle $\CS$ and the rank-$(n_1-N_c)$ quotient bundle $\CQ$ over the Grassmannian $X_+$ are related by the Euler sequence:
\be\label{SQ euler}
0 \longrightarrow \CS \longrightarrow \C^{n_1} \longrightarrow \CQ\longrightarrow 0~.
\ee
Recall that the constant modes of the fundamental chiral multiplets are essentially the homogeneous coordinates of $X_+$; they form an $N_c\times n_1$ matrix $(\phi^\alpha_i)$ which, up to ${\rm GL}(N_c)$ transformations ({\it i.e.} complexified gauge transformations), parameterise points on $X_+$. The gauge theory construct $X_+$ as the K\"ahler quotient
\be
X_+\cong \C^{N_c n_1}/\!/_{\zeta}\,  U(N_c)~.
\ee
To understand the bundle $\CE_+$, we must understand the geometric interpretation of the $n_2$ antifundamental fields $\t\phi^j_\alpha$, of the $n_3$ fundamental fermi multiplets $\bH^\alpha_k$, and of the integer 1d Chern--Simons level $Q_c$. The effect of the latter is most readily explained: it modifies Gauss's law, which allows states that are generated by homogenous polynomials in the $\phi^i_\alpha$'s.%
\footnote{More precisely, they are polynomials in the $SU(N_c)$-invariant Pl\"ucker coordinates of the Grassmannian.} The net effect is to introduce a factor of $\det(\CS)^{Q_c}$ in the bundle $\CE_+$. The gauge variables $x_\alpha$ that appear in the localisation formula~\eqref{IW SQCD JK} can be interpreted as exponentiated Chern roots for $\CS$, and we then have the Chern character:
\be
{\rm ch}(\det(\CS))=  \det(x)\equiv \prod_{\alpha=1}^{N_c} x_\alpha~.
\ee
Next, we note that the fundamental fermi multiplets $\bH^\alpha_j$ are valued in the tautological vector bundle, by their very definition. Their quantisation with $\zeta>0$ leads to a fermionic Fock space generated by the modes $\bH^\alpha_j$, in our conventions.
We then have a factor 
\be
\Lambda^\bullet(\CS^{\oplus n_3}) = \Lambda^\bullet(\CS \otimes \C^{n_3})~.
\ee
in the bundle $\CE_+$. Here $\Lambda^\bullet F$ denotes the full exterior algebra of a vector bundle $F$, and this should be understood in K-theory --- namely, we are taking into account formal signs keeping track of the exterior-algebra degree.

The role played by the antifundamental chiral multiplets when $\zeta>0$ is more subtle. Considered in isolation, the fields $\t\phi^j= (\phi^j_\alpha)$ are sections of the dual tautological vector bundle $\CS^\ast$, and the $\Gamma^j_i$ fields are valued in the trivial line bundle $\C^{n_1 n_2}$.  
In $\Gamma$-SQCD, the antifundamental chiral multiplets $\t\phi^j_\alpha$ are coupled to the gauge-neutral fermi multiplets $\Gamma^j_i$ by the $E_\Gamma$ terms which impose the constraints:
\be\label{E term constraint on tphi}
(E_\Gamma)^j_i=\t\phi^j_\alpha \phi^\alpha_i=0
\ee 
on the constant modes. The full interactions can be deduced from the superpotential~\eqref{W N02}, which we can write as:
\be\label{W GammaSQCD}
W=\phi^\alpha_i  \b\Gamma^i_j \t\phi^j_\alpha~.
\ee
On the Grassmannian, the $N_c\times n_1$ matrix $\phi^\alpha_i$ has rank-$N_c$, thus~\eqref{E term constraint on tphi} directly imposes that $\t\phi^j_\alpha=0$, removing all the non-compact directions of the classical moduli space. Moreover, at any point on the Grassmannian with fixed VEV for $\phi^\alpha_i$ of maximal rank, the superpontential~\eqref{W GammaSQCD} gives a mass to all the fields $\t\phi^j_\alpha$ and to all the gauge-neutral fermions $\Gamma^j_i$ that are parallel to the $N_c$-plane spanned by $\phi^i_\alpha$'s (the latter being viewed as $N_c$ distinct vectors in $\C^{n_1}$). In other words, for each fixed $U(n_2)$ index $j$, the only remaining massless modes are the fermions $\Gamma^{i}_j$ orthogonal to the $\phi^i_\alpha$'s viewed as sections of $\CS$. At every point $x$ on the Grassmannian, the massless $\Gamma$ fields are thus valued in the $(n_1-N_c)$-plane orthogonal to the $\C^{N_c}$ plane corresponding to $x\in X_+$. That means that the massless $\Gamma$ fields are valued in  $\CQ^\ast$, the dual of the quotient bundle $\CQ$ defined by~\eqref{SQ euler}.%
\footnote{That we have $\CQ^\ast$ instead of $\CQ$ is important, and due to the fact that it is $\bar{\Gamma}$ and not $\Gamma$ that enters in~\protect\eqref{W GammaSQCD}.}
 These massless fermions thus contribute a factor $\Lambda^\bullet((\CQ^\ast)^{\oplus n_2})$ to the vector bundle $\CE_+$.

In summary, for $\Gamma$-SQCD with positive FI parameter, we have a target space which is the total space of the vector bundle $\CE_+$ over the Grassmannian, with:
\be\label{CE total}
\CE_+= \det(\CS)^{Q_c} \otimes \Lambda^\bullet ((\CQ^\ast)^{\oplus n_2}) \otimes \Lambda^\bullet (\CS^{\oplus n_3})~.
\ee
As mentioned above, we consider $\CE_+$ as a K-theory class, with formal signs introduced when taking the exterior algebra.

\medskip
\noindent
{\bf Ground states from the Borel--Weil--Bott theorem.}
Let us compute the group states~\eqref{group states Hbullet} for the bundle~\eqref{CE total}.  This sheaf cohomology can be computed very explicitly. First, let $Y(n,m)$ denote the set of all Young tableaux that fit into an $n\times m$ rectangle:
\be
Y(n,m)\equiv \big\{ \lambda= [\lambda_1, \cdots, \lambda_n]\, \big|\, m \geq  \lambda_1\geq \lambda_2 \geq \cdots\geq \lambda_n \geq 0 \big\}~.
\ee
Then, let $\lambda\in Y(n_1-N_c, n_2)$ denote a highest weight for ${\rm GL}_{n_1-N_c}$ and let $\mu\in Y(N_c, n_3)$ denote a highest weight for ${\rm GL}_{N_c}$. The bundle $\CE_+$ then decomposes into Schur-functor vector bundles corresponding to irreducible representations of $U(n_2)\times U(n_3)$ according to:
\be\label{CE decomposition}
\CE_+= \bigoplus_{\substack{\lambda\in Y(n_1-N_c, n_2) \\ \mu\in Y(N_c, n_3)}} (-1)^{|\lambda|+|\mu|} \,  \CF_{\lambda, \mu}(-Q_c)\otimes S_{\lambda^T}(\C^{n_2}) \otimes S_{\mu^T}(\C^{n_3})~,
\ee
where we defined the vector bundle:
\be\label{defFlm}
 \CF_{\lambda, \mu}(-Q_c) \equiv \det(\CS)^{Q_c} \otimes S_{\lambda}(\CQ^\ast)\otimes  S_{\mu}(\CS)~,
\ee
and $\lambda^T$ denotes the transpose of the Young tableau $\lambda$. Note the signs in~\eqref{CE decomposition}, wherein $\CE_+$ is taken as a K-theory class.

Now, given a pair of weights $\lambda, \mu$ as in~\eqref{CE decomposition}, we define the ${\rm GL}_{n_1}$ weight:
\be
\omega = [-Q_c- \mu_{N_c}, -Q_c-\mu_{N_c-1}, \cdots, -Q_c- \mu_1, \lambda_1, \lambda_2, \cdots, \lambda_{n_1-N_c}]~,
\ee
which is a weight corresponding to~\eqref{defFlm}. 
Let $\rho= (n_1-1, n_1-2, \cdots, 1, 0)$ denote the Weyl vector of ${\rm GL}_{n_1}$. The weight $\omega$ is called regular if $\omega+\rho$ has no repeated entry. If $\omega$ is regular, their exists a minimal permutation $\sigma\in S_{n_1}$ of its entries such that $\sigma(\omega+\rho)$ has non-increasing entries. We then define a highest weight:
\be\label{def nu from om}
\nu = \sigma(\omega+\rho)- \rho~,
\ee
and we denote by $\ell(\sigma)$ the length of the permutation $\sigma$ (that is, the minimal number of transpositions it can be decomposed into). The Borel--Weil--Bott theorem then gives us the following explicit expression for the cohomology of~\eqref{defFlm}:
\be
H^k(X_+,  \CF_{\lambda, \mu}(-Q_c))= \begin{cases} 
S_\nu(\C^{n_1}) \qquad & \text{if $\omega$ is regular and } k=\ell(\sigma)~,\\
0&\text{otherwise}
\end{cases}
\ee
Putting it all together, the explicit expression for the full cohomology reads:
\be\label{HXCE full}
H^\bullet(X_+, \CE_+)= \bigoplus_{\substack{\lambda\in Y(n_1-N_c, n_2) \\ \mu\in Y(N_c, n_3) \\ \omega \; \text{regular}}} (-1)^{|\lambda|+|\mu|}\, S_\nu(\C^{n_1}) \otimes S_{\lambda^T}(\C^{n_2}) \otimes S_{\mu^T}(\C^{n_3})~.
\ee
Here, any given pair of weights $\lambda$ and $\mu$ contribute to the degree-$\ell(\sigma)$ cohomology if and only if $\omega$ is regular. Note that the fermion number of the states transforming in the representation $(\nu; \lambda^T; \mu^T)$ of $U(n_1)\times U(n_2)\times U(n_3)$ is given by
$(-1)^{\rm F}= (-1)^{\ell(\sigma)+|\lambda|+|\mu|}$.

\subsubsection{The Witten index from sheaf cohomology}
The Witten index of $\Gamma$-SQCD with $\zeta>0$ straightforwardly follows from~\eqref{HXCE full}. Indeed, the Witten index~\eqref{WIGammaSQCD} in the geometric phase is simply given by the holomorphic Euler characteristic of the bundle $\CE_+$:
\be
\IW[N_c, q_c, \boldsymbol{n}^F]_+^{\Gamma} = \chi_T(X_+, \CE_+) \equiv \sum_{k=0}^{{\rm dim}_T(X_+)} (-1)^k \, {\rm dim}_T\, H^k(X_+, \CE_+)~.
\ee
More precisely, the flavoured Witten index is given here  in terms of the equivariant Euler characteristic  $\chi_T(X_+, \CE_+)$, where $T$ is a maximal torus for the flavour group~\eqref{GF def}, with equivariant parameters $(y_I)=(y_{1,i}, y_{2,j}, y_{3,k})$. Given any of the complex vector spaces $S_\lambda(\C^{n_I})$ appearing in \eqref{HXCE full}, we define its  `equivariant dimension'  to be the Schur polynomial $\schur_\lambda$ in the $n_I$ variables $y_I$:
\be
{\rm dim}_T(S_\lambda(\C^{n_I})) \equiv \schur_\lambda(y_I)~.
\ee
Therefore, given the decomposition~\eqref{HXCE full}, we find a completely explicit expression for the Witten index:
\be\label{HB formula pos}
{\IW}_+=  \sum_{\substack{\lambda\in Y(n_1-N_c, n_2) \\ \mu\in Y(N_c, n_3) \\ \omega \; \text{regular}}} (-1)^{|\lambda|+|\mu|+ \ell(\sigma)}\, \schur_\nu(y_1^{-1}) \schur_{\lambda^T}(y_2)   \schur_{\mu^T}(y_3^{-1})~.
\ee
Unlike the Coulomb-branch formula, this gives us the index directly as a polynomial
\be
\IW \in \Z[y_I^\pm]~,
\ee
and it elucidates the exact flavour-symmetry representations spanned by the ground states. Recall also that the non-equivariant Euler characteristic can be written as an integral over $X_+$ according to the Hirzebruch-Riemann-Roch theorem:
\be\label{X index thm}
\chi(X_+, \CE_+) = \int_{X_+} {\rm Td}(X_+)\wedge {\rm ch}(\CE_+)~.
\ee
The formula~\eqref{HB formula pos} gives us an explicit expression for this integer:
\be
\chi(X_+, \CE_+) = \sum_{\substack{\lambda\in Y(n_1-N_c, n_2) \\ \mu\in Y(N_c, n_3) \\ \omega \; \text{regular}}} (-1)^{|\lambda|+|\mu|+ \ell(\sigma)}\, \dim(\nu; n_1) \dim(\lambda^T; n_2)\dim(\mu^T;n_3)~,
\ee
where $\dim(\lambda; n)=\schur_\lambda(1)$ denotes the dimension of the representation $\FR_\lambda$ of ${\rm GL}_{n}$.

\subsubsection{Special cases and examples}

It is useful to look at some special cases of this ground-state computation using sheaf cohomology. We will also present a few explicit examples of the Higgs-branch formula~\eqref{HB formula pos} for the index.

\medskip
\noindent {\bf Fundamentals only.} Setting  $n_2=n_3=0$, we are considering SQCD with only fundamental chiral multiplets, and we simply have the geometry of a line bundle:
\be
[\CE_+   \longrightarrow  X_+]  \quad = \quad  [\det(\CS)^{Q_c}  \longrightarrow   {\rm Gr}(N_c, n_1)]~.
\ee
In this case we have a single trivial pair of weights, $\lambda=\mu=0$, and so we must only consider:
\be
\omega= [\underbrace{-Q_c, \cdots, -Q_c}_{N_c \text{ times}}, \underbrace{0, 0,  \cdots, 0}_{n_1-N_c \text{ times}}]~.
\ee
If $Q_c\leq 0$, we see that $\omega$ is regular, with $\ell(\sigma)=0$ and $\nu=\omega$. If $0<Q_c<n_1$, one easily check that $\omega$ is irregular. If $Q_c \geq n_1$, $\omega$ is again regular and one finds $\ell(\sigma)=N_c(n_1- N_c)= {\rm dim}(X_+)$ and:
\bea\label{nu cohom detS}
&\nu&=&\;  [\underbrace{-N_c, \cdots, -N_c}_{n_1-N_c \text{ times}}, \underbrace{-Q_c+n_1-N_c,   \cdots, -Q_c+n_1-N_c}_{N_c \text{ times}}]\\
&&=&\; [\underbrace{Q_c-n_1,  \cdots, Q_c-n_1}_{n_1-N_c \text{ times}}, \underbrace{0,   \cdots, 0}_{N_c \text{ times}}]+(-Q_c+n_1-N_c)[1, \cdots, 1]
\eea
Thus, for $Q_c\leq 0$, the ground states are:
\be
\CH^{(Q_c\leq 0)} \cong H^0(X_+, \det(\CS)^{Q_c}) =  {\rm Sym}^{|Q_c|}(\Lambda^{N_c}\C^{n_1})~.
\ee
This corresponds to a rectangular Young tableau of size $N_c\times |Q_c|$. These global sections are the homogenous polynomials of degree $|Q_c|$ in the Pl\"ucker coordinates of the Grassmannian (modulo the Pl\"ucker relations). For $0<Q_c<n_1$ there are no ground states. For $Q_c\geq n_1$, we have:
\be
\CH^{(Q_c\geq n_1)} \cong H^{ {\rm dim}(X_+)}(X_+, \det(\CS)^{Q_c}) =  S_\nu(\C^{n_1})\cong {\rm Sym}^{Q_c-n_1}(\Lambda^{n_1-N_c}\C^{n_1})~,
\ee
with $\nu$ given by~\eqref{nu cohom detS}. This latter cohomology can be understood using Serre duality:
\be
H^{ {\rm dim}(X_+)}(X_+, \det(\CS)^{Q_c})\cong H^{0}(X_+, \det(\CS)^{n_1-Q_c})^\ast \cong H^{0}(X_+, \det(\CQ)^{Q_c-n_1})^\ast~,
\ee
using the fact that the canonical line bundle is given by $\CK_{X_+}=\det(\CS)^{n_1}$. The flavoured Witten index is then given by:
\be\label{IW detS case}
\IW=  \begin{cases}
\schur_{[(-Q_c)^{N_c},0^{n_1-N_c}]}(y_1^{-1}) \qquad\quad &\text{if }  Q_c\leq 0~,\\
 0 & \text{if }  0<Q_c < n_1~,\\
  (-1)^{\dim(X_+)} \det(y_1)^{Q_c-n_1+N_c}\, \schur_{[(Q_c-n_1)^{n_1-N_c},0^{N_c}]}(y_1^{-1})   \quad & \text{if }  Q_c \geq n_1~,\\
\end{cases}
\ee
where we used the notation $\det(y_1)\equiv \prod_{i=1}^{n_1} y_{1,i}$.

%
%
%
%
%

\medskip
\noindent
{\bf $\Gamma$-SQED.} In the abelian case, $N_c=1$, we simply have $X_+=\mathbb{P}^{n_1-1}$ and $\CS= \CO(-1)$. The decomposition~\eqref{CE decomposition} simplifies to:
\be\label{Eplus abelian}
\CE_+= \bigoplus_{\substack{\lambda\in Y(n_1-N_c, n_2) \\ p\in \{0, \cdots, n_3\} }} (-1)^{|\lambda|+p} \, S_\lambda(\CQ^\ast)(-Q_c-p)  \otimes S_{\lambda^T}(\C^{n_2}) \otimes S_{[1^p]}(\C^{n_3})~, 
\ee
where $\CF(n)\equiv \CF\otimes \CO(n)$, $\mu=[p]$, and we have
$\omega= [-Q_c-p, \lambda_1, \cdots, \lambda_{n_1-1}]$. One interesting special case is when $Q_c\leq -n_2-n_3$, in which case every $\lambda\in  Y(n_1-N_c, n_2)$ gives a regular $\omega$ and contributes to the cohomology with $\nu= [|p+Q_c|, \lambda_1, \cdots, \lambda_{n_1-1}]$. Another very special case is for $n_2=0$, in which case we have:
\be\label{E abelian special case}
\CE_+= \bigoplus_{p=0}^{n_3} (-1)^p\, \CO(-p-Q_c)\otimes S_{[1^p]}(\C^{n_3})~.
\ee
In this case, the non-flavoured Witten index is well-defined and given by the  residue formula:
\be\label{IW abel expl JK}
\IW = - \oint {d x\ov 2 \pi i x}  x^{Q_c} {(1-x)^{n_3}\ov  (1-x)^{n_1}}~,
\ee
where we pick the residue at $x=1$ in accordance with the JK prescription. One can show that formula directly follows from the index theorem~\eqref{X index thm} for $X_+=\mathbb{P}^{n_1-1}$. Indeed, expanding out the numerator of the integrand, we can rewrite~\eqref{IW abel expl JK} as:
\be
  -\sum_{p=0}^{n_3} (-1)^p \binom{n_3}{p}  \oint {d x\ov 2 \pi i x}  x^{Q_c+p} {1\ov  (1-x)^{n_1}} = \sum_{p=0}^{n_3} (-1)^p \binom{n_3}{p}  \chi(\mathbb{P}^{n-1}, \CO(-Q_c-p))~,
\ee
which precisely matches the expansion~\eqref{E abelian special case}. 
Finally, the Euler characteristic $\chi(\mathbb{P}^{n-1}, \CO(-Q_c))$ can be computed from the abelian theory with $n_2=n_3=0$, and corresponds to setting $N_c=1$ in~\eqref{IW detS case}. This case was first discussed in detail in~\cite{Hori:2014tda}.


\subsection{Dualities of 1d SQCD from dual Grassmannians}
To explain the mutation duality from the point of view of the Grassmannian target, we first consider the geometry obtained from $\Gamma$-SQCD with negative FI parameter. 

\subsubsection{Ground states for negative FI parameter}
In the case $\zeta<0$, the target geometry becomes:
\be\label{EoverX gen minus}
\CE_- \longrightarrow X_-~, \qquad \qquad X_-= {\rm Gr}(N_c, n_2)~,
\ee
in terms of the Grassmannian of $N_c$-planes inside $\C^{n_2}$, and with the bundle:
\be\label{CEminus0}
\CE_-= \det(\CS^\ast)^{Q_c} \otimes \Lambda^\bullet((\CQ^\ast)^{\oplus n_1})\otimes \Lambda^\bullet((\CS^\ast)^{\oplus n_3})~,
\ee
by a reasoning similar to that explained above for $\zeta>0$. In particular, while the $x_\alpha$'s were interpreted as the exponentiated Chern roots of the tautological bundle $\CS$ over $X_+$, the exponentiated Chern roots of $\CS$ over $X_-$ are now $x_\alpha^{-1}$, hence $x_\alpha$ are the exponentiated Chern roots of $\CS^\ast$, the dual tautological bundle.%
\footnote{In particular, we have $\prod_{\alpha=1}^{N_c} x_\alpha = {\rm ch}(\det(\CS^\ast))={\rm ch}(\det(\CS)^{-1})$ on $X_-$.} The fields $\t\phi^j_\alpha$ are now interpreted as sections of $\CS$ and the fields $\phi^\alpha_i$ are set to zero by the superpotential constraints. 

Using the identity:
\be
\Lambda^\bullet(\CS^\ast) = (-1)^{N_c} \det(\CS)^{-1}\otimes \Lambda^\bullet(\CS)~, 
\ee
we can usefully rewrite~\eqref{CEminus0} as:
\be\label{CE minus bis}
\CE_-=(-1)^{n_3 N_c} (\det y_3)^{-N_c} \det(\CS)^{-(Q_c+n_3)} \otimes \Lambda^\bullet((\CQ^\ast)^{\oplus n_1})\otimes \Lambda^\bullet(\CS^{\oplus n_3})~.
\ee
Here we kept track of the equivariant parameters, with $\det y_3\equiv \prod_{k=1}^{n_3} y_{3, k}$. Then, exactly as above, we find that the ground states in the phase with $\zeta<0$ are given by:
\be\label{HXCE full neg}
H^\bullet(X_-, \CE_-)= (-1)^{n_3 N_c} \bigoplus_{\substack{\lambda\in Y(n_2-N_c, n_1) \\ \mu\in Y(N_c, n_3) \\ \omega \; \text{regular}}} (-1)^{|\lambda|+|\mu|}\, S_\nu(\C^{n_2}) \otimes S_{\lambda^T}(\C^{n_1}) \otimes S_{\mu^T}(\C^{n_3})~,
\ee
where, for each $\lambda$, $\mu$, we define the ${\rm GL}_{n_2}$ weight
\be
\omega= [Q_c+n_3 - \mu_{N_c}, \cdots, Q_c+n_3-\mu_1, \lambda_1, \cdots, \lambda_{n_2-N_c}]~,
\ee
the Weyl vector $\rho=(n_2-1, \cdots, 1, 0)$, and the highest weight $\nu$ defined exactly as in~\eqref{def nu from om}. We therefore find the flavoured Witten index:
\be\label{HB formula neg}
{\IW}_-= (-1)^{n_3 N_c} (\det y_3)^{-N_c} \sum_{\substack{\lambda\in Y(n_2-N_c, n_1) \\ \mu\in Y(N_c, n_3) \\ \omega \; \text{regular}}} (-1)^{|\lambda|+|\mu|+ \ell(\sigma)}\, \schur_\nu(y_2) \schur_{\lambda^T}(y_1^{-1})   \schur_{\mu^T}(y_3)~,
\ee
 to be compared to~\eqref{HB formula pos}. Note the overall contact term which arises from the equivariant prefactor in~\eqref{CE minus bis}.

\subsubsection{Dual Grassmanniann and right mutation}
Under exchanging $N_c$-planes inside $\C^{n_1}$ with the orthogonal $(n_1-N_c)$-planes, we obtain an isomorphism:
\be
\varphi \; : \;X=  {\rm Gr}(N_c, n_1)\; \longrightarrow\;  X_D= {\rm Gr}(n_1-N_c, n_1)~.
\ee
Let us denote by $\CF$ and $\CF_D$ the vector bundles over $X$ and $X_D$, respectively, and by $\varphi^\ast$ the pull-back map. We then have:%
\footnote{It is important to note that, if $X= {\rm Gr}(N_c, V)$ with $V\cong \C^{n_1}$, then  $\varphi$ maps into ${\rm Gr}(n_1-N_c, V^\ast)$. Indeed the `orthogonal plane' is better defined as the set of dual vectors that annihilate a given hyperplane in $V$, without the need for choosing a pairing on $V$ itself. This explains the dualisation in the natural map~\protect\eqref{map S to Qast}.}
\be\label{map S to Qast}
\varphi^\ast(\CS_D)= \CQ^\ast~, \qquad\qquad
\varphi^\ast(\CQ_D)= \CS^\ast~,
\ee
as well as:
\be
\varphi^\ast(\det(\CS_D)) = \det(\CQ^\ast) = \det(\CQ)^{-1}=(\det y_1)^{-1} \det(\CS)~.
\ee
In the last equality we used the fact that $y_{1,i}$ are the equivariant parameters for the trivial bundle $\C^{n_1}$ in the middle of the short exact sequence~\eqref{SQ euler}.

The right mutation for $\Gamma$-SQCD corresponds to the geometric duality:
\be\label{geometric duality}
(\CE_+\rightarrow X) \qquad \longleftrightarrow \qquad (\CE_-^D \rightarrow X_D)~,
\ee
where we have:
\bea
&\CE_+ &=& \; \det(\CS)^{Q_c} \otimes \Lambda^\bullet ((\CQ^\ast)^{\oplus n_2}) \otimes \Lambda^\bullet (\CS^{\oplus n_3})~, \\
&\CE_-^D &=& \; (\det y_1)^{Q_c} \det(\CS_D)^{Q_c} \otimes \Lambda^\bullet ((\CQ_D^\ast)^{\oplus n_3}) \otimes \Lambda^\bullet (\CS_D^{\oplus n_2})~.
\eea
Here we used the duality map $N_c^D= n_1-N_c$ and $Q_c^D= -Q_c-n_2$ together with the fact that the contact terms in the dual theory (namely  $Q^F_1= Q_c$, $Q^F_2=N_c^D$, and the overall shift~\eqref{shift fermion nbr right mut} of the fermion number) contribute the prefactor:
\be
(-1)^{n_3 N_c^D} (\det y_1)^{Q_c} (\det y_2)^{N_c^D}~,
\ee
which partially cancels out against the prefactors in~\eqref{CE minus bis} (with the appropriate substitution of the parameters according to the duality map). Hence, we precisely have:
\be
\varphi^\ast(\CE_-^D) = \CE_+~.
\ee
This proves the $\Gamma$-SQCD right mutation beyond the index computation, since it shows that the geometric target spaces~\eqref{geometric duality} are indeed isomorphic in the infrared, and therefore that the supersymmetric ground states are exactly the same in the two 1d $\CN=2$ supersymmetric gauge theories, at least in the limit $\zeta \rightarrow \infty$.

Similarly, the left mutation corresponds to the Grassmannian duality:
\be
\t\varphi \; : \;X=  {\rm Gr}(N_c, n_2)\; \longrightarrow\;  X_D= {\rm Gr}(n_2-N_c, n_2)~,
\ee
with:
\bea
&\CE_- &=& \;{(-1)^{n_3 N_c} \ov (\det y_3)^{N_c}}\det(\CS)^{-(Q_c+n_3)} \otimes \Lambda^\bullet ((\CQ^\ast)^{\oplus n_1}) \otimes \Lambda^\bullet (\CS^{\oplus n_3})~, \\
&\CE_+^D &=& \; {(-1)^{n_3 N_c} \ov (\det y_3)^{N_c}} (\det y_2)^{Q_c+n_3} \det(\CS_D)^{-(Q_c+n_3)} \otimes \Lambda^\bullet ((\CQ_D^\ast)^{\oplus n_3}) \otimes \Lambda^\bullet (\CS_D^{\oplus n_1})~.
\eea
Note that the equivariant parameters for the trivial bundle $\C^{n_2}$ over $X$ are now $y_2^{-1}$, hence we have
\be
\varphi^\ast(\det(\CS_D)) = \det y_2\, \det(\CS)
\ee 
in this case. We then easily check that:
\be
\t\varphi^\ast(\CE_+^D) = \CE_-~.
\ee
This completes the proof of the dualities in the Higgs phase.

\section{Abelian gauge theories in the Coulomb phase}\label{sec:Coulomb}

In this section, we study the abelian theory, $N_c=1$, in its Coulomb phase, following the logic explained in section~\ref{subsec:U1 EFT} to explicitly compute the perturbative ground states. This then allow us to recompute the flavoured Witten index as:
\be
\IW = \sum_{\Psi} \bra{\Psi} (-1)^{\rm F} Y \ket{\Psi}~,
\ee
where $\ket{\Psi}$ form a basis of orthonormal ground states, and $Y= \prod_{I=1}^3 y_I^{{\bf J}_I^F}$ denotes the exponentiated flavour charges. In general, as anticipated in  section~\ref{subsec:U1 EFT}, there are more perturbative ground states than true ground states, but of course the Witten index still gives us the correct answer when computed from the perturbative ground states.

\subsection{$U(1)_{q_c}$ with $n_1$ fundamental chirals}\label{subsec:fund only}
Consider first the case $n_2=n_3=0$. The $U(1)_{q_c}$ theory with $n_1$ chiral multiplets of gauge charge $+1$ has the bare CS level $Q_c= q_c+{n_1\ov 2}$. In the Higgs phase with $\zeta>0$, this gives us the $\mathbb{P}^{n_1-1}$ model with $\CE= \CO(-Q_c)$. We then find the ground states:%
\footnote{Note that $\Lambda^{n_1-1}\C^{n_1}\cong \C^{n_1}$, but it is useful to leave the exterior product explicit in the third line as it matches the cohomological degree.}
\be\label{states Psi U1q geom}
\ket{\Psi}\cong \begin{cases}
    H^0(\mathbb{P}^{n_1-1},\CO(-Q_c))\cong {\rm Sym}^{|Q_c|}(\C^{n_1})  \qquad &\text{if }\, Q_c\leq 0~,\\
    0  &\text{if }\, 0<Q_c< n_1~,\\
    H^{n_1-1}(\mathbb{P}^{n_1-1},\CO(-Q_c)) \cong {\rm Sym}^{Q_c-n_1}(\Lambda^{n_1-1}\C^{n_1}) \quad &\text{if }\, Q_c\geq n_1~.
\end{cases}
\ee
On the other hand, there are no supersymmetric ground states when $\zeta<0$. 

Let us see how one reproduces these results from the Coulomb-branch analysis. At constant value of $\sigma$, the quantisation of the free chiral multiplets with mass $\omega= |\sigma|$ gives us standard bosonic and fermionic Fock spaces populated by the states:
\be
\ket{s_{1},  t_{1}, f_{1}}\equiv \prod_{i=1}^{n_1}\frac{(a_{1,i}^{\dag})^{s_{1,i}}}{\sqrt{s_{1,i}!}}\frac{(b_{1,i}^{\dag})^{ t_{1,i}}}{{\sqrt{t_{1,i}!}}} {\b\psi}_{1,i}^{f_{1,i}}\ket{0}_+~, \qquad s_{1,i}~,\;t_{1,i}\in\mathbb{Z}_{\geq 0} \; \text{ and }\; f_{1,i}\in\{0,1\}~,
\ee
if $\sigma>0$,  and similarly for $\sigma<0$. 
Recall the form of the supercharges in this free theory:
\bea\label{Q_n1}
  &  \mathcal{Q}=-\sqrt{2}\sum_{i=1}^{n_1}(\Pi_{\phi_{1,i}}\psi_{1,i}-i\b{\phi}_{1,i}\sigma\psi_{1,i})=\begin{cases}
        2i\sqrt{\omega}\sum_{i=1}^{n_1}b_{1,i}^\dag\psi_{1,i}\quad &\text{if } \sigma>0~,\\
        -2i\sqrt{\omega}\sum_{i=1}^{n_1}a_{1,i}\psi_{1,i}\quad &\text{if }\sigma<0~,
    \end{cases}
    \\
&    \b{\mathcal{Q}}=-\sqrt{2}\sum_{i=1}^{n_1}(\b{\psi}_{1,i}\b{\Pi}_{\b{\phi}_{1,i}}+i\b{\psi}_{1,i}\sigma\phi_{1,i})=\begin{cases}
        -2i\sqrt{\omega}\sum_{i=1}^{n_1}b_{1,i}\b{\psi}_{1,i}\quad &\text{if } \sigma>0~,\\
        2i\sqrt{\omega}\sum_{i=1}^{n_1}a_{1,i}^{\dag}\b{\psi}_{1,i}\quad &\text{if }\sigma<0~.
    \end{cases} 
\eea
For simplicity of notation, let us denote by $\ket{0} \equiv \ket{0}_+$ the `empty' ground state for $\sigma>0$, and we then have
\be\label{ket0pm U1 case1}
\ket{0}_- = \prod_{i=1}^{n_1}\b{\psi}_{1,i}\ket{0}~,
\ee
so that we might write down all the states in terms the flavour-neutral ket $\ket{0}$.
 One easily finds the supersymmetric ground states to be of the form:
\be
    \mathcal{F}^0=\left\{\begin{cases}
       \ket{s_{1}}_+\equiv  \prod_{i=1}^{n_1}\frac{1}{\sqrt{s_{1,i}!}}(a^{\dag}_{1,i})^{s_{1,i}}\ket{0}\qquad &\text{if } \sigma>0~,\\
        \ket{t_{1}}_- \equiv \prod_{i=1}^{n_1}\frac{1}{\sqrt{t_{1,i}!}}(b^{\dag}_{1,i})^{t_{1,i}} \b{\psi}_{1,i}\ket{0}\qquad\quad &\text{if } \sigma<0~.
    \end{cases}\right\}~.
\ee
The next step is to make the $U(1)$ gauge field dynamical. Firstly, we must impose Gauss's law: all physical states should be gauge invariants. Since the vacuum $\ket{0}$ carries $U(1)$ charge $Q_c$, one finds the conditions:
\begin{align}\label{FPhys Pn model}
    \mathcal{F}^0_{\rm phys}=\left\{ \begin{cases}
          \ket{s_{1}}_+ \quad | \quad  \sum_{i=1}^{n_1} s_{1,i} + Q_c=0 \quad &\text{if } \sigma>0~,\\
       \ket{t_{1}}_- \quad |\quad  -\sum_{i=1}^{n_1} t_{1,i} -n_1+ Q_c=0 \quad  &\text{if } \sigma<0~.
    \end{cases}\right\}~.
\end{align}
The state at $\sigma>0$ only exist if $Q_c\leq 0$, while the states at $\sigma<0$ only exist if $Q_c\geq n_1$. There are no ground states for $0< Q_c<n_1$. Finally, we need to consider the dynamics of the $\sigma$ field itself, as discussed in section~\ref{subsec:U1 EFT}. The effective theory depends on the free-field ground state~\eqref{FPhys Pn model}. In each non-trivial sector, we have the superpotential:
\be
h(\sigma)= -\zeta \sigma  +N_{\phi} \sign(\sigma) \log|\sigma|~,\qquad \text{with}\quad 
N_{\phi} =\begin{cases}
    n_1-Q_c \qquad &\text{if } Q_c\leq 0~,\\
   Q_c \qquad &\text{if } Q_c\geq n_1~.
\end{cases} 
\ee
Note that $N_\phi \geq n_1$ in all cases. We then have the scalar potential:%
\footnote{Here and in the following, we consider the rescaled FI parameter $\t\zeta$ introduced around~\protect\eqref{Usigma general}, but we still call it $\zeta$ to avoid clutter. Thus, when we consider large $\zeta$ later one this really means $M_C \gg 0$, while we are still in the Coulomb-branch scaling limit.}
\begin{align}
    U(\sigma)=\frac{1}{2}\left(\frac{N_{\phi}}{2 |\sigma|}-\zeta\right)^2~.\label{n_2=0 potential}
\end{align}
We see that a critical point at $|\sigma_\ast|=N_\phi/(2\zeta)$ can only exist if $\zeta>0$, which confirms that there are no ground state if $\zeta<0$. Otherwise, for $Q_c\leq 0$ and $Q_c\geq n_1$, we find the ground states
\be
\ket{\Psi_+; s_1}= \ket{s_1}_+  \qquad\text{and} \qquad \ket{\Psi_-; t_1}=  \b\lambda \ket{t_1}_-~,
\ee
respectively. The $\ket{t_1}_-$ states are dressed by a gaugino due to the fact that these states live at $\sigma_\ast<0$ such that $h''(\sigma_\ast)>0$, as reviewed in section~\ref{subsec:U1 EFT}. 
These Coulomb-branch states match exactly the cohomology classes~\eqref{states Psi U1q geom}.

\begin{figure}
    \centering
    \begin{tabular}{c c}
         \includegraphics[width=0.48\linewidth]{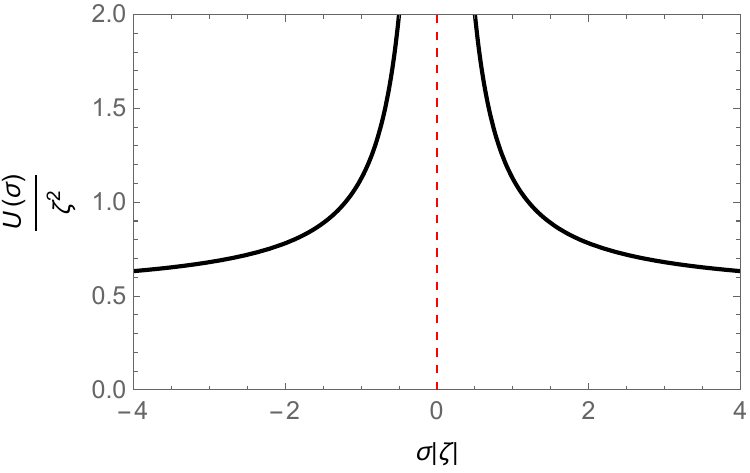}& \includegraphics[width=0.48\linewidth]{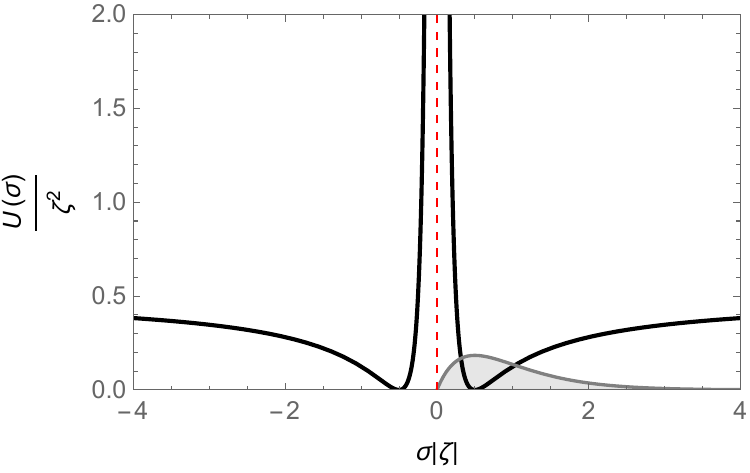} \\         
    \end{tabular}
    \caption{ The scalar potential~\eqref{n_2=0 potential} with $N_\phi=1$ in both cases, $\zeta<0$ (\textsc{Left}) and $\zeta>0$ (\textsc{Right}). Note that both axes are dimensionless. Depending on $Q_c$, we should only look at $\sigma>0$ (for $Q_c\leq 0$) or $\sigma<0$ (for $Q_c\geq n_1$), with no state living on the other side of the Coulomb branch. Here we also indicated the absolute value squared of the wave function if $Q_c \leq 0$.}
\end{figure}

\medskip
\noindent {\bf Recovering the flavoured Witten index.} Finally, we note that the ground states carry the flavour charges:
\be
Y \ket{\Psi_+; s_1} = \prod_{i=1}^{n_1} y_{1,i}^{-s_i} \ket{\Psi_+; s_1}~, \qquad \qquad
Y\ket{\Psi_-; t_1}= \prod_{i=1}^{n_1} y_{1,i}^{s_i+1}\ket{\Psi_-; t_1}~,
\ee
and the fermion numbers:
\be
(-1)^{\rm F}\ket{\Psi_+; s_1} = \ket{\Psi_+; s_1} ~, \qquad \qquad
(-1)^{\rm F}\ket{\Psi_-; t_1}= (-1)^{n_1-1}\ket{\Psi_-; t_1}~.
\ee
Hence, we can directly compute the Witten index. For $Q_c\leq 0$, we need to sum over all states   $\ket{\Psi_+; s_1}$ indexed by the sets $s_1 = (s_{1,i}) \in C_{n_1}(-Q_c; s_1)$, where we defined the set of $n$-tuples:
\be
 C_{n}(K; k)\equiv \left\{ k= (k_1, \cdots, k_n) \, \Big| \, \sum_{i=1}^n k_i = K\right\}~. 
\ee
The index is thus the fully symmetric homogeneous polynomial of degree $|Q_c|$ in the $n_1$ variables $y_1^{-1}$:
\be
\IW^{(Q_c\leq 0)} = \sum_{s_1 \in C_{n_1}(-Q_c; s_1)} \frac{1}{y_{1}^{s_{1}}}  = \schur_{[-Q_c,\,  0^{n_1-1}]}(y_1^{-1})~, 
\ee
with the obvious shorthand notation $y_1^{s_1}\equiv  \prod_{i=1}^{n_1} y_{1,i}^{s_{1,i}}$. Similarly, for $Q_c\geq n_1$ we have:
\be
\IW^{(Q_c\geq n_1)} =(-1)^{n_1-1}\det(y_1) \sum_{t_1 \in C_{n_1}(Q_c-n_1; t_1)} y_1^{t_1}  = \schur_{[Q_c-n_1,\,  0^{n_1-1}]}(y_1)~. 
\ee
Noting the identity,
\be
\schur_{[Q_c-n_1,\,  0^{n_1-1}]}(y_1)= \det(y_1)^{Q_c-n_1} \schur_{[(Q_c-n_1)^{n_1-1},\,  0]}(y_1^{-1})~,
\ee
this reproduces the result~\eqref{IW detS case} for $N_c=1$. Finally,  the numerical Witten index reads:
\begin{align}
  \IW|_{y_1=1} =\begin{cases}
        \binom{|Q_c|+n_1-1}{n_1-1} &\text{if }\, Q_c\leq 0~,\\
        0&\text{if }\,  0<Q_c<n_1~,\\
        (-1)^{n_1-1} \binom{Q_c-1}{n_1-1} \qquad &\text{if }\,  Q_c\geq n_1~.
    \end{cases}\label{Un_flav_index}
\end{align}
Note that this $(\mathbb{P}^{n_1-1}, \CO(-Q_c))$ model and its Coulomb branch states were already studied in detail in~\cite{Hori:2014tda}; of course we reproduce their results here.

 \subsection{$U(1)_{q_c}$ with $n_1$ fundamental chirals and $n_3$ fundamental fermis}\label{subec:U1n1n3}

The next level of difficulty is a slight generalisation of the previous case, where we add $n_3$ fundamental fermi multiplets, corresponding to $\Gamma$-SQED with $n_2=0$. We then have $Q_c= q_c+{n_1\ov 2}-{n_3\ov 2}$. From the Higgs branch analysis, we know that the ground states correspond to the cohomology:
\be
\ket{\Psi} \cong \bigoplus_{p=0}^{n_3} (-1)^p H^\bullet(\mathbb{P}^{n_1-1}, \CO(-p-Q_c))\otimes \Lambda^p(\C^{n_3})~.
\ee
As before, we denote by $\ket{0}=\ket{0}_+$ the `empty' ground state at $\sigma>0$, and we then have
\be\label{ket0pm U1 case2}
\ket{0}_- = \prod_{i=1}^{n_1}\b{\psi}_{1,i} \prod_{k=0}^{n_3}\eta_{3,k}\ket{0}
\ee
at $\sigma<0$, generalising~\eqref{ket0pm U1 case2}. 
The Coulomb branch analysis is straightforward in this case because we do not have any superpotential interactions and the supercharges remain the same as in~\eqref{Q_n1}. We then find the free-field ground states:
\be
    \mathcal{F}^0=\left\{\begin{cases}
       \ket{s_{1}, f_{3}}_+\equiv  \prod_{i=1}^{n_1}\frac{1}{\sqrt{s_{1,i}!}}(a^{\dag}_{1,i})^{s_{1,i}} \prod_{k=0}^{n_3} \eta_{3,k}^{f_{3,k}}\ket{0}\qquad &\text{if } \sigma>0~,\\
        \ket{t_{1}, f_3}_- \equiv \prod_{i=1}^{n_1}\frac{1}{\sqrt{t_{1,i}!}}(b^{\dag}_{1,i})^{t_{1,i}} \b{\psi}_{1,i}\prod_{k=0}^{n_3} \eta_{3,k}^{f_{3,k}}\ket{0}\qquad\quad &\text{if } \sigma<0~.
    \end{cases}\right\}~,
\ee
where $f_{3,k}\in \{0,1\}$. Here we used the fact that, for $\sigma<0$, all the oscillators $\eta_{3,k}$ should be turned on (in our conventions), but we obviously have that:
\be
\prod_{k=1}^{n_3} \b\eta_{3,k}^{\t f_{3,k}}\eta_{3, k}\ket{0} = \prod_{k=0}^{n_3} \eta_{3,k}^{f_{3,k}}\ket{0} \qquad \text{for }\, f_{3,k}= \t f_{3,k}+1 \; \text{mod } 2~.
\ee
Imposing Gauss's law gives us the states:
\begin{align}\label{FPhys Pn model}
    \mathcal{F}^0_{\rm phys}=\left\{ \begin{cases}
          \ket{s_{1},f_3}_+ \quad | \quad  \sum_{i=1}^{n_1} s_{1,i} +\sum_{k=1}^{n_3} f_{3,k}+ Q_c=0 \qquad &\text{if } \sigma>0~,\\
       \ket{t_{1},f_3}_- \quad | \quad  -\sum_{i=1}^{n_1} t_{1,i}+\sum_{k=1}^{n_3} f_{3,k} -n_1+ Q_c=0 \quad &\text{if } \sigma<0~.
    \end{cases}\right\}~.
\end{align}
The scalar potential $U(\sigma)$ is unchanged compared to the $n_3=0$ case, hence we only find ground states for $\zeta>0$. We thus have $n_3+1$ distinct instances of the $n_3=0$ result, with $Q_c$ shifted to $Q_c+p$ with:
\be\label{def p n3}
p \equiv \sum_{k=0}^{n_3} f_{3,k} \in \{0, \cdots, n_3\}~.
\ee
The true supersymmetric ground states are therefore
\be
\ket{\Psi}\cong  \begin{cases}
    \ket{\Psi_+; s_1, f_3}= \ket{s_1}_+  \quad & \text{if }Q_c+p\leq 0~,\\
    \ket{\Psi_+; t_1, f_3}=  \b\lambda \ket{t_1}_-\quad  & \text{if }Q_c+p\geq n_1~,
\end{cases}  
\ee
with no ground states whenever $0<Q_c+p<n_1$. (In particular, the theory has no supersymmetric ground states at all if $|q_c|< {n_1-n_3\ov 2}$.) 
 Note the $U(n_3)$ fugacities:
 \be
Y_3\ket{\Psi} = \prod_{k=1}^{n_3} y_{3,k}^{- f_{3,k}}\ket{\Psi}~.
 \ee
At fixed $p$, we find states indexed by elements $f_3$ of the set:
\be
 F(p, n_3; f_3)\equiv \left\{(f_{3,k})\in\{0,1\}^{n_3} \; \Big|\; \sum_{k=1}^{n_3}f_{3,k}=p\right\}~,\qquad \text{with }\; 
\left|F(p,n_3; f_3)\right|= \binom{n_3}{p}~,
\ee
which thus contribute to the flavoured index a factor of the Schur polynomial for the fully antisymmetric representation $[1^p]$ of $U(n_3)$, namely:
\be
\sum_{f_3\in  F(p, n_3; f_3)} \prod_{k=1}^{n_3} y_{3,k}^{- f_{3,k}} = \schur_{[1^p,0^{n_3-p}]}(y_3^{-1})~.
\ee 
We thus find the Witten index:%
\be
\IW[1, Q_c, (n_1,0, n_3)] =  \sum_{p=0}^{n_3}(-1)^p  \schur_{[1^p,0^{n_3-p}]}(y_3^{-1})\;  \IW[1, Q_c+p, (n_1,0, 0)]~,
\ee
in terms of the $n_3=0$ index computed above. This exactly reproduces the Higgs-phase computation.

\subsection{$U(1)_{q_c}$ with $n_1$ fundamentals, $n_2$ antifundamentals and no $\Gamma$ fields}

We turn next to the theory with both fundamental and antifundamental multiplets, while setting $n_3=0$ for simplicity. Let us first consider the SQED theory without $\Gamma$ field. In the case of vanishing flavour masses, we expect to have a gapless sector related to the non-compact target space due to the gauge-invariant field $\t\phi\phi$. In order to lift this flat direction, we turn on a real mass $m$ along $U(n_1)\times U(n_2)$ so that the fundamental and antifundamental chiral multiplets obtain a common mass $\pm\sigma+m$, respectively, on the Coulomb branch. Equivalently, we turn on a complexified fugacity:
\be\label{def by flavour}
y_{1, i} = \by^{-1}~, \qquad y_{2,j}=\by~, \qquad \forall i, j~,
\ee
with $m\propto -\log \by$. The distinct frequencies of the fundamental and antifundamental chiral multiplets are then denoted by:
\be\label{mass to omegas}
\omega= |\sigma+m|~, \qquad\qquad \t\omega =|-\sigma+m|~,
\ee
respectively. Let us choose $m>0$ for definiteness. The supercharges of the free theory are now given by:
\bea\label{N1N2Q}
  &  \mathcal{Q}={\begin{cases}
        2i\sqrt{\omega}\sum_{i=1}^{n_1}b_{1,i}^\dag\psi_{1,i}-2i\sqrt{\t\omega}\sum_{j=1}^{n_2}a_{2,j}\psi_{2,j}\qquad & \text{if } \,\sigma>m~,\\
        2i\sqrt{\omega}\sum_{i=1}^{n_1}b_{1,i}^\dag\psi_{1,i}+2i\sqrt{\t\omega}\sum_{j=1}^{n_2}b^\dag_{2,j}\psi_{2,j}\qquad & \text{if } |\sigma|<m~,\\
        -2i\sqrt{\omega}\sum_{i=1}^{n_1}a_{1,i}\psi_{1,i}+2i\sqrt{\t\omega}\sum_{j=1}^{n_2}b^\dag_{2,j}\psi_{2,j}\qquad & \text{if }\sigma<-m~,
    \end{cases}}
    \\
  &  \bar{\mathcal{Q}}={\begin{cases}
        -2i\sqrt{\omega}\sum_{i=1}^{n_1}b_{1,i}\bar{\psi}_{1,i}+2i\sqrt{\t\omega}\sum_{j=1}^{n_2}a^\dag_{2,j}\bar{\psi}_{2,j}\qquad & \text{if } \sigma>m~,\\
        -2i\sqrt{\omega}\sum_{i=1}^{n_1}b_{1,i}\bar{\psi}_{1,i}-2i\sqrt{\omega}\sum_{j=1}^{n_2}b_{2,j}\bar{\psi}_{2,j}\qquad & \text{if } |\sigma|<m~,\\
        2i\sqrt{\omega}\sum_{i=1}^{n_1}a^\dag_{1,i}\bar{\psi}_{1,i}-2i\sqrt{\omega}\sum_{j=1}^{n_2}b_{2,j}\bar{\psi}_{2,j}\qquad & \text{if }\sigma<-m~.
    \end{cases} }
\eea
There are now three distinct regions on the Coulomb branch. Similarly to the previous cases, we now define 
\be
\ket{0}= \ket{0}_{++}\qquad \text{such that}\quad  \psi_{1,i}\ket{0}= \psi_{2,j}\ket{0}=0~,
\ee
which is the `empty' state in the region $|\sigma|<m$.
The ground states of the free theory, before imposing Gauss's law, are given by: 
\be\label{F0n3eq0}
    \mathcal{F}^0=\left\{\begin{cases}
       \ket{s_1, t_2}_{+-}\equiv \prod_{i=1}^{n_1}\frac{(a_{1,i}^\dag)^{s_{1,i}}}{\sqrt{s_{1,i}!}}\prod_{j=1}^{n_2}\frac{(b_{2,j}^\dag)^{ t_{2,j}}}{\sqrt{t_{2,j}!}}\bar{\psi}_{2,j}\ket0 \qquad &\text{if }\, \sigma>m\\
        \ket{s_1, s_2}_{++}\equiv \prod_{i=1}^{n_1}\frac{(a_{1,i}^\dag)^{ s_{1,i}}}{\sqrt{s_{1,i}!}}\prod_{j=1}^{n_2}\frac{(a_{2,j}^\dag)^{s_{2,j}}}{\sqrt{s_{2,j}!}}\ket0 \qquad &\text{if }\,|\sigma|<m\\
         \ket{t_1, s_2}_{-+}\equiv\prod_{i=1}^{n_1}\frac{(b_{1,i}^\dag)^{ t_{1,i}}}{\sqrt{t_{1,i}!}}\bar{\psi}_{1,i}\prod_{j=1}^{n_2}\frac{(a_{2,j}^\dag)^{ s_{2,j}}}{\sqrt{s_{2,j}!}}\ket0\quad \qquad &\text{if }\, \sigma<-m
    \end{cases}\right\}~,
\ee
and we then obtain the would-be physical states:
\be\label{FPhys Pn model}
    \mathcal{F}^0_{\rm phys} =
    \left\{\begin{cases}
       \ket{s_1, t_2}_{+-} \quad\, | \quad  |s_1|+|t_2|+n_2+Q_c=0 \qquad\quad &\text{if }\, \sigma>m\\
        \ket{s_1, s_2}_{++} \quad |\quad  |s_1|-|s_2|+Q_c=0\qquad &\text{if }\,|\sigma|<m\\
         \ket{t_1, s_2}_{-+} \quad\, | \quad  -|t_1|-|s_2|-n_1+Q_c=0 \quad \qquad &\text{if }\, \sigma<-m
    \end{cases}\right\}~.
\ee
Here we introduced the notation:
\be
|s_1|\equiv \sum_{i=1}^{n_1} s_{1,i}~, \qquad
|s_2|\equiv \sum_{j=1}^{n_2} s_{2,j}~, \qquad 
|t_1|\equiv \sum_{i=1}^{n_1} t_{1,i}~, \qquad 
|t_2|\equiv \sum_{j=1}^{n_2} t_{1,j}~.
\ee
Note that the states at $\sigma>0$ can only exist if $Q_c\leq -n_2$, and those at $\sigma<-m$ can only exist if $Q_c\geq n_1$. In particular, we never have states at the same time on both external regions of the $\sigma$ line.

Last but not least, let us see which perturbative ground states can arise depending on the sign of the FI parameter. The scalar potential reads:
\be
    U(\sigma)=\frac{1}{2}\left(\frac{N_\phi}{2|\sigma+m|}-\frac{N_{\t \phi}}{2|-\sigma+m|}-\zeta\right)^2~,\label{potental}
\ee
with $N_\phi \geq n_1$ and $N_{\t\phi}\geq n_2$ for all states in each region. We analyse this potential in the limit of large $\zeta$, which means that we consider some arbitrarily large Coulomb-branch scale $M_C$. We then find that, for $\zeta>0$, there can be one vacuum in the region $\sigma<-m$ and one vacuum in the region $|\sigma|<0$, while for $\zeta<0$ there is one vacuum with $\sigma>0$ and one with $|\sigma|<0$. Let us explore these perturbative ground states in more detail. 
\begin{figure}
    \centering
    \begin{tabular}{c c}
         \includegraphics[width=0.48\linewidth]{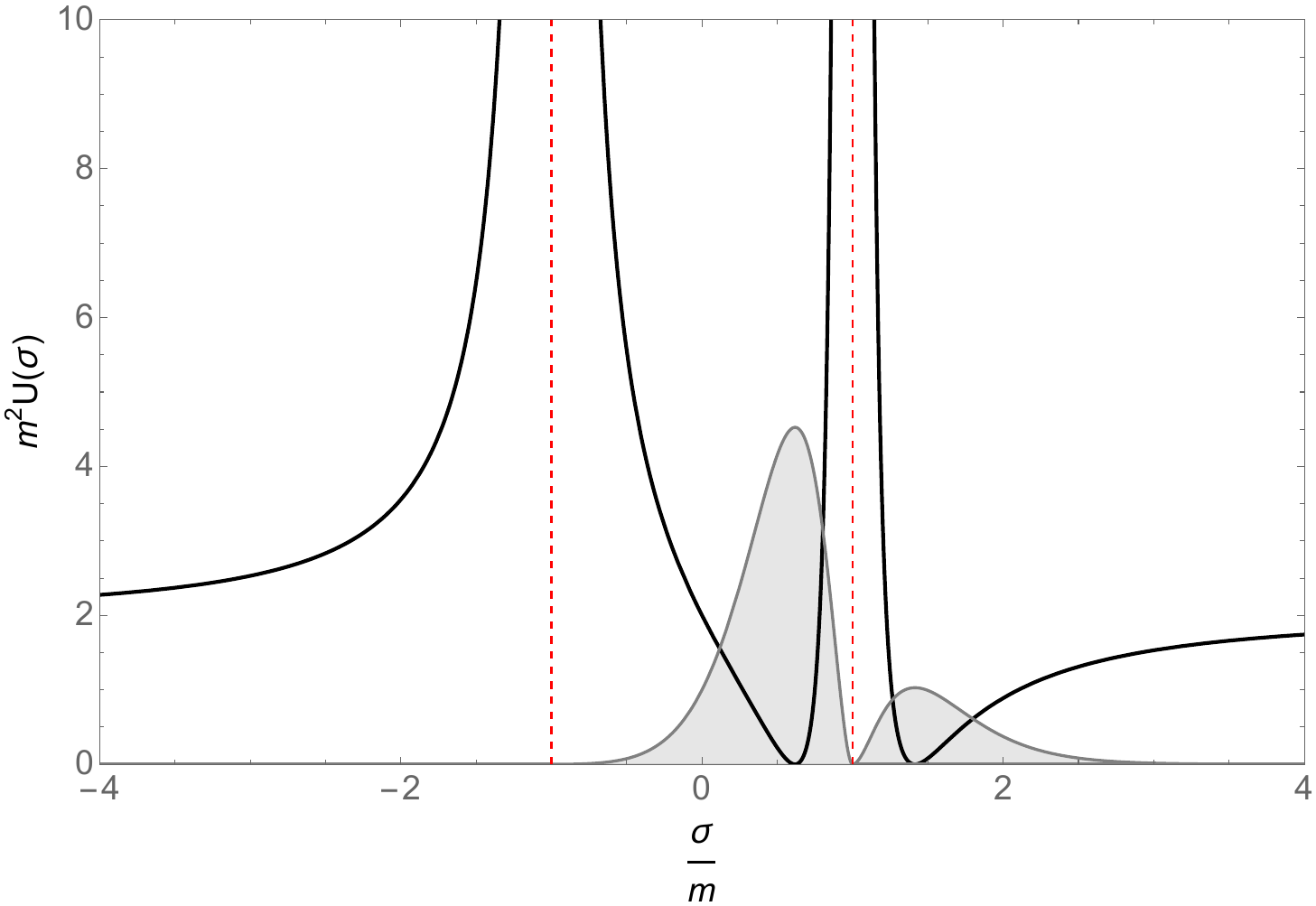}& \includegraphics[width=0.48\linewidth]{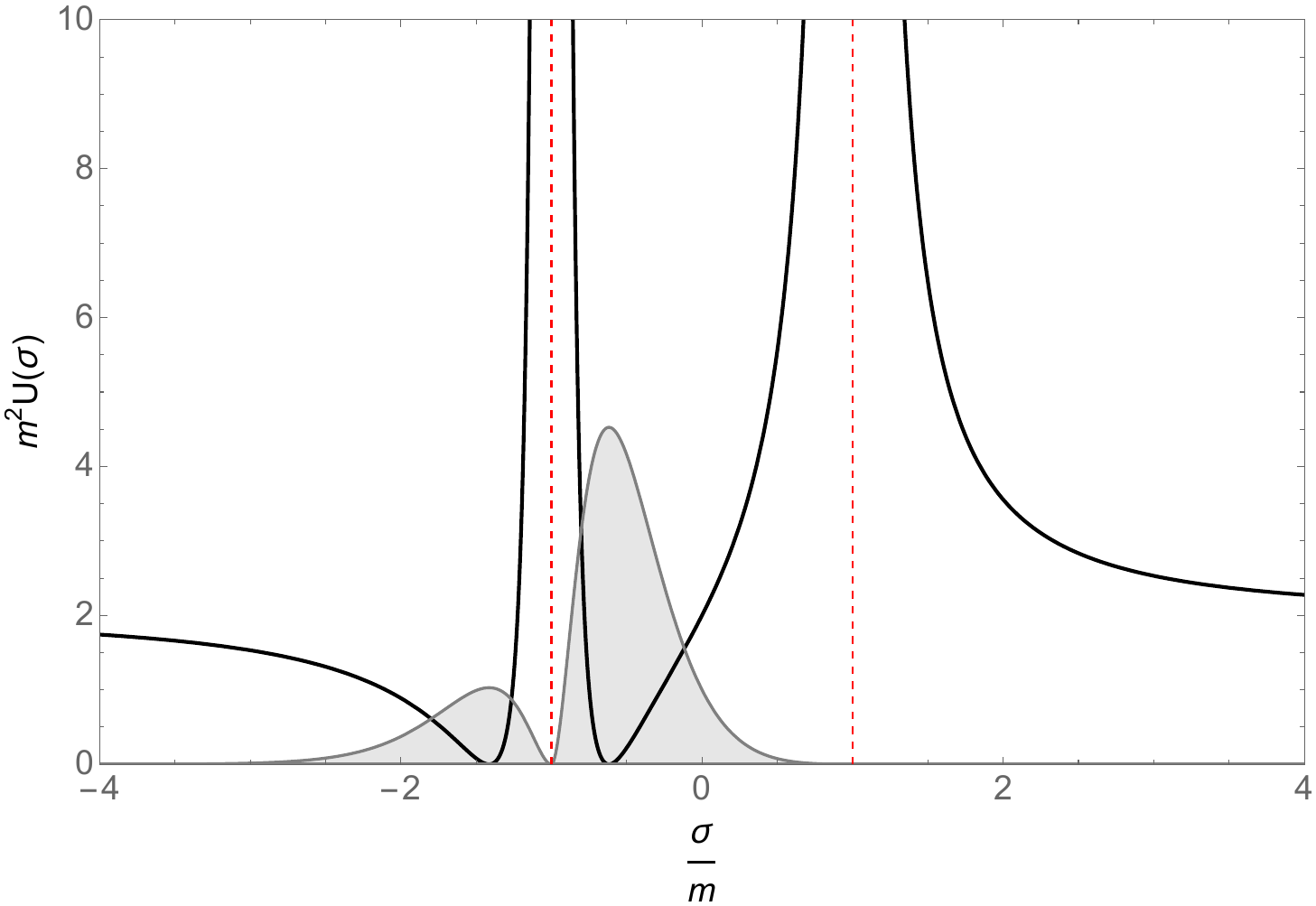} \\         
    \end{tabular}
    \caption{The scalar potential~\protect\eqref{potental} with $N_\phi=N_{\tilde{\phi}}=2$ in both cases, $\zeta m=-2$ (\textsc{Left}) and $\zeta m=2$ (\textsc{Right}). We indicated the modulus squared of the wavefunction as well (with the one for the state located in the external region scaled up by $10^3$ for effect). }
\end{figure}
\subsubsection{Perturbative ground states with $|\sigma|>m$}
We begin by analysing the `exterior' regions at $|\sigma|>m$. 
 In the region $\sigma>m$ and in the limit $|\zeta|\rightarrow \infty$, we have a critical point if and only if $\zeta$ is negative:
\be
\sigma_\ast \approx  m - {N_{\t\phi}\ov 2\zeta}~, \qquad h''(\sigma_\ast) >0~.
\ee
Similarly, in the region $\sigma<-m$, a critical point exists if and only if $\zeta$ is positive:
\be
\sigma_\ast \approx -m - {N_{\phi}\ov 2\zeta}~, \qquad h''(\sigma_\ast) >0~.
\ee
We therefore find the perturbative ground states:
\be
\ket{\Psi}\cong  \begin{cases}
    \ket{\Psi_{+-}; s_1, t_2}= \b\lambda\ket{s_1,t_1}_{+-}  \quad & \text{if }\, \zeta<0\, \text{ and }\, Q_c\leq -n_2~,\\
    \ket{\Psi_{-+}; t_1, s_2}=  \b\lambda \ket{t_1, s_2}_{-+}\quad  & \text{if }\, \zeta>0\, \text{ and }\, Q_c\geq n_1.
\end{cases}  
\ee
Note the gaugino insertion due to the fact that $h''(\sigma_\ast)>0$. 
There are only a finite number of perturbative ground states in the external regions, and we easily compute their contribution to the Witten index of the full theory. For simplicity of presentation, we only turn on the specific flavour fugacity~\eqref{def by flavour}. The ground states  located at $\sigma > m$ for $\zeta<0$ then contribute:
\bea
&(\IW)_{+-}^{\zeta<0}=\\
&\; {(-1)^{n_2+1} \sum_{|s_1|, |t_2|=0}^\infty   \delta_{|s_1|+|t_2|+n_2+Q_c, 0}\binom{n_1+|s_1|-1}{n_1-1}\binom{n_2+|t_2|-1}{n_2-1} \by^{|s_1|-|t_2|-n_2}}~.
\eea
This Witten index for these perturbative vacua can be usefully written as a contour integral. Let us define the matter integrand:
\be\label{Zxyb matter}
Z(x,\by)\equiv {x^{Q_c}\ov (1-x \by)^{n_1}(1-x^{-1}\by)^{n_2}}~.
\ee
Then one easily checks that:
\be\label{IWpmwithby}
(\IW)_{+-}^{\zeta<0} = -\oint_{(x=0)} {dx\ov 2\pi i x} Z(x,\by)~,
\ee
where we pick the residue at $x=0$. Indeed, picking out the $x^0$ factor in the integrand precisely imposes Gauss's law on the free-field partition function expanded at small $x$, which indeed corresponds to the $\sigma>m$ region. 
Similarly, the ground states located at $\sigma < -m$ for $\zeta>0$ contribute:
\bea
&(\IW)_{-+}^{\zeta>0}=\\
&\; {(-1)^{n_1+1} \sum_{|t_1|, |s_2|=0}^\infty   \delta_{|t_1|+|s_2|+n_1-Q_c, 0}\binom{n_1+|t_1|-1}{n_1-1}\binom{n_2+|s_2|-1}{n_2-1} \by^{-|t_1|+|s_2|-n_1}}~,
\eea
which can also be written as the contour integral:
\be
(\IW)_{-+}^{\zeta>0} = \oint_{(x=\infty)} {dx\ov 2\pi i x} Z(x,\by)~,
\ee
where we pick the residue at infinity. This naturally imposes Gauss's law in the $\sigma<-m$ region by picking out the $(1/x)^0$ factor when expanding the integrand in $1/x$.

\subsubsection{Perturbative ground states with $|\sigma|<m$}

In the interior region at large $\zeta$, we find the critical points
\be
\sigma_\ast \approx \begin{cases}
    -m + {N_\phi\ov 2\zeta} \quad &\text{if }\, \zeta>0~,\\
     m + {N_{\t\phi}\ov 2\zeta} \quad &\text{if }\, \zeta<0~,
\end{cases}
\qquad 
\text{with }\; h''(\sigma_\ast)<0~.
\ee
Therefore, for any $\zeta\neq 0$, we find the perturbative ground states:
\be\label{psipp states}
\ket{\Psi_{++}; s_1, s_2}\equiv \ket{s_1, s_2}_{++}~, \qquad \text{with }\quad |s_1|-|s_2|+Q_c=0~,
\ee
giving us an infinite number of ground states. They contribute to the index as:
\be\label{WIpp series}
(\IW)_{++} =  \sum_{|s_1|, |s_2|=0}^\infty   \delta_{|s_1|-|s_2|+Q_c, 0}\binom{n_1+|s_1|-1}{n_1-1}\binom{n_2+|s_2|-1}{n_2-1} \by^{|s_1|+|s_2|}~.
\ee
This infinite series can be resummed explicitly by writing it as a hypergeometric function. For $Q_c< n_1$, let us define:
\bea
&F_-(\by) &\equiv &\; { \sum_{l=0}^\infty    \binom{n_1+l-Q_c-1}{n_1-1}\binom{n_2+l-1}{n_2-1} \by^{2l-Q_c}}\\
& &=&\; {\binom{n_1-Q_c-1}{n_1-1}} \by^{-Q_c}\, _2F_1\left(n_2,n_1-Q_c;-Q_c+1;\by^2\right)~.
\eea
while for $Q_c>-n_2$, we define:
\bea
&F_+(\by)  &\equiv&\; { \sum_{l=0}^\infty    \binom{n_1+l-1}{n_1-1}\binom{n_2+l+Q_c-1}{n_2-1} \by^{2l+Q_c}}\\
& &=&\; {\binom{n_2+Q_c-1}{n_2-1}} \by^{Q_c}\, _2F_1\left(n_1,n_2+Q_c;Q_c+1;\by^2\right)~.
\eea
The two functions coincide for the values of $Q_c$  for which they are both defined:
\be\label{Fpm identity}
F_-(\by) =F_+(\by)  \qquad \text{iff}\; \quad  -n_2 <Q_c <n_1~.
\ee
We can then rewrite the Witten index~\eqref{WIpp series} as:
\be
(\IW)_{++} =  \begin{cases}
    F_-(\by)\qquad &\text{if }\; Q_c<n_1~,\\
    F_+(\by)\qquad &\text{if }\; Q_c>-n_2~,\\
\end{cases}
\ee
Note that this expression holds for either sign of $\zeta\neq 0$. 
The Witten index for these perturbative ground states can again be written as a contour integral of the same integrand~\eqref{Zxyb matter} as for the external regions. Indeed, one easily checks that:
\be\label{Fpm as residue}
 F_-(\by) =- \oint_{(x=\by^{-1})} {dx\ov 2\pi i x} Z(x,\by)~,\qquad\qquad
 F_+(\by) = \oint_{(x=\by)} {dx\ov 2\pi i x} Z(x,\by)~.
\ee
where one picks the residue at $x=\by^{-1}$ and $x=\by$, respectively. 

\subsubsection{Compairing to the JK residue: perturbative versus true ground states}

Adding up the perturbative contributions for the exterior and interior regions, we find the Witten index for the full theory. It is interesting to check this explicitly against the expected full result. First, recall that the full index can be obtained by the JK residue formula~\eqref{IW SQCD JK}, namely:
\be\label{fullWIexpln1n2}
(\IW)^{\zeta>0} = - \oint_{(x=\by^{-1})} Z(x,\by)= F_-(\by)~, \qquad\quad
(\IW)^{\zeta<0} =  \oint_{(x=\by)} Z(x,\by)= F_+(\by)~,
\ee
for positive or negative $\zeta$, respectively. For $\zeta>0$, adding up the perturbative contributions gives us:
\be
(\IW)^{\zeta>0} =\begin{cases}
    - \oint_{(x=\by^{-1})} Z(x,\by) \qquad &\text{if }\; Q_c\leq -n_2~, \\
      \oint_{(x=\by)} Z(x,\by) \qquad &\text{if }\; -n_2 <Q_c< n_1~, \\
      \oint_{(x=\infty)} Z(x,\by)  + \oint_{(x=\by)} Z(x,\by)  \qquad &\text{if }\; Q_c\geq n_1~.
\end{cases}
\ee
Similarly, the sum of the perturbative Witten indices for $\zeta<0$ gives us:
\be
(\IW)^{\zeta<0} =\begin{cases}
    - \oint_{(x=\by^{-1})} Z(x,\by)-    \oint_{(x=0)} Z(x,\by) \qquad &\text{if }\; Q_c\leq -n_2~, \\
    -  \oint_{(x=\by^{-1})} Z(x,\by) \qquad &\text{if }\; -n_2 <Q_c< n_1~, \\
      \oint_{(x=\by)} Z(x,\by)  \qquad &\text{if }\; Q_c\geq n_1~.
\end{cases}
\ee
This reproduces the exact index~\eqref{fullWIexpln1n2}, as expected. The non-trivial identities are those where we need to sum over perturbative ground states from two different regions (that is, for $\zeta>0$ with $Q_c\geq n_1$ and for $\zeta<0$ with $Q_c\leq -n_2$). The identity between the perturbative approach and the JK residue formula is easily shown by deforming the contour in the $x$-plane. 

The key physics, in these cases, is that we have more perturbative ground states than true ground states, resulting in non-trivial cancellations between distinct perturbative contributions to the index. If the effective theory for the $(\sigma, \lambda)$ supermultiplet were exact with a finite scalar potential, the lifting of any pair of states with the same flavour quantum numbers in the interior and exterior region would be expected, because the two states in such a pair differ by a $\b\lambda$ insertion. This is precisely the condition for the existence of a SQM instanton contribution lifting the pair to non-zero energy~\cite{Witten:1982im}. In the present situation, we can only trust the scalar potential $U(\sigma)$ up to $|U(\sigma)|\approx M_C$, yet it is natural to conjecture that a similar phenomenon occurs; here, it must be the case that the relevant `instantons' are really the 1d gauge theory effects which smoothen out the singularities of the effective theory at $\sigma=\pm m$. Similar cancellations between perturbative states are a common occurrence in supersymmetric QFT --- see {\it e.g.}~\cite{Bullimore:2019qnt} for some 3d GLSM related to the present work, and~\cite{Pedder:2007ff} for a discussion in $\CN=4$ SQM.

Incidentally, the above discussion also gives us an explicit expression for the wall-crossing of the Witten index between positive and negative FI parameter:
\be
(\IW)^{\zeta>0} - (\IW)^{\zeta<0} =   (\IW)^{\zeta>0}_{-+}- (\IW)^{\zeta<0}_{+-} = \oint_{(x=\infty)} Z(x,\by)+ \oint_{(x=0)} Z(x,\by)~.
\ee
Indeed the perturbative states in the interior region exist for any $\zeta\neq 0$, and therefore the wall-crossing comes entirely from the perturbative states in the exterior region. In particular,  the wall-crossing is trivial if $-n_2<Q_c<n_1$. We will discuss wall-crossing further in section~\ref{sec:wc}. 

\subsection{Perturbative ground states in $\Gamma$-SQED with $n_3=0$}\label{subsec:pert theory GammaSQED}

We now consider adding the gauge-neutral fields $\Gamma^j_i$ to the theory together with the $E$-term~\eqref{EGamma SQCD}, while keeping $n_3=0$ for simplicity. We turn on the same real mass $m$ as in~\eqref{mass to omegas}, which gives a common real mass $2m>0$ to the $\Gamma$ fields. Therefore, we can quantise these fermi multiplets in the usual way, which gives us:
\be
\b\Gamma^i_j \ket{\Psi}=0~,\qquad \forall i, j~,
\ee
where $\Psi$ denotes the free-field ground states~\eqref{F0n3eq0} of the theory without $\Gamma$ fields. Once we add in $\Gamma$, we must consider a larger set of free-field ground states obtained by tensoring the previous results with the full fermionic Fock space generated by the  raising operators $\Gamma^j_i$. That is, we find:
\be\label{F0n3eq0 with Gamma}
    \mathcal{F}^0=\left\{\begin{cases}
       \ket{s_1, t_2, \gamma}_{+-}\equiv \prod_{i=1}^{n_1}\prod_{j=1}^{n_2} (\Gamma_i^j)^{\gamma_{i,j}} \ket{s_1, t_2}_{+-} \qquad &\text{if }\, \sigma>m\\
        \ket{s_1, s_2, \gamma}_{++}\equiv\prod_{i=1}^{n_1}\prod_{j=1}^{n_2} (\Gamma_i^j)^{\gamma_{i,j}} \ket{s_1, s_2}_{++} \qquad &\text{if }\,|\sigma|<m\\
         \ket{t_1, s_2, \gamma}_{-+}\equiv \prod_{i=1}^{n_1}\prod_{j=1}^{n_2} (\Gamma_i^j)^{\gamma_{i,j}}  \ket{t_1, s_2}_{-+} \quad \qquad &\text{if }\, \sigma<-m
    \end{cases}\right\}~,
\ee
with $\gamma_{i,j}\in \{0,1\}$. Note that we need to specify an ordering of the $\Gamma$ fields, since they are fermionic; we choose to order them according to:
\be\label{orderGammas}
\prod_{i=1}^{n_1}\prod_{j=1}^{n_2} (\Gamma_i^j)^{\gamma_{i,j}} \equiv (\Gamma_1^{1})^{\gamma_{1,1}} (\Gamma_1^{2})^{\gamma_{1,2}}\cdots (\Gamma_1^{n_2})^{\gamma_{1,n_2}} (\Gamma_2^{1})^{\gamma_{2,1}}\cdots (\Gamma_{n_1}^{n_2})^{\gamma_{n_1,n_2}}~.
\ee
Since the $\Gamma$'s are gauge-neutral, the Gauss law constraints remain the same as in~\eqref{FPhys Pn model}.
 The $E$-term interaction
\be\label{EtermN1N2Gamma}
E_{\Gamma^j_i}= \t \phi^j \phi_i
\ee
will pair together many of the states in~\eqref{F0n3eq0 with Gamma} and give them a positive energy. To see this explicitly, we treat this interaction in perturbation theory. Since we are only interested in the ground states, we can focus on the supercharges:
\be\label{Q deformed}
\CQ= \CQ_0+ \varepsilon \CQ_1~, \qquad \qquad
\b\CQ= \b\CQ_0+ \b\varepsilon \CQ_1~, 
\ee
where $\CQ_0$, $\b\CQ_0$ denote the unperturbed supercharges~\eqref{N1N2Q} and $\varepsilon$ is a supersymmetric coupling constant which we can take to be arbitrarily small. The $E$-term~\eqref{EtermN1N2Gamma} introduces the perturbation:
\bea
&\CQ_1 = -{i\ov \sqrt{2\omega\t\omega}} \sum_{i=1}^{n_1}\sum_{j=1}^{n_2} \Gamma^j_i (a_{1,i}+b_{1,i}^\dag)(a_{2,j}+b_{2,j}^\dag)~, \\
&\b\CQ_1 = {i\ov \sqrt{2\omega\t\omega}} \sum_{i=1}^{n_1}\sum_{j=1}^{n_2} \b\Gamma^i_j  (a_{1,i}^\dag+b_{1,i})(a_{2,j}^\dag+b_{2,j})~,\\
\eea
which contributes terms of order $\varepsilon$ and $|\varepsilon|^2$ to the Hamiltonian. 

\medskip
\noindent {\bf Ground-state perturbation theory.} We wish to find the exact ground states after deforming the supercharges as in~\eqref{Q deformed}. 
Let us expand the new would-be ground states as
\be
\ket{\Psi}= \ket{\Psi_0}+ \epsilon \ket{\Psi_1}+ \epsilon^2 \ket{\Psi_2}+\cdots~,
\ee
where $\ket{\Psi_0}$ denotes the ground states at $\varepsilon=0$. Let us try and solve the new ground-state equations
\be
\CQ \ket{\Psi}=0~, \qquad\qquad \b\CQ\ket{\Psi}=0
\ee
in perturbation theory --- in fact, we only need to consider the first-order effect.
 At first order in $\varepsilon$, this gives us:
\be\label{Q1Q0 rel}
\CQ_1\ket{\Psi_0} = -\CQ_0  \ket{\Psi_1}~,\qquad \b\CQ_1\ket{\psi_0} = -\b\CQ_0  \ket{\Psi_1}~.
\ee
Recall that the ground states $\ket{\Psi_0}$ are harmonic representatives of the $\CQ_0$-cohomology. Here, in particular, we see that the $\CQ_0$-cohomology class with representative $\ket{\Psi_0}$ still determines a ground states, at first order in $\epsilon$, if and only if $\CQ_1\ket{\Psi_0}$ is trivial in $\CQ_0$-cohomology. 
  Solving this cohomological problem determines $\ket{\Psi_1}$, in principle. 

More concretely, we can present this as a linear algebra problem. Consider an orthonormal basis of ground states $\{\ket{\Psi_0; \alpha}\}$ of the undeformed theory, and define the matrices
\be\label{def M matrices}
 {M_{\alpha\beta} = \bra{\Psi_0;\alpha}  {\boldsymbol{Q}}_1  \ket{\Psi_0; \beta}~,}\qquad\quad
 {\t M_{\alpha\beta} = \bra{\Psi_0;\alpha} \b{\boldsymbol{Q}}_1  \ket{\Psi_0; \beta}~,}\\
\ee
where we chose a new normalisation of $\CQ_1$ for future convenience:
\be
\boldsymbol{Q}_1 \equiv i \sqrt{2\omega\t\omega} \,\CQ_1~,\qquad\quad
\b{\boldsymbol{Q}}_1 \equiv -i \sqrt{2\omega\t\omega}\, \b\CQ_1~,
\ee
 Then, if follows from~\eqref{Q1Q0 rel} that any zeroth-order ground state
\be
\ket{\Psi_0} \equiv \sum_\alpha c_\alpha \ket{\Psi_0; \alpha}
\ee
remains a ground state at first order if and only if%
\footnote{
Note that this is equivalent to requiring that the vector $c=(c_\alpha)$ be in the kernel of both the matrix $M$ and of its transpose $M^T$, namely $Mc=M^Tc=0$.}
\be
\sum_\beta  M_{\alpha\beta}\, c_\beta=0~, \qquad 
\sum_\beta  \t M_{\alpha\beta} \, c_\beta=0~, \qquad \forall \beta~.
\ee
All the other states are lifted by the interactions. This procedure is easily implemented on a computer.  Note that, since the supercharges commute with the flavour symmetry, the perturbation theory can be carried out in each flavour superselection sector separately --- that is, we can fix the integer vectors $F_1=(F_{1,i})$ and $F_2=(F_{2,j})$ which determine the flavour eigenvalues according to:
\be\label{def F1 F2}
Y \ket{\Psi_0; F_1, F_2} = \prod_{i=1}^{n_1} y_{1,i}^{F_{1,i}}\prod_{j=1}^{n_2} y_{2,j}^{F_{2,j}}\ket{\Psi_0; F_1, F_2}~,
\ee
and then carry out the perturbation theory on the set of perturbative ground states $\ket{\Psi_0; F_1, F_2}$.

\subsubsection{Deformed perturbative ground states in the exterior region} \label{subsec:defpertext}
Let us first consider the exterior region $\sigma>m$, where the orthonormal basis for the free theory is: 
\be\label{F0physbeforedefpossigma}
\CF^0_{\rm phys}= \big\{\ket{s_1, t_2, \gamma}_{+-}\;\; \big|\; \; |s_1|+|t_2|= -n_2-Q_c\big\}~.
\ee
Recall that these states only exist for $Q_c\leq -n_2$ and with $\zeta<0$. 
Up to terms that create excitations orthogonal to the zeroth-order ground states and drop out from~\eqref{def M matrices}, we have
\be
{\boldsymbol{Q}}_1 ^{(+-)} = \sum_{i,j}\Gamma^j_i a_{1, i} b^\dagger_{2,j}~,
\ee
which is an operator that acts as
\be
{\boldsymbol{Q}}_1 ^{(+-)} \ket{s_1, t_2, \gamma}_{+-}=\sum_{i,j}  \eps_{i,j}^\gamma  \ket{s_1-\delta_i, t_2+\delta_j, \gamma+\delta_{i,j}}_{+-}~.
\ee
Here we used the obvious notation that $s_1-\delta_i \equiv (s_{1,k}-\delta_{ki})$, and similarly for the shifts of $t_2$ and $\gamma$. We also defined the important sign:
\be
\eps_{i,j}^\gamma  = (-1)^{\left(\sum_{k=1}^{i-1}\sum_{l=1}^{n_2}\gamma_{k,l} + \sum_{l=1}^{j-1}\gamma_{i,l}\right)}~,
\ee
which corresponds to the ordering~\eqref{orderGammas}. 
Note also that any would-be state on the right-hand-side with some $\gamma_{i,j}=2$ or $s_{1,i}=-1$ actually vanishes. One can explicitly check that ${\boldsymbol{Q}}_1$ so defined is a nilpotent operator. The complex-conjugate supercharge acts as:
\be
\b{\boldsymbol{Q}}_1 ^{(+-)} \ket{s_1, t_2, \gamma}_{+-}=\sum_{i,j} \eps_{i,j}^\gamma \ket{s_1+\delta_i, t_2-\delta_j, \gamma-\delta_{i,j}}_{+-}~,
\ee
where any state with $\gamma_{i,j}=-1$ or $t_{2,j}=-1$ on the right-hand-side actually vanishes. 

The Witten index of these perturbative ground states can be computed by a residue at $x=0$, as before. Restoring all the flavour fugacities, let us define the matter integrand:
\be\label{Zxyexplicit}
Z(x, y)= x^{Q_c} {\prod_{i=1}^{n_1}\prod_{j=1}^{n_2}\left(1-{y_{2,j}\ov y_{1, i}}\right)\ov  \prod_{i=1}^{n_1} \left(1- {x  \ov y_{1,i}}\right) \prod_{j=1}^{n_2} \left(1- {y_{2,j} \ov x }\right)}~.
\ee
Then the fully flavoured perturbative Witten index reads:
\be\label{IWpmn1n2 I}
(\IW)_{+-}^{\zeta<0} = -\oint_{(x=0)}{dx\ov 2\pi i x} Z(x, y)~,
\ee
generalising~\eqref{IWpmwithby}. Of course this captures precisely all the states~\eqref{F0physbeforedefpossigma}. The lifting of massive fermion-boson pairs due to the $E$-term is seen in the index as cancellations between fermion and boson contributions, which can be worked out explicitly by expanding the integrand~\eqref{Zxyexplicit} around $x=0$ before picking out all the $x^0$ terms. One can check in examples that this indeed matches the first-order perturbation theory result.

Similarly, in the exterior region $\sigma <-m$, we have the zeroth-order ground states:
\be
\CF^0_{\rm phys}= \big\{ \ket{t_1, s_2, \gamma}_{-+}\;\; \big|\; \; |t_1|+|s_2|= -n_1+Q_c\big\}~,
\ee
which only exist for $Q_c\geq n_1$ and $\zeta>0$.  The supercharges now act as:
\be
{\boldsymbol{Q}}_1^{(-+)} = \sum_{i,j}\Gamma^j_i b^\dagger_{1, i} a_{2,j}~,\qquad\quad
{\b{\boldsymbol{Q}}}_1^{(-+)} = \sum_{i,j}\b\Gamma^i_j b_{1, i} a^\dagger_{2,j}~,
\ee
so that:
\bea
&{{\boldsymbol{Q}}_1 ^{(-+)} \ket{t_1, s_2, \gamma}_{-+}=\sum_{i,j} \eps_{i,j}^\gamma\ket{t_1+\delta_i, s_2-\delta_j, \gamma+\delta_{i,j}}_{-+}~,}\\
&{{\b{\boldsymbol{Q}}}_1 ^{(-+)} \ket{t_1, s_2, \gamma}_{-+}=\sum_{i,j}\eps_{i,j}^\gamma  \ket{t_1-\delta_i, s_2+\delta_j, \gamma-\delta_{i,j}}_{-+}~.}\\
\eea
The flavoured Witten index for those perturbative ground states is given by:
\be\label{IWpmn1n2 II}
(\IW)_{-+}^{\zeta>0} = \oint_{(x=\infty)}{dx\ov 2\pi i x} Z(x, y)~,
\ee
where we now expand the integrand~\eqref{Zxyexplicit} in $1/x$.

\subsubsection{Deformed perturbative ground states in the interior region} 
In the interior region $|\sigma|<m$, we have the orthonormal basis of perturbative ground states:
\be\label{F0physbeforedefpossigma}
\CF^0_{\rm phys}= \big\{\ket{s_1, s_2, \gamma}_{++}\;\; \big|\; \; |s_1|-|s_2|+Q_c = 0\big\}~,
\ee
and:
\be
{\boldsymbol{Q}}_1^{(++)} = \sum_{i,j}\Gamma^j_i a_{1, i} a_{2,j}~,\qquad\quad
{\b{\boldsymbol{Q}}}_1^{(++)} = \sum_{i,j}\b\Gamma^i_j a^\dagger_{1, i} a^\dagger_{2,j}~.
\ee
We thus find the action:
\bea
&{{\boldsymbol{Q}}_1 ^{(++)} \ket{s_1, s_2, \gamma}_{-+}=\sum_{i,j} \eps_{i,j}^\gamma\ket{s_1-\delta_i, s_2-\delta_j, \gamma+\delta_{i,j}}_{-+}~,}\\
&{{\b{\boldsymbol{Q}}}_1 ^{(++)} \ket{s_1, s_2, \gamma}_{-+}=\sum_{i,j} \eps_{i,j}^\gamma\ket{s_1+\delta_i, s_2+\delta_j, \gamma-\delta_{i,j}}_{-+}~.}\\
\eea
The Witten index for the perturbative ground states can again be captured by a residue formula, directly generalising~\eqref{Fpm as residue}. We have:
\be\label{IWpmn1n2 III}
(\IW)_{++} =  \begin{cases}
     -\sum_{i=1}^{n_1}\oint_{(x=y_{1,i})} {dx\ov 2\pi i x} Z(x,y) \qquad &\text{if }\; Q_c<n_1~,\\
     \sum_{j=1}^{n_2}\oint_{(x=y_{2,j})} {dx\ov 2\pi i x} Z(x,y) \qquad &\text{if }\; Q_c>-n_2~,\\
\end{cases}
\ee
where we picked a contour that encloses all the poles from fundamental or antifundamental chiral multiplets, depending on the value of $Q_c$. As in the case without the $\Gamma$ fields, the JK residue formula for the full Witten index can be recovered by summing up the perturbative contributions in the exterior and interior regions. We also find, as before, that non-perturbative effects must lift certain fermion-boson pairs of ground states  located in distinct regions of the Coulomb branch. 

\subsection{General case: $\Gamma$-SQED with $n_3>0$}
Finally, we consider the general case of $\Gamma$-SQED with chiral and fermi multiplets. Adding the effect of the $n_3$ fundamental fermi multiplets to the previous discussion is straightforward, following the analysis in subsection~\ref{subec:U1n1n3}, since they do not contribute the supercharges. Before turning on the $E$-term interactions, we have the physical ground states:
\be 
    \mathcal{F}^0_{\rm phys} =
    \left\{\begin{cases}
       \ket{s_1, t_2, f_3, \gamma}_{+-} \quad | \quad  |s_1|+|t_2|+|f_3|+n_2+Q_c=0 \quad &\text{if }\, \sigma>m\\
        \ket{s_1, s_2, f_3, \gamma}_{++} \quad |\quad  |s_1|-|s_2|+|f_3|+Q_c=0\quad &\text{if }\,|\sigma|<m\\
         \ket{t_1, s_2, f_3, \gamma}_{-+} \quad | \quad  -|t_1|-|s_2|+|f_3|-n_1+Q_c=0 \quad &\text{if }\, \sigma<-m
    \end{cases}\right\}~,
\ee
with the obvious notation generalising~\eqref{F0n3eq0 with Gamma} --- for instance:
\be
\ket{s_1, t_2, f_3, \gamma}_{+-} \equiv \prod_{k=1}^{n_3} \eta_{3,k}^{f_{3,k}} \ket{s_1, t_2,  \gamma}_{+-}~.
\ee
We can treat this effective theory as a direct sum of $n_3+1$ theories with $n_3=0$, indexed by the effective CS level:
\be
\t Q_c(p) = Q_c+p~, \qquad p\equiv |f_3|~,
\ee
with $p$ defined as in~\eqref{def p n3}. In particular, the flavour Witten index of the full theory takes the form:
\be\label{IW expand in p full}
\IW[1, Q_c, (n_1,n_1, n_3)] =  \sum_{p=0}^{n_3}(-1)^p  \schur_{[1^p,0^{n_3-p}]}(y_3^{-1})\;  \IW[1, Q_c+p, (n_1,n_2, 0)]~,
\ee
for either sign of $\zeta$. 
The Witten indices for the perturbative ground states are given exactly as in~\eqref{IWpmn1n2 I}, \eqref{IWpmn1n2 I} and~\eqref{IWpmn1n2 I}, with the more general integrand:
\be\label{Zxyfull}
Z(x, y)= x^{Q_c} {\prod_{i=1}^{n_1}\prod_{j=1}^{n_2}\left(1-{y_{2,j}\ov y_{1, i}}\right)  \prod_{k=1}^{n_3}(1- {x \ov y_{3,k}})\ov  \prod_{i=1}^{n_1} \left(1- {x  \ov y_{1,i}}\right) \prod_{j=1}^{n_2} \left(1- {y_{2,j} \ov x }\right)}~.
\ee
Indeed, from the point of view of the residue formulas, the expansion~\eqref{IW expand in p full} simply follows from the identity:
\be
 \prod_{k=1}^{n_3}\left(1- {x \ov y_{3,k}}\right) = \sum_{p=0}^{n_3}(-1)^p  \schur_{[1^p,0^{n_3-p}]}(y_3^{-1})\;  x^p~.
\ee
This concludes our general discussion of $\Gamma$-SQED on the Coulomb branch. As an illustration of the formalism, we will explore several examples in fuller detail in section~\ref{sec:ab expls}.


\section{Wall-crossing, trialities and confinement}\label{sec:wc}

As we vary the FI parameters of a gauged $\CN=2$  SQM, the number of supersymmetric ground states can jump discontinuously, leading to discrete jumps in the Witten index. This is known as the wall-crossing phenomenon, which has been most studied in the case of 1d $\CN=4$ systems seen as the worldline theories of half-BPS particles on the Coulomb branch of 4d $\CN=2$ SQFTs. In this section, we explore aspects of wall-crossing for our 1d $\CN=2$ SQCD.

\subsection{Trivial wall-crossing and trialities}

For gauged $\CN=2$ SQM, a wall-crossing formula was first derived in~\cite{Hori:2014tda} together with the Coulomb-branch localisation result for the Witten index. Focussing on unitary SQCD, we are interested in the change in the Witten index as we go from positive to negative FI parameter:
\be
\Delta \IW[N_c, Q_c, \boldsymbol{n}^F]\equiv  \IW[N_c, Q_c, \boldsymbol{n}^F]^{\zeta>0} \,-\,  \IW[N_c, Q_c, \boldsymbol{n}^F]^{\zeta<0}~.
\ee
Of particular interest to us are the cases with trivial wall-crossing (TWC), in which case the supersymmetric ground states can be continuously transported through $\zeta=0$. We will show that trivial wall-crossing always occurs when $q_c$ sits in a particular window determined by the parameter:
\be
\CA= {n_1+n_2-n_3\ov 2}- N_c~,
\ee
according to:
\be\label{TWC cond}
-\CA-1 < q_c < \CA+1   \qquad \Leftrightarrow \qquad
 \Delta \IW[N_c, Q_c, \boldsymbol{n}^F]=0 \quad \text{(TWC).}
\ee
Note that, since $q_c+\CA\in \Z$, the condition on $q_c$ stated in~\eqref{TWC cond} is equivalent to $|q_c|\leq \CA$. We will shortly demonstrate that this condition is sufficient for having TWC, and we further conjecture that it is also a necessary condition, as claimed in~\eqref{TWC cond}, except in trivial cases where $\IW[N_c, q_c, \boldsymbol{n}^F]_\pm=0$. Note that the inequalities in~\eqref{TWC cond} can be written as:
\be\label{TWC cond Qc}
-n_2 +N_c-1 < Q_c < n_1-n_3-N_c+1~.
\ee
For $N_c=1$, the condition for TWC is $-n_2< Q_c< n_1-n_3$, which we can rewrite as $-n_2 <Q_c+p <n_1$ for all $p=0, \cdots, n_3$ --- this condition therefore follows from the Coulomb-branch discussion of the previous section, as the condition that there are no perturbative ground states from the `exterior' regions $|\sigma|>m$ which could disappear to infinity as we cross $\zeta=0$.

\medskip
\noindent
{\bf Trialities from wall-crossing.} Consider the SQCD theory with $\sign(\zeta)=\pm$, which we denote here by $\CT_\pm$, and let us denote by $\CT'_-$ and $\CT''_+$ the theories obtained by right and left mutation, respectively. Assuming theory $\CT$ has trivial wall-crossing, which means that $\CA\geq 0$ and $|q_c|\leq \CA$, we may view  this as an `infrared' duality:
\be
\CT_- \qquad \overset{\text{TWC}}{\longleftrightarrow}\qquad \CT_+~,
\ee
since the supersymmetric ground states can be continuously deformed from positive to negative $\zeta$ and so the two theories have isomorphic zero-energy ground states.
Combining TWC with mutations, we then obtain at least a triality of 1d gauge theories:
\be\label{triality}
\CT''_+ \qquad \overset{\text{LM}}{\longleftarrow} \qquad  \Big(\CT_- \qquad \overset{\text{TWC}}{\longleftrightarrow} \qquad \CT_+\Big)\qquad \overset{\text{RM}}{\longrightarrow} \qquad \CT'_-~,
\ee
where LM and RM denote the left and right mutations, respectively.
Here the three gauge groups involved in the triality are:
\be\label{triality case one}
U(n_2-N_c)_{-q_c}^{(\zeta''>0)} \qquad \overset{\text{LM}}{\longleftarrow} \qquad U(N_c)_{q_c}
\qquad \overset{\text{RM}}{\longrightarrow} \qquad U(n_1-N_c)_{-q_c}^{(\zeta'<0)}~.
\ee
The next natural question is whether we can go further in cases where the wall-crossing for the theory $\CT'$ or $\CT''$ is trivial as well. Note that we have:
\be
q_c' = -q_c'~, \quad \CA' = -\CA~, \qquad\text{and} \qquad q_c''=-q_c~, \quad \CA''=-\CA
\ee
for the mutated theories, hence the absence of wall crossing in either theory is equivalent to $|q_c|\leq -\CA$. This is in contradiction with the condition for TCW of theory $\CT$ except in the special case $q_c=\CA=0$, which we shall discuss momentarily.
 Hence, for $\CA>0$, we have exactly the three distinct gauge theories~\eqref{triality case one} and nothing more.

\medskip
\noindent
{\bf Connection to the Gadde--Gukov--Putrov triality.} In the special case $q_c=0$ with $\CA=0$, as argued above, all three theories appearing in~\eqref{triality} enjoy trivial wall-crossing. The condition $\CA=0$ ensures that repeated mutations give rise to an order-three operation, with the three gauge groups:
\be
 U({n_2+n_3-n_1\ov 2})_{0}  \qquad \longleftrightarrow \qquad U({n_1+n_2-n_3\ov 2})_{0}
\qquad \longleftrightarrow \qquad U({n_3+n_1-n_2\ov 2})_{0}~.
\ee
This can be understood as the dimensional reduction of the triality between 2d $\CN=(0,2)$ gauge theories discovered in~\cite{Gadde:2013lxa}. In this context, the condition $\CA=0$ is the anomaly-free condition for the 2d gauge group.

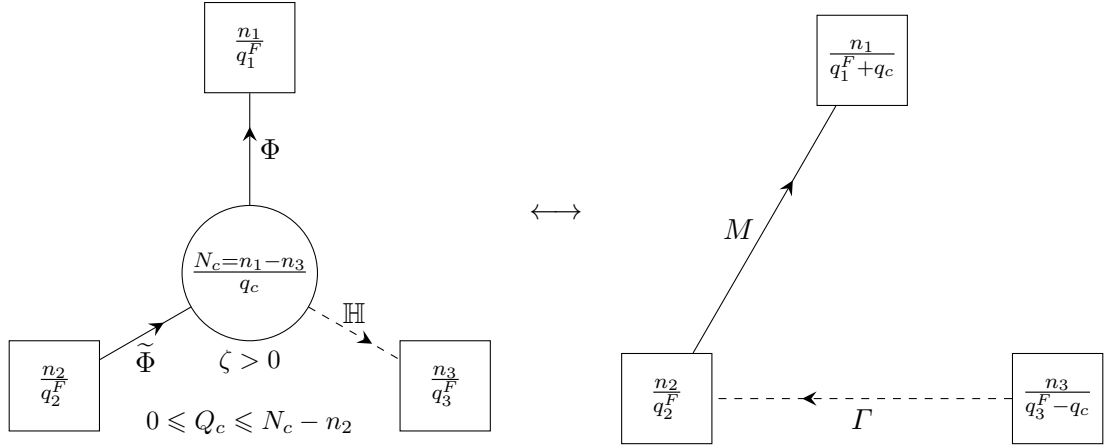
\begin{figure}
    \centering
\begin{tikzpicture}[>=stealth,scale=0.75]
 \def\SquareSize{1.2}  
    \def\CircleRadius{1.2}
    \def\NodeRadius{4}
    \draw (0,0) circle (\CircleRadius cm);

    \node[] (Nc) at (0,0) {${N_c=n_1-n_3\ov q_c}$};
    \node[draw, rectangle, shape aspect=1, minimum width= \SquareSize cm, minimum height=\SquareSize cm] (N1) at (90:\NodeRadius cm) {${n_1\ov q_1^F}$};
    \node[draw, rectangle, shape aspect=1, minimum width=\SquareSize cm, minimum height=\SquareSize cm] (N3) at (-30:\NodeRadius cm) {${n_3\ov q_3^F}$};
    \node[draw, rectangle, shape aspect=1, minimum width=\SquareSize cm, minimum height=\SquareSize cm] (N2) at (210:\NodeRadius cm) {${n_2\ov q_2^F}$};
    \node[below =0.48cm of Nc] {\small $\zeta > 0$};
        \node[below =1.3cm of Nc] {\small $0\leq Q_c\leq N_c-n_2$};

    \draw[->-=0.7] (90:\CircleRadius cm) --node[midway, right] {$\Phi$} (N1);
    \draw[->-=0.7] (N2) -- node[midway, below]{${\t\Phi}$} (210:\CircleRadius cm);

    \draw[->-=0.7, dashed] (-30:\CircleRadius cm) -- node[midway, above]{$\bH$} (N3);

\end{tikzpicture}
\quad
\raisebox{3cm}{$\longleftrightarrow$}
\quad
\begin{tikzpicture}[>=stealth,scale=0.75]
 \def\SquareSize{1.2}  
    \def\CircleRadius{1}
    \def\NodeRadius{4}
    \node[draw, rectangle, shape aspect=1, minimum width= \SquareSize cm, minimum height=\SquareSize cm] (N1) at (90:\NodeRadius cm) {${n_1\ov q_1^F+q_c}$};
    \node[draw, rectangle, shape aspect=1, minimum width=\SquareSize cm, minimum height=\SquareSize cm] (N3) at (-30:\NodeRadius cm) {${n_3\ov q_3^F-q_c}$};
    \node[draw, rectangle, shape aspect=1, minimum width=\SquareSize cm, minimum height=\SquareSize cm] (N2) at (210:\NodeRadius cm) {${n_2\ov q_2^F}$};
 
    \draw[->-=0.7, dashed] (N3) --node[midway,below]{$\Gamma$} (N2);
  \draw[->-=0.7] (N2) --node[midway,left]{$M$} (N1);
\end{tikzpicture}
    \caption{ Confinement duality for the $U(n_1-n_3)_{q_c}$ SQCD theory with $\zeta>0$ whenever $|q_c|\leq -\CA$. Note that $\CA=-{n_1-n_2-n_3\ov 2}={n_2-N_c\ov 2}$ in this case. Note also the shifts in $q_I^F$.}
    \label{fig:confinement case two}
\end{figure}

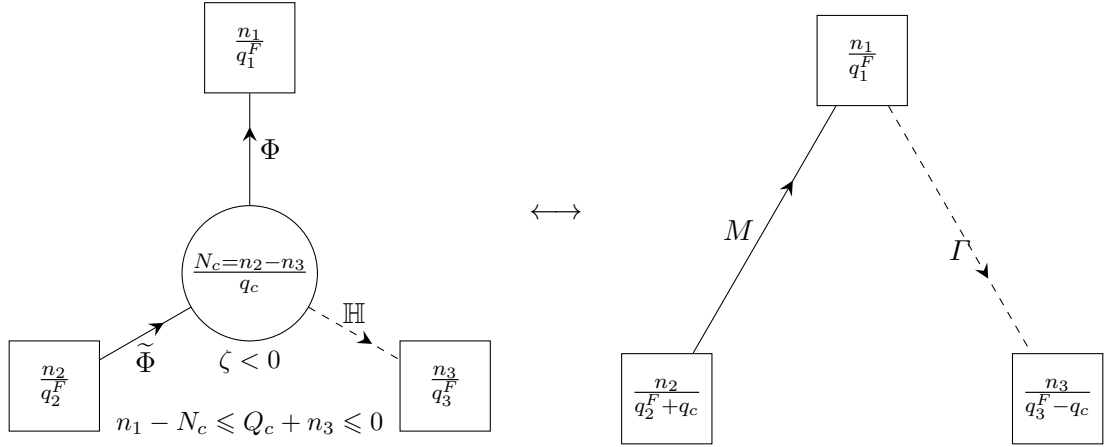
\begin{figure}
    \centering
\begin{tikzpicture}[>=stealth,scale=0.75]
 \def\SquareSize{1.2}  
    \def\CircleRadius{1.2}
    \def\NodeRadius{4}
    \draw (0,0) circle (\CircleRadius cm);

    \node[] (Nc) at (0,0) {${N_c=n_2-n_3\ov q_c}$};
    \node[draw, rectangle, shape aspect=1, minimum width= \SquareSize cm, minimum height=\SquareSize cm] (N1) at (90:\NodeRadius cm) {${n_1\ov q_1^F}$};
    \node[draw, rectangle, shape aspect=1, minimum width=\SquareSize cm, minimum height=\SquareSize cm] (N3) at (-30:\NodeRadius cm) {${n_3\ov q_3^F}$};
    \node[draw, rectangle, shape aspect=1, minimum width=\SquareSize cm, minimum height=\SquareSize cm] (N2) at (210:\NodeRadius cm) {${n_2\ov q_2^F}$};
    \node[below =0.48cm of Nc] {\small $\zeta < 0$};
        \node[below =1.3cm of Nc] {\small $n_1-N_c\leq Q_c+n_3\leq 0$};

    \draw[->-=0.7] (90:\CircleRadius cm) --node[midway, right] {$\Phi$} (N1);
    \draw[->-=0.7] (N2) -- node[midway, below]{${\t\Phi}$} (210:\CircleRadius cm);

    \draw[->-=0.7, dashed] (-30:\CircleRadius cm) -- node[midway, above]{$\bH$} (N3);

\end{tikzpicture}
\quad
\raisebox{3cm}{$\longleftrightarrow$}
\quad
\begin{tikzpicture}[>=stealth,scale=0.75]
 \def\SquareSize{1.2}  
    \def\CircleRadius{1}
    \def\NodeRadius{4}
    \node[draw, rectangle, shape aspect=1, minimum width= \SquareSize cm, minimum height=\SquareSize cm] (N1) at (90:\NodeRadius cm) {${n_1\ov q_1^F}$};
    \node[draw, rectangle, shape aspect=1, minimum width=\SquareSize cm, minimum height=\SquareSize cm] (N3) at (-30:\NodeRadius cm) {${n_3\ov q_3^F-q_c}$};
    \node[draw, rectangle, shape aspect=1, minimum width=\SquareSize cm, minimum height=\SquareSize cm] (N2) at (210:\NodeRadius cm) {${n_2\ov q_2^F+q_c}$};
 
    \draw[->-=0.7, dashed] (N1) --node[midway,below]{$\Gamma$} (N3);
  \draw[->-=0.7] (N2) --node[midway,left]{$M$} (N1);
\end{tikzpicture}
    \caption{ Confinement duality for the $U(n_2-n_3)_{q_c}$ SQCD theory with $\zeta<0$ whenever $|q_c|\leq -\CA$. Note that we have $\CA={n_1-n_2+n_3\ov 2}={n_1-N_c\ov 2}$ in this case. }
    \label{fig:confinement case three}
\end{figure}

\medskip
\noindent
{\bf Further trialities and confinement.} Up to relabelling of the flavour groups, shifting $N_c$ and adding or removing mesons ({\it i.e.} gauge-invariant fields), the triality~\eqref{triality case one} is the most general one. Nonetheless, it is useful to spell out the equivalent trialities that arise for $\CA<0$ and $|q_c|\leq -\CA$. Since the original SQCD theory $\CT$ has non-trivial wall-crossing, we have two distinct cases to consider. If $\zeta>0$, we have:
\be\label{triality bis}
\CT_+ \qquad \overset{\text{RM}}{\longrightarrow} \qquad  \Big(\CT'_- \qquad \overset{\text{TWC}}{\longleftrightarrow} \qquad \CT'_+\Big)\qquad \overset{\text{RM}}{\longrightarrow} \qquad \h\CT'_-~,
\ee
corresponding to the gauge groups:
\be\label{triality case two}
U(N_c)_{q_c}^{(\zeta>0)} \qquad \overset{\text{RM}}{\longrightarrow} \qquad U(n_1-N_c)_{-q_c}
\qquad \overset{\text{RM}}{\longrightarrow} \qquad U(n_3-n_1+N_c)_{q_c}^{(\h \zeta'<0)}~.
\ee
If $\zeta<0$, on the other hand, we have:
\be\label{triality ter}
\CT_- \qquad \overset{\text{LM}}{\longrightarrow} \qquad  \Big(\CT_+'' \qquad \overset{\text{TWC}}{\longleftrightarrow} \qquad \CT_-''\Big)\qquad \overset{\text{LM}}{\longrightarrow} \qquad \h\CT''_+~,
\ee
with the gauge groups:
\be\label{triality case three}
U(N_c)_{q_c}^{(\zeta<0)} \qquad \overset{\text{LM}}{\longrightarrow} \qquad U(n_2-N_c)_{q_c}
\qquad \overset{\text{LM}}{\longrightarrow} \qquad U(n_3-n_2+N_c)_{q_c}^{(\h \zeta''<0)}~.
\ee
Interesting special cases of the trialities~\eqref{triality case two} and~\eqref{triality case three} is when the third gauge theory is trivial. This gives non-trivial examples of {\it confinement} dualities, where the low-energy theory consists of free fields only, the so-called mesons $M$ and $\Gamma$. For $N_c=n_1-n_3$ and $\zeta>0$, we have $q_c= Q_c+\CA$ and a confinement duality:
\be
\Big(U(N_c=n_1-n_3)^{(\zeta>0)}_{q_c} \qquad \longleftrightarrow\qquad M\oplus \Gamma \Big)  \qquad \Leftrightarrow \qquad 0 \leq Q_c \leq n_1-n_2-n_3~.
\ee
This duality is shown in more detail in figure~\ref{fig:confinement case two}. Note that, since $N_c\geq n_2$ in this case, the same gauge theory with $\zeta<0$ has no supersymmetric ground state (if $N_c>n_2$) or it is confining as seen from the left mutation (if $N_c=n_2$).

Similarly, in the case $N_c=n_2-n_3$ and $\zeta<0$, we have the confinement duality shown in figure~\ref{fig:confinement case three} whenever the condition $n_1-N_c\leq Q_c+n_3\leq 0$ holds. In this case, the same gauge theory with $\zeta>0$ has either no ground state (if $N_c>n_1$) or is confining as seen by a right mutation (if $N_c=n_1$).

\subsection{Wall-crossing formulas for $\Gamma$-SQCD}
To conclude this section, let us write down the wall-crossing formula for 1d $\CN=2$ SQCD more explicitly in order to derive the condition~\eqref{TWC cond}. We can consider either SQCD or $\Gamma$-SQCD, as the contribution from the $\Gamma$ fields factor out of the JK residue and, therefore, of the wall-crossing formula.

\begin{figure}
    \centering
    \begin{tabular}{c c c}
    \begin{tikzpicture}[scale=1]

 \node at (1.3, 0) {$-$};
 
    \foreach \i/\label in {1/1} {
        \coordinate (X) at (\i*3, 0);
        
        \draw (X) circle(1);
        
        \node at (\i*3, -1.3) {$x_{\label} = 0$};
        \node at (\i*3, 1.3) {$x_{\label} = \infty$};
        
        \fill[blue] (\i*3 - 0.5, -0.5) circle(0.06);
        \fill[blue] (\i*3 - 0.6, 0) circle(0.06);
        \fill[blue] (\i*3 - 0.4, 0.5) circle(0.06);
        
        \fill[red] (\i*3 + 0.5, -0.5) circle(0.06);
        \fill[red] (\i*3 + 0.6, 0) circle(0.06);
        \fill[red] (\i*3 + 0.4, 0.5) circle(0.06);
        
        \fill[black] (\i*3, 1) circle(0.06); 
        \fill[black] (\i*3, -1) circle(0.06); 
        
        \draw[black, thick,
            decoration={markings, mark=at position 0.5 with {\arrow{<}}},
            postaction={decorate}]
            (\i*3 - 0.8, -0.6)
            .. controls (\i*3 - 1, 0) and (\i*3 - 0.8, 0.8) .. (\i*3 - 0.4, 0.8)
            .. controls (\i*3 - 0.2, 0.8) and (\i*3 - 0.2, -0.8) .. (\i*3 - 0.4, -0.8)
            .. controls (\i*3 - 0.6, -0.8) and (\i*3 - 0.8, -0.8) .. (\i*3 - 0.8, -0.6)
            -- cycle;
    \node at (4.5, 0) {$=$};
    }\end{tikzpicture} &\begin{tikzpicture}[scale=1]

    \coordinate (X1) at (0, 0);
    \draw (X1) circle(1);
    
    \node at (0, -1.3) {$x_1 = 0$};
    \node at (0, 1.3) {$x_1 = \infty$};
    
    \fill[blue] (-0.5, -0.5) circle(0.06);
    \fill[blue] (-0.6, 0) circle(0.06);
    \fill[blue] (-0.4, 0.5) circle(0.06);
    
    \fill[red] (0.5, -0.5) circle(0.06);
    \fill[red] (0.6, 0) circle(0.06);
    \fill[red] (0.4, 0.5) circle(0.06);
    
    \fill[black] (0, -1) circle(0.06);   
    \fill[black] (0, 1) circle(0.06);    
    
    \draw[black, thick,
        decoration={markings, mark=at position 0.5 with {\arrow{>}}},
        postaction={decorate}]
        (0, -1) circle(0.2);
    
    \draw[black, thick,
        decoration={markings, mark=at position 0.5 with {\arrow{>}}},
        postaction={decorate}]
        (0, 1) circle(0.2);
    \node at (1.5, 0) {$+$};
\end{tikzpicture}
& \begin{tikzpicture}[scale=1]

    \coordinate (X1) at (0, 0);
    \draw (X1) circle(1);

    \node at (0, -1.3) {$x_1 = 0$};
    \node at (0, 1.3) {$x_1 = \infty$};

    \fill[blue] (-0.5, -0.5) circle(0.06);
    \fill[blue] (-0.6, 0) circle(0.06);
    \fill[blue] (-0.4, 0.5) circle(0.06);

    \fill[red] (0.5, -0.5) circle(0.06);
    \fill[red] (0.6, 0) circle(0.06);
    \fill[red] (0.4, 0.5) circle(0.06);

    \fill[black] (0, 1) circle(0.06);   
    \fill[black] (0, -1) circle(0.06);  

    \draw[black, thick,
        decoration={markings, mark=at position 0.6 with {\arrow{>}}},
        postaction={decorate}]
        (0.8, -0.7)
        .. controls (0.9, -0.5) and (0.9, 0.5) .. (0.8, 0.7)
        .. controls (0.6, 0.9) and (0.2, 0.9) .. (0, 0.7)
        .. controls (-0.2, 0.5) and (-0.2, -0.5) .. (0, -0.7)
        .. controls (0.2, -0.9) and (0.6, -0.9) .. (0.8, -0.7)
        -- cycle;
\end{tikzpicture}
    \end{tabular}
    \caption{ Contour manipulation for the SQED wall-crossing formula. The poles from the fundamental and antifundamental chiral multiplets are indicated in blue and red, respectively.}
    \label{U1:cont}
\end{figure}
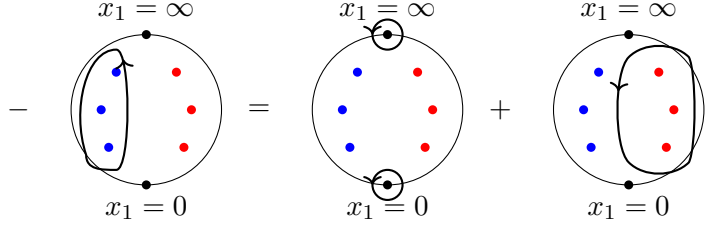

\medskip
\noindent
{\bf The case $N_c=1$.} Consider the abelian case first.  By deformation of the integration contours involved in the JK residue (see figure~\ref{U1:cont}), we directly see that:
\be
\Delta \IW = \left[\oint_{(x=0)}+\oint_{(x=\infty)}\right]{dx\ov 2\pi i x} Z(x,y)~,
\ee
with the integrand given in~\eqref{Zxyfull}. Consider SQED (striping off the $\Gamma$ field contributions) for simplicity of notation. The integrand reads:
\be
{1\ov x} Z(x,y) =x^{Q_c+n_2-1}  x^{Q_c} {\prod_{i=1}^{n_1}\prod_{j=1}^{n_2}\left(1-{y_{2,j}\ov y_{1, i}}\right)  \prod_{k=1}^{n_3}(1- {x \ov y_{3,k}})\ov  \prod_{i=1}^{n_1} \left(1- {x  \ov y_{1,i}}\right) \prod_{j=1}^{n_2} \left(x- y_{2,j} \right)}~,
\ee
which makes it clear that the residue at $x=0$ is non-trivial only if $Q_c\leq n_2$. Similarly, one easily checks that the residue at $x=\infty$ is non-trivial only if $Q_c\geq n_1-n_3$. This establishes the TWC condition~\eqref{TWC cond Qc} for $N_c=1$.

\medskip
\noindent
{\bf The general case.} For the $U(N_c)$ SQCD theory, the JK residue expressions gives us the wall-crossing formula:
\be\label{DeltaIW nonab}
\Delta \IW  = \left[(-1)^{N_c} \oint_{(x=y_1)}-  \oint_{(x=y_2)}\right] {d^{N_c}x\ov (2\pi i)^{N_c} }  {Z(x,y)\ov \det(x)}~,
\ee
where the integrand can be read off from \eqref{IW SQCD JK} and \eqref{Imatter full}. Here $x=y_1$ is a schematic notation to denote taking residues at $x_a=y_{1,i}$ for all possibilities (assuming that $n_1\geq N_c$), and similarly for $x=y_2$. This is a highly redundant description of the integration contour since the residue at any such pole with $x_\alpha=x_\beta$ for $\alpha\neq \beta$ vanishes due to the Vandermonde determinant from the W-bosons that appears in~\eqref{IW SQCD JK}, but it is a convenient description for our purpose. We can then mimic the contour deformation of the abelian case in each $x_\alpha$ plane, leading to a more involved `contour at infinity'.

Let us adopt the shorthand notation $(a_1,a_2,...a_{N_c})$ to denote a residue at the location $(x_\alpha)=(a_\alpha)$, namely:
\be
(a_1,a_2,...a_{N_c}) \equiv \oint_{(x_\alpha=a_\alpha)}  {d^{N_c}x\ov (2\pi i)^{N_c}  } {Z(x,y)\ov \det(x)}~.
\ee
We also denote by $a_\alpha= F$ or $a_\alpha=A$ the $x_\alpha$-plane contour that encloses all fundamental poles at $x_\alpha=y_{1,i}$ or antifundamental poles  at $x_\alpha=y_{2,j}$, respectively. In addition,  $a_\alpha= \zinf$ denotes the contour around $x_\alpha=0$ plus the contour around  $x_\alpha=\infty$. Therefore, we have defined residues such as:
\bea
&{(A,...,\underbrace{\zinf}_{=x_\beta},F,...,F) 
\equiv }\\
&\, {\left(\sum_{i=1}^{n_1}\res_{x_{N_c}=y_{1,i}}\right)\circ\cdots \circ\left(\res_{x_{\beta}=0}+\res_{x_\beta=\infty}\right)\circ\cdots \circ\left(\sum_{j=1}^{n_2}\res_{x_{1}=y_{2,j}}\right)\circ \frac{Z(x,y)}{\det(x)}}~.
\eea
In this notation, the wall-crossing formula~\eqref{DeltaIW nonab} reads:
\be\label{DeltaIW nonab}
\Delta \IW  = (-1)^{N_c}   (F,F,...,F)- (A,A,...,A)~.
\ee
By iterated contour deformation in each $x_\alpha$-plane, one can massage this into:
\be
\Delta \IW  =\sum_{\alpha=1}^{N_c} (-1)^{N_c-\alpha}(A,...,\underbrace{\zinf}_{=x_\alpha},F,...,F)~.
\ee
This is a sum over residues that all include taking one $x_\alpha$ to either zero or infinity. Therefore, a sufficient condition for the wall-crossing to vanish is that there are no pole at infinity in each $x_\alpha$ plane (all other $x_\beta$ variables being held fixed). Expanding the integrand in $x_\alpha^{\pm 1}$ gives us
\be
 Z(x,y) \sim x_{\alpha}^{Q_c+n_2-N_c+1}~,\qquad\quad
  Z(x,y) \sim \left({1\ov x_{\alpha}}\right)^{-Q_c+n_1-n_3-N_c+1}~,
\ee
near $x_\alpha=0$ and $\infty$, respectively. This directly gives us the necessary condition~\eqref{TWC cond Qc} for having trivial wall-crossing.

\section{More on 1d SQED: abelian examples}\label{sec:ab expls}

In this section, we present the ground states in a few examples of increasing complexity, in order to illustrate our general discussion above in a pedagogical manner. We focus on the case of an abelian gauge group, so that we can explicitly compare the Higgs-branch results of section~\ref{sec:Higgs} to the Coulomb-branch approach of section~\ref{sec:Coulomb}. 

\subsection{Case $(n_1. n_2, n_3)=(1,1,0)$: an almost trivial model}
Consider 1d $\Gamma$-SQED ($N_c=1$) with $n_1=n_2=1$ and $n_3=0$, for any CS level $Q_c\in \Z$. Since $n_1-N_c=0$ and $n_2-N_c=0$,  we know from our dualities that this theory is essentially trivial, with a single ground state for either sign of $\zeta$. For $Q_c\neq 0$, the ground state actually depends on the sign of $\zeta$, as is apparent from the Witten index:
\be\label{WI almost trivial expl}
\IW^{\zeta>0} = y_{1,1}^{Q_c}~, \qquad \qquad \IW^{\zeta<0} = y_{2,1}^{Q_c}~.
\ee
The two states are distinguished by their flavour charge, and therefore we have non-trivial wall-crossing for any $Q_c\neq 0$, while for $Q_c=0$ we have a unique flavourless ground state with trivial wall-crossing.

From the Higgs-branch perspective, we have a single ground state because the Grassmannian geometry is a point in this case, $X_\pm=\mathbb{P}^0$, for either sign of $\zeta$, therefore the tautological bundle $\CS\cong \CO(-1)$ is a trivial line and the quotient bundle $\CQ$ is non-existent. Nonetheless, the equivariant cohomology 
\be
H_T^0(\mathbb{P}^0, \CO(-Q_c)) = S_{[-Q_c]}\big(\C_{y_1^{-1}}\big)~.
\ee
does distinguish between distinct values of $Q_c$. In the language of section~\ref{sec:Higgs}, we have the representation $\nu=[-Q_c]$ of $G^F_1= U(1)$ for $\zeta>0$, and similarly we have $\nu=[Q_c]$ for the symmetry $G^F_2= U(1)$ when  $\zeta<0$. The corresponding Schur polynomials reproduce the flavoured index~\eqref{WI almost trivial expl}.

From the Coulomb branch perspective, there are several cases to consider separately. Consider first $Q_c=0$, where we have no wall-crossing. This means that the only states are the perturbative states~\eqref{psipp states} in the interior region $\sigma<|m|$ together with the $\Gamma$ oscillators. For any $Q_c$, we have an infinite tower of free-field states:
\be
\ket{\Psi}_{++}=  \ket{s_1, s_2, \gamma}_{++}= (\Gamma^1_1)^\gamma {(a_{1,1}^\dagger)^{s_1}\ov \sqrt{s_1!}}{(a_{2,1}^\dagger)^{s_2}\ov \sqrt{s_2!}}\ket{0}~, \quad\qquad s_1-s_2 = -Q_c~.
\ee
Turning on the $E$-term interaction lifts all these states except for one. For $Q_c\geq 0$, we have the states $\ket{s_1, s_1+Q_c, \gamma}$ and the boson-fermion pairing:
\be
\ket{s_1, s_1+Q_c, 0}_{++}  \overset{\boldsymbol{Q}_1}{\longrightarrow} \ket{s_1-1, s_1-1+Q_c, 1}_{++}~,
\ee
where all pairs are lifted to non-zero energy, and the only unpaired state is the one with $s_1=0$. Similarly, for $Q_c\leq 0$, we have the pairing
\be
\ket{s_2-Q_c, s_2, 0}_{++} \overset{\boldsymbol{Q}_1}{\longrightarrow} \ket{s_2-1-Q_c, s_2-1, 1}_{++}~,
\ee
with $s_2=0$ being the unpaired state. Therefore, in the interior region we have the perturbative ground states:
\be
|\sigma|<m\quad : \quad
\begin{cases}
\ket{0,|Q_c|,0}_{++}= {(a_{2,1}^\dagger)^{Q_c}\ov \sqrt{Q_c!}} \ket{0}  \qquad &\text{if }\, Q_c\geq 0~, \\
\ket{|Q_c|,0,0}_{++} = {(a_{1,1}^\dagger)^{|Q_c|}\ov \sqrt{|Q_c|!}}\ket{0}  \qquad &\text{if }\, Q_c\leq 0~. \\
\end{cases}
\ee
For $Q_c=0$, we have a unique ground state $\ket{0}$ and trivial wall-crossing. For $Q_c\neq 0$, we also need to consider the exterior regions. Consider $Q_c>0$ for definiteness. If furthermore $\zeta<0$, there is no perturbative ground state at $|\sigma|>m$ and therefore we only need to consider the interior ground state $\ket{0,Q_c,0}_{++}$, which indeed reproduces the Witten index~\eqref{WI almost trivial expl}.  If $\zeta>0$ instead, we have a finite number of perturbative ground states at $\sigma<-m$ (before turning on the $E$-term), namely:
\be
\ket{\Psi}_{-+}= \b\lambda\ket{t_1, s_2, \gamma}_{-+}=\b\lambda (\Gamma^1_1)^\gamma {(b_{1,1}^\dagger)^{t_1}\ov \sqrt{t_1!}}{(a_{2,1}^\dagger)^{s_2}\ov \sqrt{s_2!}} \bar{\psi}_{1,1}\ket{0}~, \quad\qquad  
t_1+s_2= Q_c-1~.
\ee
Turning on the $E$-term lifts most of the states due to the pairing:
\be
\ket{t_1, s_2, \gamma}_{-+}  \overset{\boldsymbol{Q}_1}{\longrightarrow}\ket{t_1+1, s_2-1, \gamma+1}_{-+}~,
\ee
but the following two states remain unpaired and survive as perturbative ground states:
\be\nn
\b\lambda\ket{Q_c-1, 0, 0}_{-+}=\b\lambda {(b_{1,1}^\dagger)^{Q_c-1}\ov \sqrt{(Q_c-1)!}} \bar{\psi}_{1,1}\ket{0}~,\quad \b\lambda\ket{0,Q_c-1,1}_{-+}=\b \lambda \Gamma^1_1 {(a_{2,1}^\dagger)^{Q_c-1}\ov \sqrt{(Q_c-1)!}}\bar{\psi}_{1,1}  \ket{0}~.
\ee
 Thus, for $Q_c>0$ and $\zeta>0$, we find three perturbative ground state --- one in the interior region and two in the exterior region, with the respective contributions to the Witten index:
\be
(\IW)_{++}^{\zeta>0} = y_{2,1}^{Q_c}~, \qquad \quad (\IW)_{-+}^{\zeta>0} =  y_{1,1}^{Q_c} - y_{2,1}^{Q_c}~.
\ee
The cancellation in the Witten index suggests that the flavour-degenerate pair
\be
\ket{\text{fer}}=\b\lambda\ket{0,Q_c-1,1}_{-+}~, \qquad  \ket{\text{bos}}=\ket{0,Q_c,0}_{++}  \label{(1,1,0)_pair_lifting}
\ee
must be lifted by non-perturbative effects, as they would be if the scalar potential were smoothened out around $\sigma=-m$~\cite{Witten:1982im}. The case $Q_c<0$ and $\zeta<0$ is similar, with a non-perturbative cancellation between the state at $|\sigma|<m$ and one of the two states at $\sigma>m$. Thus we see that this ``almost trivial'' example already captures most of the subtleties of our abelian Coulomb branch analysis, including both perturbative and non-perturbative effects.

\subsection{Case $(n_1,n_2,n_3)=(2,1,0)$: a less trivial duality}

The $\Gamma$-SQED model with $(n_I)=(2,1,0)$ with 1d CS level $Q_c$ is instructive. Since $N_c-n_2=0$, we know from the left-mutation that this model with $\zeta<0$ has a unique ground state; the description of that ground state is similar to the previous example. In the following we focus on the case $\zeta>0$, which has more than one ground state except if $Q_c\in \{0,1\}$, which are the value with trivial wall-crossing. 

The Higgs-branch geometry for this model is
\be
X_+= \mathbb{P}^1~, \qquad\qquad \CE_+ = \bigoplus_{q=0}^1 \CO(-Q_c-q)\otimes \Lambda^q(\C_{y_2})~,
\ee
where $\lambda=[q]$ in the notation of section~\ref{subsec:HBpos}.  Note that~\eqref{Eplus abelian} simplifies significantly for $n_1=2$ because $\CQ^\ast\cong \CO(-1)$ in this case. There are thus two cohomology groups to consider for $q=0,1$, giving us the states
\be\label{Hcoh 210}
H^{\bullet}( \mathbb{P}^1, \CO(-Q_c)) \oplus y_{2,1} H^{\bullet}( \mathbb{P}^1, \CO(-Q_c-1))
\ee
for $q=0$ and $q=1$, respectively. 
Following the computation of section~\ref{subsec:HBpos}, we have the weight $\omega=[-Q_c,q]$ and therefore the weights $\nu$ that determine the $U(2)_{I=1}$ flavour representations are given by:
\be
\nu = \begin{cases}
    [-Q_c, q]\qquad &\text{if }\, Q_c<0~,\\
    [0,0] \qquad &\text{if } Q_c=q=0 \in \{0,1\}~,\\
    [q-1, -Q_c+1] \qquad &\text{if } Q_c>1~.
\end{cases}
\ee
We also see that $\ell(\sigma)=1$ for $Q_c<0$ and $Q_c=1$, while $\ell(\sigma)=0$ for $Q_c>1$ and $Q_c=0$; the cases $(Q_c,q)=(0,1), (1,0)$ give us an irregular $\omega$ and therefore do not contribute any states. We thus find the flavoured Witten index:
\be\label{IW full 210}
(\IW)^{\zeta>0} = \begin{cases}
    \schur_{[-Q_c,0]}(y_1^{-1})- y_{2,1}(y_{1,1}y_{1,2})^{-1} \schur_{[-Q_c-1,0]}(y_1^{-1})\quad &\text{if }\, Q_c<0~,\\
  1  \quad &\text{if }\, Q_c=0~,\\
    y_{2,1}   \quad &\text{if }\, Q_c=1~,\\ 
    -y_{1,1}y_{1,2}  \schur_{[Q_c-2,0]}(y_1)+ y_{2,1} \schur_{[Q_c-1,0]}(y_1)    \quad &\text{if }\, Q_c>1~.\\
\end{cases}
\ee
Let us now see how we can recover these states in the Coulomb phase, and how the duality is realised in that phase. This duality between two abelian theories can be summarised as:
\bea
&{\left[U(1)~, \, (n_I)=(2,1,0)~,\, \begin{array}{c}
  Q_c  \\
 \zeta>0    
\end{array} \right]\qquad \longleftrightarrow} \\
&\qquad\qquad {- (y_{1,1}y_{1,2})^{Q_c} y_{2,1}\times \left[U(1)~, \, (n_I')=(0,2,1)~,\, \begin{array}{c}
  Q_c'=-Q_c-1  \\
 \zeta'<0   
\end{array}\right]\Bigg|_{\begin{array}{c}
     y_2\rightarrow y_1  \\
      y_3\rightarrow y_2
\end{array}}~,}
\eea
where we included the non-trivial relative fermion number and flavour Chern--Simons terms, and indicated the change of flavour parameters.

\medskip
\noindent 
{\bf Coulomb-branch states when $Q_c<0$.} For $\zeta>0$ and $Q_c<0$ in the original theory, there are no states in the exterior region and we can thus focus on the interior states
\be
\ket{s_1, s_2, \gamma}_{++} \equiv  \ket*{(s_{1,1},s_{1,2}), s_2, (\gamma_{1,1},\gamma_{2,1})}_{++}~,\qquad  |s_1|-s_2=-Q_c~.\label{(2,1,0)_non_purt_GS}
\ee
Most of this infinite tower of states are lifted by the $E$-term interactions. The supercharge $\boldsymbol{Q}_1$ acts as:
\bea
&{\ket*{(s_{1,1},s_{1,2}), s_2, (0,0)}_{++} \longrightarrow} && {\ket*{(s_{1,1}-1,s_{1,2}), (s_2-1), (1,0)}}_{++} \\
&&&{\quad + \ket*{(s_{1,1},s_{1,2}-1), (s_2-1), (0,1)}_{++}~,}\\
&{\ket*{(s_{1,1},s_{1,2}), s_2, (1,0)}_{++} } \longrightarrow &&{-\ket*{(s_{1,1},s_{1,2}-1), (s_2-1), (1,1)}_{++}~,}\\
&{\ket*{(s_{1,1},s_{1,2}), s_2, (0,1)}_{++}} \longrightarrow &&{\ket*{(s_{1,1}-1,s_{1,2}), (s_2-1), (1,1)}_{++}~,}\\
&{\ket*{(s_{1,1},s_{1,2}), s_2, (1,1)}_{++}} \longrightarrow && {0~,}\\ \label{(2,1,0) Q_1 pairings}
\eea
Let $F_1=(F_{1,1},F_{1,2})\in \Z^2$ and $F_2\in \Z$ denote the flavour superselection sectors as in~\eqref{def F1 F2}. One can check can the only flavour sectors that are not entirely lifted by the $E$-term are the sectors $F_1=(Q_c+k,-k)$ with $0\leq k\leq -Q_c$, $F_2=0$,  and the sectors $F_1=(Q_c+k,-1-k)$ with $0\leq k\leq -Q_c-1$, $F_2=1$, which precisely matches the $U(2)\times U(1)$ representations that appear on the first line of \eqref{IW full 210}. The first set of flavour sectors are one-dimensional even before the interactions are turned on, and therefore contribute the states:
\be\label{states Qcneg 201 i}
\ket*{(s_{1,1},s_{1,2}), 0, (0,0)}_{++} = {(a_{1,1}^\dagger)^{s_{1,1}}(a_{1,2}^\dagger)^{s_{1,2}}\ov \sqrt{s_{1,1}!s_{1,2}!}}\ket{0}~, \qquad s_{1,1}+s_{1,2}=-Q_c~.
\ee
The second set of flavour sectors form three-dimensional Hilbert spaces in the free theory:
\bea
&{\ket{b_1}\equiv \ket*{(1+s_{1,1},1+s_{1,2}), 1, (0,0)}_{++} }~,\qquad
&& {\ket{f_1}\equiv \ket*{(s_{1,1},1+s_{1,2}), 0, (1,0)}_{++} }~,\\
&&&  {\ket{f_2}\equiv \ket*{(1+s_{1,1},s_{1,2}), 0, (0,1)}_{++} }~,
\eea
for all $(s_{1,1}, s_{1,2})$ such that $s_{1,1}+ s_{1,2}=|Q_c|-1$. 
We have $\boldsymbol{Q}_1 \ket{b_1}= \ket{f_1}+\ket{f_2}$, $\boldsymbol{Q}_1 \ket{f_1}=0$ and $\boldsymbol{Q}_1 \ket{f_2}=0$, so the $E$-term deformation leaves us with a single ground state in each flavour sector:
\be\label{states Qcneg 201 ii}
\ket{f_1}-\ket{f_2} =  {(a_{1,1}^\dagger)^{s_{1,1}}(a_{1,2}^\dagger)^{s_{1,2}}\ov \sqrt{s_{1,1}!s_{1,2}!}}\left(\Gamma^1_1 a_{1,2}^\dagger-\Gamma^1_2 a_{1,1}^\dagger\right)\ket{0}~, \qquad s_{1,1}+s_{1,2}=-Q_c-1~.
\ee
The states~\eqref{states Qcneg 201 i} and~\eqref{states Qcneg 201 ii} obviously match the cohomology classes~\eqref{Hcoh 210} for $Q_c<0$. 

Now consider the dual theory with $\zeta'<0$,  $(n_I')=(0,2,1)$ and $Q_c'=-Q_c-1$. Since we have a single fundamental fermi multiplet, we just need to consider two sets of theories with $(n_I'')=(0,2,0)$ with shifted $Q_c'$, which gives us the states:
\be
\eta_{3,1}^p \ket{\Psi; Q_c''=-Q_c-1+p}~, \qquad p=0,1~.
\ee
Note that, while we keep $\eta_{3,k}$ as the notation for the fundametanl fermis in the dual theory, they are now labelled by the $U(n_2)$ indices (here with $n_2=1$). Consider the case $p=0$ first. We have $Q_c'' =-Q_c-1 \geq n_1''=0$ and therefore we are in the situation discussed in subsection~\ref{subsec:fund only} up to a charge conjugation operation. In the $(0,2,1)$ model, this provides us with the following perturbative states at $\sigma<0$:
\be\label{dual states 021 i}
 {(a_{2,1}^\dagger)^{s_{2,1}}(a_{2,2}^\dagger)^{s_{2,2}}\ov \sqrt{s_{2,1}!s_{2,2}!}} \ket{0'}~,\qquad\quad s_{2,1}+s_{2,2}=-Q_c-1~,
\ee
which precisely matches the states~\eqref{states Qcneg 201 ii} once we recall the non-trivial charge of the dual vacuum:
\be
(-1)^{\rm F} Y \ket{0'} = - (y_{1,1}y_{1,2})^{Q_c} y_{2,1}\ket{0}~\label{(2,1,0)_dual_vac},
\ee
and take into account that the oscillators in~\eqref{dual states 021 i} are for antifundamental fields (despite the relabelling $y_{2,j}\rightarrow y_{1,i}$). At the level of the Witten index, one reproduces the expected result once we use the identity
\be
(y_{1,1}y_{1,2})^{Q_c} \schur_{[-Q_c-1,0]}(y_1) = (y_{1,1}y_{1,2})^{-1} \schur_{[-Q_c-1,0]}(y_1^{-1})~.
\ee
Similarly, when $p=1$ we have $Q_c''=-Q_c$ and this gives us the states:
\be
 {(a_{2,1}^\dagger)^{s_{1,1}}(a_{2,2}^\dagger)^{s_{2,2}}\ov \sqrt{s_{2,1}!s_{2,2}!}} \eta_{3,1}\ket{0'}~,\qquad\quad s_{2,1}+s_{2,2}=-Q_c~,
 \ee
which precisely matches the states~\eqref{states Qcneg 201 i}. We have thus checked the duality at the level of the Coulomb-phase ground states.

\medskip
\noindent 
{\bf Coulomb-branch states when $Q_c>1$.} For $\zeta>0$ and $Q_c>1$,  the original theory has Coulomb-branch states at $|\sigma|<m$ and at $\sigma<-m$. Let us consider those in turn.

In the interior region, we have the states~\eqref{(2,1,0)_non_purt_GS} with $Q_c>0$, which means that $s_2$ is fully fixed for any choice of $s_{1,i}\geq 0$ since $s_2= |s_1|+Q_c$. Then one can check using the pairing \eqref{(2,1,0) Q_1 pairings}  that the only state not lifted by the $E-$term interaction is $s_1=(0,0)$, namely:
\be\label{interior state 210 F2Qc}
    \ket{(0,0),Q_c,(0,0)}_{++}~, \qquad \text{with }\, F_1=(0,0)~, \; F_2=Q_c~. 
\ee
In the exterior region with  $\sigma<-m$, we have the states:
\be
    \b\lambda\ket{(t_{1,1},t_{1,2}),s_2,(\gamma_{1,1},\gamma_{1,2})}_{-+},\qquad |t_1|+s_2=Q_c-2~.
\ee
Turning on the $E$-term, the supercharge $\boldsymbol{Q}_1$ acts as:
\bea
&{\ket*{(t_{1,1},t_{1,2}), s_2, (0,0)}_{-+} \longrightarrow} && {\ket*{(t_{1,1}+1,t_{1,2}), (s_2-1), (1,0)}}_{-+} \\
&&&{\qquad +\, \ket*{(t_{1,1},t_{1,2}+1), (s_2-1), (0,1)}_{-+}~,}\\
&{\ket*{(t_{1,1},t_{1,2}), s_2, (1,0)}_{-+} } \longrightarrow &&{-\ket*{(t_{1,1},t_{1,2}+1), (s_2-1), (1,1)}_{-+}~,}\\
&{\ket*{(t_{1,1},t_{1,2}), s_2, (0,1)}_{-+}} \longrightarrow &&{\ket*{(t_{1,1}+1,t_{1,2}), (s_2-1), (1,1)}_{-+}~,}\\
&{\ket*{(t_{1,1},t_{1,2}), s_2, (1,1)}_{-+}} \longrightarrow && {0~.}\\
\eea
The ground states which survive the $E-$term deformation fall into three flavour sectors indexed by $F_2$:~%
\footnote{Here the notation for the $F_2=1$ states is such that, for the states with either $t_{1,1}=0$ or $t_{1,2}=0$, there is only one state in the linear combination, while there are two states contributing otherwise.}
\bea
 & F_2=0~:&&\; {\b\lambda\ket{(t_{1,1},t_{1,2}),0,(0,0)}_{-+}}~,\qquad & t_{1,1}+t_{1,2}=Q_c-2~,\\
 & F_2=1~:&&\;{\begin{array}{l}
  { \b\lambda\ket{(t_{1,1},t_{1,2}-1),0,(1,0)}_{-+}}   \\  
   {\qquad - \, \b\lambda\ket{(t_{1,1}-1,t_{1,2}),0,(0,1)}_{-+}~,}
 \end{array}} 
 \qquad & t_{1,1}+t_{1,2}=Q_c-1~,\\
 & F_2=Q_c~:&&\; {\b\lambda\ket{(0,0),Q_c-2,(1,1)}_{-+}}~.\qquad & \\
\eea
The $F_2=0$ sector has $F_1=(t_{1,1}+1, t_{1,2}+1)$ and reproduces the first term in the Witten index on the last line of~\eqref{IW full 210}, while the $F_2=1$ sector has $F_1=(t_{1,1},t_{1,2})$ and reproduces the second term in the Witten index. The third state with $F_2=Q_c-2$ has the same flavour charges as the unique interior state~\eqref{interior state 210 F2Qc}, but it is fermionic while the latter is bosonic due to the $\bar\lambda$ oscillator. Therefore, we expect that the pair
\be
 \ket{(0,0),Q_c,(0,0)}_{++}~, \qquad \b\lambda\ket{(0,0),Q_c-2,(1,1)}_{-+}
\ee
is lifted by non-perturbative effects. This is corroborated with the fact that there are no  ground states matching these two perturbative ground states in the dual description, which we discuss next.

In the dual abelian gauge theory with $(n_I')=(0,2,1)$ and $\zeta'<0$, we have $Q_c'+p=-Q_c-2+p \leq -2=-n_2'$ for $p=0,1$, hence we only find states in the $\sigma>0$ sector. They take the form
\be
 \b\lambda  \eta_{3,1}^p \ket{(t_{2,1},t_{2,2})}_+\qquad \text{with }\qquad \ket{(t_{2,1},t_{2,2})}_+ \equiv \frac{(b^\dag_{2,1})^{t_{2,1}}}{\sqrt{t_{2,1}!}} \frac{(b^\dag _{2,2})^{t_{2,2}}}{\sqrt{t_{2,2}!}}\b\psi_{2,1}\b\psi_{2,2}\ket{0'}~,
\ee
with the Gauss law constraint $|t_2|+2+p+Q_c'$. We thus find states in the two sectors $F_2= p=0,1$, namely:
\bea\label{021dualstates Qcpos}
&F_2= 0\; : \; \quad  && {\b\lambda \ket{(t_{2,1},t_{2,2})}_+~,} \qquad  &t_{2,1}+t_{2,2}= Q_c-1~,\\
&F_2= 1\; : \; \quad  && {\b\lambda\eta_{3,1} \ket{(t_{2,1},t_{2,2})}_+~,} \qquad  &t_{2,1}+t_{2,2}= Q_c-2~.
\eea
Recalling the shift of quantum numbers~\eqref{(2,1,0)_dual_vac}, one easily checks that the states~\eqref{021dualstates Qcpos} precisely matches the states in the original theory, with the duality map:
\bea
  & {\b\lambda\ket{t_1,0,(0,0)}_{-+}} \, &\rightsquigarrow\quad  & 
 {\b\lambda  \eta_{3,1} \ket{t_1}_+~,} \quad & |t_1|=Q_c-2~,\\
    &{\b\lambda\ket{t_1-\delta_2,1,(1,0)}_{-+}-\b\lambda\ket{t_1-\delta_1,1,(0,1)}_{-+} } \, &\rightsquigarrow\quad  & 
  {\b\lambda \ket{t_1}_+~,} \quad & |t_1|=Q_c-1~,\\
\eea
where we relabelled the flavour indices $t_2\equiv (t_{2,1}, t_{2,2})\rightarrow t_1\equiv (t_{1,1},t_{1,2})$ in the dual theory, and used the same shorthand notation as in subsection~\ref{subsec:defpertext}.

\subsection{Case $(n_1,n_2,n_3)=(2,2,2)$ with trivial wall-crossing and triality}
Consider the case $n_1=n_2=n_3=2$, where the dualities are self-similar and amount to permuting the three $U(2)$ flavour groups. We also choose $Q_c=-1$ (that is, $q_c=0$), which is the only value for which we have trivial wall-crossing, and we have a full-fledged triality. The full Witten index reads:
\bea\label{WI222}
&\IW[1,-1,(2,2,2)] =\\
&\quad \IW[1,-1,(2,2,0)]-\left({1\ov y_{3,1}}+{1\ov y_{3,2}}\right)\IW[1,0,(2,2,0)]+{1\ov y_{3,1}y_{3,2}}\IW[1,1,(2,2,0)]~,
\eea
where we decomposed the index as in~\eqref{IW expand in p full}, and we have:
\bea
&(p=0)\quad&& \IW[1,-1,(2,2,0)]&=&\;\, \frac{1}{y_{1,1}}+\frac{1}{y_{1,2}}  -\frac{y_{2,1}+y_{2,2}}{y_{1,1} y_{1,2}}~,\\
&(p=1)\quad&& \IW[1,0,(2,2,0)]&=&\; \, 1-\frac{y_{2,1} y_{2,2}}{y_{1,1} y_{1,2}}~,\\
&(p=2)\quad&& \IW[1,0,(2,2,1)]&=&\; \, y_{2,1}+y_{2,2}-\left(\frac{1}{y_{1,1}}+\frac{1}{y_{1,2}}\right)y_{2,2} y_{2,1}~.
\eea
The triality of the $(2,2,2)$ model is reflected in the following identities for the Witten index viewed as a function of the flavour fugacities:
\be\label{IW identity 222}
\IW(y_1, y_2, y_3)= {\det(y_2)\ov \det(y_1)}\IW(y_3, y_1, y_2)=
{\det(y_2)\ov \det(y_3)}\IW(y_2, y_3, y_1)~.
\ee
From the Higgs-branch perspective, and setting $\zeta>0$ without any real loss of generality, we have the geometry:
\be\label{XE222}
X_+= \mathbb{P}^1~, \qquad \CE_+ = \bigoplus_{q, p=0}^2  \CO(1-p-q)\otimes \Lambda^q\big(\C^2_{y_2}\big)\otimes \Lambda^p\big(\C^2_{y_3}\big)~.
\ee
 One easily checks that the equivariant sheaf cohomology
\be
H_T^\bullet(\mathbb{P}^1,\CE_+) \cong \bigoplus_{\substack{q,p=0\\ \omega \text{ regular}}}^2 H^{\ell(\sigma)}(\mathbb{P}^1,\CO(1-p-q)) \otimes \Lambda^q\big(\C^2_{y_2}\big)\otimes \Lambda^p\big(\C^2_{y_3^{-1}}\big)
\ee
gives us states that reproduce the Witten index~\eqref{WI222}. They are all listed in table~\ref{tab: states 222 model}.

\begin{table}[t]
\renewcommand{\arraystretch}{1.2}
\centering
\be\nn
 \begin{array}{|c||c|c|}
 \hline
\begin{array}{c} (q,p)\\ \nu= [\nu_1,\nu_2] \end{array}&~  ~\text{ Higgs phase } & ~ \text{ Coulomb phase } \\ \hline \hline
\begin{array}{c} (0,0)\\ \nu= [1,0] \end{array} & 
H^0_T(\mathbb{P}^1, \CO(1))\cong \C^2_{y_1^{-1}}  & 
a_{1,i}^\dagger \ket{0} \\ \hline
\begin{array}{c} (1,0)\\ \nu= [1,1] \end{array} &   
\left(H^0_T(\mathbb{P}^1, \CO) \cong \Lambda^2(\C^2_{y_1^{-1}}) \right)\otimes  \C^2_{y_2}   & 
\left(\Gamma_1^j a_{1,2}^\dagger-\Gamma_2^j a_{1,1}^\dagger\right) \ket{0} \\ \hline
\begin{array}{c} (0,1)\\ \nu= [0,0] \end{array} &   
\left(H^0_T(\mathbb{P}^1, \CO) \cong 1 \right)\otimes \C^2_{y_3^{-1}}   & 
\eta_{3,k} \ket{0} \\ \hline
\begin{array}{c} (2,1)\\ \nu= [1,1] \end{array} &   
\begin{array}{c}  \left(H^1_T(\mathbb{P}^1, \CO(-2)) \cong  \Lambda^2(\C^2_{y_1^{-1}}) \right)\\
\qquad\otimes \Lambda^2(\C^2_{y_2})\otimes\C^2_{y_3^{-1}} \end{array}  & 
\begin{array}{c}  \Big(\Gamma_2^2 a_{1,1}^\dagger a_{2,1}^\dagger-\Gamma_2^1 a_{1,1}^\dagger a_{2,2}^\dagger\\
\quad - \Gamma_1^2 a_{1,2}^\dagger a_{2,1}^\dagger+\Gamma_1^1 a_{1,2}^\dagger a_{2,2}^\dagger\Big) \eta_{3,k}\ket{0}\;\;\end{array} \\ \hline
\begin{array}{c} (1,2)\\ \nu= [0,0] \end{array} &   
\left(H^1_T(\mathbb{P}^1, \CO(-2)) \cong  1 \right)\otimes\C^2_{y_2} \otimes \Lambda^2(\C^2_{y_3^{-1}})  & 
a^\dagger_{2,j} \eta_{3,1}\eta_{3,2}\ket{0} \\ \hline
\begin{array}{c} (2,2)\\ \nu= [1,0] \end{array} &   
\begin{array}{c}  \left(H^1_T(\mathbb{P}^1, \CO(-3)) \cong  \C^2_{y_1^{-1}} \right) \\
\qquad \otimes\Lambda^2(\C^2_{y_2}) \otimes \Lambda^2(\C^2_{y_3^{-1}})\end{array}  & 
\left(\Gamma_i^2 a^\dagger_{2,1}-\Gamma_i^1 a^\dagger_{2,2}\right) \eta_{3,1}\eta_{3,2}\ket{0}  \\ \hline
\end{array}
\ee
\caption{Ground states for the $U(1)_0$, $(n_I)=(2,2,2)$ model ($Q_c=-1$), indexed by $(\lambda,\mu)=(q,p)$ in the notation of the geometric phase $(X_+,\CE_+)$ at $\zeta>0$. Note that the fermion number in the geometric description in terms of $H_T^k(\mathbb{P}^1, \CF_{q,p}(1))$ is $(-1)^{p+q+k}$. The Coulomb-branch ground states shown here are those that survive the $E$-term deformation; they all live in the interior region $|\sigma|<m$.}
\label{tab: states 222 model}
\end{table}

In the Coulomb phase, this model only admits perturbative ground states in the interior region $|\sigma|<m$, which are also listed in table~\ref{tab: states 222 model}. These states are derived by a straightforward application of the formalism of section~\ref{subsec:pert theory GammaSQED}. The fermi multiplets $\eta_{3,k}$ dress the states of the $n_3=0$ theory according to:
\be
 \ket{s_1, s_2, \gamma}_{++} \; (p=0)~, \qquad 
  \eta_{3,k}\ket{s_1, s_2, \gamma}_{++} \; (p=1)~, \qquad 
  \eta_{3,1}\eta_{3,2}\ket{s_1, s_2, \gamma}_{++} \; (p=2)~, 
\ee
and we simply have to work out the ground states $\ket{s_1, s_2, \gamma}_{++}$ for the $\Gamma$-SQED model with $(n_I)=(2,2,0)$ with $Q_c=-1+p$. Let us illustrate this computation in the case $p=1$, namely for the $(n_I)=(2,2,0)$ model with $Q_c=0$. The interior ground states are indexed by $s_1$, $s_2$ and $\gamma$ according to:
\be
\ket{s_1, s_2, \gamma}_{++}\equiv  \ket*{{\footnotesize (s_{1,1}, s_{1,2}),(s_{2,1}, s_{2,2}), \mat{\gamma_{1,1}& \gamma_{1,2}\\ \gamma_{2,1}&\gamma_{2,2}}}}_{++}~, \qquad |s_1|=|s_2|~.
\ee
One can check that most states are lifted by the $E$-term interactions, with only two states remaining (they are listed as $(q,p)=(0,1), (2,1)$ in table~\ref{tab: states 222 model}). The state $(q,p)=(0,1)$ arises from the flavour sector $F_1=F_2=(0,0)$, which is already one-dimensional:
\be
\ket*{{\footnotesize (0,0),(0,0), \mat{0&0\\ 0&0}}}_{++}~.
\ee
The state $(q,p)=(2,1)$ is more interesting. It arises from the sector $F_1=(-1,-1)$, $F_2=(1,1)$, which contains seven states:
\bea
&{\ket{b_1}\equiv \ket*{{\footnotesize (1,1),(1,1), \mat{0&0\\ 0&0}}}_{++}~,}
\qquad&{\ket{f_1}\equiv \ket*{{\footnotesize (1,0),(1,0), \mat{0&0\\ 0&1}}}_{++}~,}\\
&{\ket{b_2}\equiv \ket*{{\footnotesize (0,0),(0,0), \mat{1&0\\ 0&1}}}_{++}~,} 
\qquad& {\ket{f_2}\equiv \ket*{{\footnotesize (1,0),(0,1), \mat{0&0\\ 1&0}}}_{++}~,}\\
&{\ket{b_3}\equiv \ket*{{\footnotesize (0,0),(0,0), \mat{0&1\\ 1&0}}}_{++}~,}
\qquad&{\ket{f_3}\equiv \ket*{{\footnotesize (0,1),(1,0), \mat{0&1\\ 0&0}}}_{++}~,}\\
& &{\ket{f_4}\equiv \ket*{{\footnotesize (0,1),(0,1), \mat{1&0\\ 0&0}}}_{++}~,}\\
\eea
The supercharge $\boldsymbol{Q}_1$ acts as:
\bea
 &{\ket{b_1} \longrightarrow  \ket{f_1}+\ket{f_2}+\ket{f_3}+\ket{f_4}~,} \qquad &
&{ \ket{f_1}\longrightarrow \ket{b_2}~,} \\
 &{\ket{b_2} \longrightarrow  0}~, \qquad & 
& { \ket{f_2}\longrightarrow \ket{b_3}~,} \\
  &{\ket{b_3} \longrightarrow  0}~, \qquad & 
 & { \ket{f_3}\longrightarrow -\ket{b_3}~,} \\
  & &&  { \ket{f_4}\longrightarrow -\ket{b_2}~,}
\eea
and the only surviving state is then  
 $\ket{f_1} - \ket{f_2}- \ket{f_3}+ \ket{f_4}$, 
which is what is shown in table~\ref{tab: states 222 model}. Similar computations for $p=0$ and $p=2$ determine all the other states of the  $(n_I)=(2,2,2)$ model.  

Finally, it is interesting to note the explicit map of Coulomb-branch states under the triality. Using the notation of table~\ref{tab: states 222 model}, we have the map:
\be
\begin{array}{l||c|c|c|c|c|c}
(q,p) & (0,0) & (1,0) & (0,1) & (2,1)& (1,2)&(2,2) \\ \hline
\text{right-mutated}& (1,2) & (0,1) & (2,2) & (0,0)& (2,1)&(1,0) \\ \hline
\text{left-mutated} & (2,1) & (2,2) & (1,0) & (1,2)& (0,0)&(0,1)\\
\end{array}
\ee
This is easily derived from the identity~\eqref{IW identity 222} for the flavoured Witten indices. Note that the left and right mutations are inverse of each other, as the $(2,2,2)$ model is self-dual and we have $q_c=0$. Hence we have a genuine triality (a duality operation of order $3$) that permutes the ground states --- indeed this model can be viewed as a dimensional reduction of a 2d $\CN=(0,2)$ supersymmetric gauge theory  which itself admits a triality~\cite{Gadde:2013lxa}. The duality map is quite non-trivial, as for instance we have
\be
a_{1,i}^\dagger \ket{0} \, \rightsquigarrow\, a^\dagger_{2,j} \eta_{3,1}\eta_{3,2}\ket{0}~,\qquad\quad
 \eta_{3,k} \ket{0}\, \rightsquigarrow\, \left(\Gamma_i^2 a^\dagger_{2,1}-\Gamma_i^1 a^\dagger_{2,2}\right) \eta_{3,1}\eta_{3,2}\ket{0}~,
\ee
which suggests a very non-linear map of the fields in the path integral, as it were. Yet it might be possible to derive our dualities directly in the UV, perhaps along the line of known derivations for abelian dualities~\cite{Kapustin:1999ha}). This point deserves further investigation. 

\acknowledgments 
We are grateful to Wei Gu, Osama Khlaif, Heeyeon Kim, Simone Rota, Eric Sharpe, Hao Zhang and Hao Zou for useful correspondence and discussions. CC also acknowledges helpful interactions with Claude.AI related to the Borel--Weil--Bott theorem. CC is a Royal Society University Research Fellow. The work of JW is supported by the School of Mathematics at the University of Birmingham.

\appendix

\section{The flavoured Witten index and dualities}\label{app:Witten index}

\subsection{The Witten index of gauged SQM}
The 1d $\CN=2$ supersymmetric index was obtained in~\cite{Hori:2014tda} by localising the path integral of the SQM on the circle. The resulting JK residue is closely related to the one for the elliptic genus of 2d $\mathcal{N}=(0,2)$ supersymmetric gauge theories~\cite{Benini:2013xpa}, and the former can be formally obtained from the latter by dimensional reduction on a circle. 

Consider a generic 1d $\CN=2$ gauge theory with a vector multiplet transforning in the adjoint representation of a gauge group $G$, chiral multiplets $\Phi$ in representations $\FR_\Phi$ of $G$, and fermi multiplets $\bH$ in representations $\FR_\bH$. The one-loop determinant for the matter fields read:
\be
Z^{\rm matter}(x,y)= \prod_{\Phi}\prod_{\rho_\Phi\in\mathcal{R}_\Phi} \frac{1}{1-x^{\rho_\Phi}y_{\Phi}^{-1}}\prod_{\bH}\prod_{\rho\in\mathcal{R}_{\bH}}(1-x^{\rho_\bH} y_{\bH}^{-1})~,
\ee
where $x^{\rho}\equiv \prod_{\alpha=1}^{\text{rank}(G)} x_\alpha^{\rho^\alpha}= e^{2\pi i \rho(u)}$. We used the conventions of subsection~\ref{subsec: quantize free fields}, which differ from those of~\cite{Hori:2014tda} in how we treat the 1d parity anomaly. Here $y_\Phi$ and $y_\bH$ denote fugacities for the various distinct chiral and fermi multiplets (transforming in irreducible representations of the gauge group).

To fully specify the 1d $\CN=2$ supersymmetric gauge theory with gauge group $G$, we must specify a Wilson line insertion for some representation $\FR_0$ of $G$, which is the insertion of some Wilson line $W_{\FR_{0}}$ in the (Euclidean) path integral:
\be
W_{\FR_{0}}(x) = {\rm Tr}\, P e^{-i \int d\tau (A_\tau -i \sigma)} \cong \sum_{\rho_0 \in \FR_0} x^{\rho_0}~.
\ee
In this paper, we focus on $G=U(N_c)$ and on the one-dimensional representation $\FR_0= {\bf det}^{Q}$, which gives us the 1d Chern--Simons interaction~\eqref{contact term Q}.  The Witten index is obtained as the JK residue formula~\cite{Hori:2014tda}:
\be
  \IW(y)=   \frac{1}{|W_G|}\int_{\rm JK} \prod_{\alpha=1}^{\text{rank}(G)} \left[- \frac{dx_{\alpha}}{2\pi i x_\alpha}\right] W_{\FR_{0}}(x) \prod_{\delta\in \Delta_G}(1-x^{\delta}) Z^{\rm matter}(x,y)~,
\ee
where $\Delta_G$ is the root system of the gauge group $G$ and $|W_G|$ is the order of the Weyl group of $\Fg= {\rm Lie}(G)$. The JK residue prescription generally depends on the Fayet--Iliopoulos parameters.

. 
\subsection{Proving the equality of the Witten indices}\label{subsec:proof WI}
The flavour Witten index for $\Gamma-$SQCD as defined in section~\ref{subec:defSQCD} is given by:
\bea\label{app:IW SQCD}
&{\IW[N_c, Q_c, (n_1, n_2, n_3),(0,0,0)]^{\sign(\zeta)=\pm}_\Gamma=}\\
&\qquad\qquad\quad {\frac{(-1)^{N_c}}{N_c!}\int_{{\rm JK}_\pm}\prod_{\alpha=1}^{N_c}\left[ {dx_\alpha\ov 2\pi i x_\alpha} x_\alpha^{Q_c} \right] \prod_{\substack{\alpha, \beta=1\\ \alpha\neq \beta}}^{N_c}(1-x_\alpha x_{\beta}^{-1}) Z^{\rm matter}(x,y)~,}
\eea
with the matter one-loop determinant being:
\be
Z^{\rm matter}(x,y) = \prod_{i=1}^{n_1}\prod_{j=1}^{n_2}\left(1-y_{2,j}y_{1,i}^{-1}\right) 
\prod_{\alpha=1}^{N_c}\left[ {\prod_{k=1}^{n_3} \left(1- {x_\alpha \ov y_{3,k}}\right) \ov \prod_{i=1}^{n_1} \left(1- {x_\alpha \ov y_{1,i}}\right) \prod_{j=1}^{n_2} \left(1- {y_{2,j} \ov x_\alpha}\right)}\right]~.
\ee
The JK residue picks all the iterated residues at $x_\alpha= y_{1,i}$ if $\zeta>0$, and all the poles at $x_\alpha=y_{2,j}$ if $\zeta<0$. When considering generic flavour fugacities, all these poles from the chiral multiplets are simple poles. The W-boson contribution in~\eqref{app:IW SQCD} means that any choice of simple poles such that $x_\alpha=x_\beta$ for $\alpha\neq \beta$ gives a vanishing contribution. This together with the $S_{N_c}$ Weyl symmetry that permutes the $x_\alpha$'s implies that we can write the index as a sum over the sets 
\begin{align}
    \CS_{n_l,  N_c}:=\{S\subset[1,n_l]\cap \mathbb{Z}:|S|=N_c\},
\end{align}
where $l=1$ for $\zeta>0$ and $l=2$ for $\zeta<0$. 
The expression for the index for a theory with $\zeta>0$ can then be written as a sum over $S\in \CS_{n_1,N_c}$ of the integrand with iterated residues (for $\alpha=1,2,...N_c$) at $x_\alpha=y_{1,S_\alpha}$, where $S_\alpha$ denotes the $\alpha^{th}$ element of the ordered set $S$. Explicitly, for $\zeta>0$, we can write the index as:
\begin{align}
    \IW[N_c, q_c, (n_1, n_2, n_3),(0,0,0)]_\Gamma^+=\\ \prod_{i=1}^{n_1}\prod_{j=1}^{n_2}\left(1-y_{2,j}y_{1,i}^{-1}\right) 
    \sum_{S\in \mathcal{S}_{n_1,N_c}}\prod_{\alpha=1}^{N_c}y_{1,i_\alpha}^{Q_c}\Bigg(\prod_{i=1,i\neq i_\alpha}^{n_1}&\frac{1}{1-y_{1,i_\alpha}y_{1,i}^{-1}}\prod_{j=1}^{n_2}\frac{1}{1-y_{2,j}y_{1,i_\alpha}^{-1}}\notag \\
     \prod_{k=1}^{n_3}&(1-y_{1,i_\alpha}y_{3,k}^{-1})\prod_{\beta\neq \alpha}^{N_c}(1-y_{1,i_\alpha}y_{1,i_{\beta}}^{-1})\Bigg)~.\notag
\end{align}
First, consider the case where $n_1=N_c$, in which case the right-mutated theory is a free theory; in the original theory, $S_{n_1,n_1}$ then only contains the set $[1,n_1]$, and therefore $i_\alpha=\alpha$, and the our expression for the index can be written as:
\begin{align}
    &\IW[N_c=n_1, Q_c, (n_1, n_2, n_3),(0,0,0)]^\Gamma_+\\ &=\prod_{i=1}^{n_1}\prod_{j=1}^{n_2}\left(1-y_{2,j}y_{1,i}^{-1}\right)\left(\frac{1}{1-y_{2,j}y_{1,i}^{-1}}\right)\\
    &\times \prod_{\alpha=1}^{n_1}y_{1,\alpha}^{Q_c}\prod_{i=1,i\neq \alpha}^{n_1}\left(\frac{1}{1-y_{1,\alpha}y_{1,i}^{-1}}(1-y_{1,\alpha}y_{1,i}^{-1})\right)\prod_{k=1}^{n_3}(1-y_{1,\alpha}y_{3,k}^{-1})\\
    &=\prod_{i=1}^{n_1}y_{1,i}^{Q_c}\prod_{k=1}^{n_3}(1-y_{1,i}y_{3,k}^{-1})
    \\
    &=\IW[0,Q_c',(n_3,n_1,n_2),(0,Q_c,0)]^{\Gamma'}_-
\end{align}
as claimed in equation\eqref{duality indentity right mutation} with $n_1=N_c$, noting that the parameter $Q_c'$ is superfluous in the free theory --- the dual theory contains only $n_1n_3$ free mesonic fermi multiplets and the shifted flavour CS level.

For the cases $n_l>N_c$, we can use the one to one correspondence 
\begin{align}
    S\in \CS_{n_l,N_c}\leftrightarrow S^c\in \CS_{n_l,n_l-N_c} \label{Set_parings}
\end{align}
alongside the notation 
\begin{align}
    S^c= \{i_{\alpha'}^c\}_{\alpha'=1}^{n_1-N_c} \in \CS_{n_1, n_1-N_c}~.
\end{align}
to concisely combine the fundamental chiral contributions with that the vector multiplet contribution, as follows:
\begin{align}
    \prod_{\alpha=1}^{N_c}\left[\prod_{i=1,i\neq i_\alpha}^{n_1}\frac{1}{1-y_{1,i_\alpha}y_{1,i}^{-1}}\prod_{\beta\neq \alpha }^{N_c}(1-y_{1,i_\alpha}y^{-1}_{1,i_\beta})\right]\\=
    \prod_{\alpha=1}^{N_c}\prod_{\alpha'=1}^{n_1-N_c}\frac{1}{1-y_{1,i_\alpha}y_{1,i^c_{\alpha'}}^{-1}}~.
\end{align}
Similarly, combining the fermi mesons contributions with that of the antifundermental chiral multiplets, we find: 
\begin{align}
    \prod_{i=1}^{n_1}\prod_{j=1}^{n_2}(1-y_{2,j}y_{1,i}^{-1})\prod_{\alpha=1}^{N_c}\prod_{j=1}^{n_2}\frac{1}{1-y_{2,j}y_{1,i_\alpha}^{-1}}=\prod_{\alpha'=1}^{n_1-N_c}\prod_{j=1}^{n_2}(1-y_{2,j}y_{1,i_{\alpha'}^c}^{-1})~.
\end{align}
We can then massage the expression for the index of the original $\zeta>0$ theory for $n_1>N_c$ into:
\begin{align}
    \IW[N_c, q_c, (n_1, n_2, n_3),(0,0,0)]^\Gamma_+=\sum_{S\in\CS_{n_1,N_c}}\prod_{\alpha=1}^{N_c}y_{1,i_\alpha}^{Q_c}\prod_{\alpha'=1}^{n_1-N_c}\frac{1}{1-y_{1,i_\alpha}y_{1,i_{\alpha'}^c}^{-1}}\\\prod_{j=1}^{n_2}\prod_{\alpha'=1}^{n_1-N_c}(1-y_{2,j}y_{1,i_{\alpha'}^c}^{-1})
    \prod_{\alpha=1}^{N_c}\prod_{k=1}^{n_3}(1-y_{1,i_\alpha}y_{3,k}^{-1})~.
\end{align}
In a similar manner, one can manipulate the expression for the index of $\Gamma-$SQCD with $\zeta<0$, which gives:
\begin{align}
    \IW&[N_c ,q_c ,(n_1 ,n_2 ,n_3 ),(0,0,0)]^{\Gamma }_-=\\
    &\frac{(-1)^{N_c }}{N_c !}\prod_{i =1}^{n_1 }\prod_{j =1}^{n_2 }\left(1-\frac{y _{2,j }}{y _{1,i }}\right) \int_{\gamma(\zeta <0)}\prod_{\alpha =1}^{N_c }x_{\alpha }^{Q_c -1}dx_{\alpha }\Bigg(\
    \\&\prod_{i =1}^{n_1 }\frac{1}{1-x_{\alpha } y_{1,i } ^{-1}}\prod_{j =1}^{n_2 }\frac{1}{1-x_{\alpha }^{-1}y_{2,j }} \prod_{k =1}^{n_3 }(1-x_{\alpha } y_{3,k } ^{-1})\prod_{\beta \neq \alpha }^{N_c }(1-x_{\alpha } x_{\beta }^{-1})\Bigg )=\notag\\&
    \prod_{i =1}^{n_1 }\prod_{j =1}^{n_2 }(1-y_{2,j } y_{1,i } ^{-1})\sum_{S \in \CS_{n_2 ,N_c }}\prod_{\alpha =1}^{N_c }y_{2,j _{\alpha }} ^{Q_c }\prod_{i =1}^{n_1 }\frac{1}{1-y _{2,j_{\alpha }^c} y_{1,i } ^{-1}}\\&
    \prod_{j \neq j _{\alpha }}^{n _2}\frac{1}{1-y_{2,j_{\alpha } } ^{-1}y_{2,j } }\prod_{k =1}^{n_3 }(1-y_{2,j_{\alpha } } y_{3,k } ^{-1})\prod_{\beta \neq\alpha }^{N_c }(1-y_{2,j _{\alpha }} y_{2,j _{\beta }} ^{-1})~.
\end{align}
Then, using the identities
\begin{align}
    \prod_{i =1}^{n_1 }&\prod_{j =1}^{n _2}(1-y_{2,j } y_{1,i } ^{-1})\prod_{\alpha =1}^{N_c }\prod_{i =1}^{n_1 }\frac{1}{1-y_{2,j _{\alpha }} y_{1,i } ^{-1}}\\&=\prod_{i =1}^{n_1 }\prod_{\alpha'=1}^{n_2 -N_c }(1-y_{2,j ^c_{\alpha'}} y_{1,i } ^{-1})
    \\\prod_{\alpha =1}^{N_c }&\prod_{j\neq j _{\alpha }}^{n_2 }\frac{1}{1-y_{2,j _{\alpha }} ^{-1}y _{2,j }}\prod_{\alpha =1}^{N_c }\prod_{\beta \neq\alpha }^{N_c }(1-y_{2,j _{\alpha }} y_{2,j _{\beta }}^{-1})\\&=(-1)^{N_c (N_c -1)}\prod_{\alpha =1}^{N_c }\prod_{\alpha'=1}^{n_2 -N_c }\frac{1}{1-y_{2,j _{\alpha }} ^{-1}y_{2,j_\alpha'^c} }\\
\end{align}
and 
\begin{align}
    \prod_{\alpha =1}^{N_c }&\prod_{k =1}^{n_3 }(1-y_{2,j _{\alpha }} y_{3,k } ^{-1})=\\&(-1)^{n_3 N_c }\prod_{\alpha =1}^{N_c }y_{2,j _{\alpha }} ^{n_3}\prod_{k =1}^{n_3 }y_{3,k } ^{-N_c }\prod_{\alpha =1}^{N_c }\prod_{k =1}^{n_3 }(1-y_{2,j_{\alpha } } ^{-1}y_{3,k} )
\end{align}
alongside the isomorphism \eqref{Set_parings}, we can write the index of the $\zeta<0$ theory as:
\begin{align}
    \IW[&N_c',q_c',(n'_1,n'_2,n'_3),(0,0,0)]^{\Gamma'}_-=\prod_{k'=1}^{n_3'}y_{3,k'}'^{-N_c'}\sum_{S'\in\CS_{n_2',N_c'}}\prod_{\alpha'=1}^{N_c'}y_{2,j'_{\alpha'}}'^{Q_c'}\\
    &\prod_{i'=1}^{n_1'}\prod_{\alpha=1}^{n_2'-N_c'}(1-y_{2,j'^c_\alpha}'y_{1,i'}'^{-1})\prod_{\alpha=1}^{n_2'-N_c'}\prod_{\alpha'=1}^{N_c'}\frac{1}{1-y_{2,j_{\alpha'}}'^{-1}y_{2,j'^c_\alpha}'}\prod_{\alpha'=1}^{N_c'}\prod_{k'=1}^{n_3'}(1-y_{2,j_{\alpha'}'}'^{-1}y_{3,k'}')~,
\end{align}
where we used the notation 
\begin{align}
    S'=\{j'_{\alpha'}\}_{\alpha'=1}^{N_c'}\in\CS_{n_2',N_c'}\;\;\text{and}\;\;S'^c=\{j'^c_\alpha\}_{\alpha=1}^{n_2'-N_c'}~.
\end{align}
By setting $N_c'=n_1-N_c$, $(n_1',n_2',n_3')=(n_3,n_1,n_2)$ and $q_c'=-q_c$ and correspondingly $(y_1',y_2',y_3')=(y_3,y_1,y_2)$, and noting that 
\begin{align}
    \prod_{i=1}^{n_1}y_{1,i}^{Q_c}\prod_{\alpha'=1}^{n_1-N_c}y_{1,i_{\alpha'}^c}^{-Q_c}=\prod_{\alpha=1}^{N_c}y_{1,i_{\alpha}}^{Q_c}
\end{align}
we then conclude that:
\begin{align}
&\IW[N_c, q_c, (n_1, n_2, n_3),(0,0,0)]_\Gamma^+ \\
&\qquad=(-1)^{n_2(n_1-N_c)}\, \prod_{i=1}^{n_1}y_{1,i}^{Q_c}\prod_{j=1}^{n_2}y_{2,j}^{n_1-N_c}\IW[n_1-N_c, -q_c,(n_3, n_1, n_2), (0, 0, 0)]^-_{\Gamma'}~.\\
&\qquad=(-1)^{n_2(n_1-N_c)}\, \IW[n_1-N_c, -q_c,(n_3, n_1, n_2), (0, Q_c, n_1-N_c)]^-_{\Gamma'}~.
\end{align}
By relabelling the variables $(n_3,n_1,n_2)\rightarrow(n_1,n_2,n_3)$, $n_1-N_c\rightarrow N_c$ and $-q_c$ one can prove the equality of the index under left mutation. Noting that the relation between $Q_c$ and $q_c$ implies
\begin{align}
    Q_c=q_c+\frac{n_1}{2}-\frac{n_2}{2}-\frac{n_3}{2}\rightarrow-q_c+\frac{n_2}{2}-\frac{n_3}{2}-\frac{n_1}{2}=-Q_c-n_3
\end{align}
we find 
\begin{align}
    &\IW[N_c,q_c,(n_1,n_2,n_3),(0,0,0)]^-_\Gamma\\
    &\qquad=(-1)^{n_3 N_c}\prod_{j=1}^{n_2}y_{2,j}^{Q_c+n_3}\prod_{k=1}^{n_3}y_{3,k}^{-N_c}\IW[n_2-N_c,-q_c,(n_2,n_3,n_1),(0,0,0)]_+^\Gamma\\
    &\qquad=(-1)^{n_3 N_c}\IW[n_2-N_c,-q_c,(Q_c+n_3,-N_c,0)]^+_\Gamma~.
\end{align}

\bibliography{1dbib} 
\bibliographystyle{JHEP}

\end{document}